\documentclass[sponsored]{acmsiggraph}

\usepackage{amsmath}
\usepackage{amssymb}
\usepackage{array}

\TOGvolume{0}
\TOGnumber{0}

\TOGarticleDOI{1111111.2222222}

\TOGprojectURL{}
\TOGvideoURL{}
\TOGdataURL{}
\TOGcodeURL{}

\title{Snaxels on a Plane}

\author{Kevin Karsch \hspace{.5in} John C. Hart \\[2ex]
University of Illinois at Urbana-Champaign \\
\{karsch1,jch\}@illinois.edu}

\pdfauthor{Kevin Karsch, John C. Hart}

\keywords{contour, line/vector art, planar map, SVG}

\begin{document}

%%% A ``teaser'' image appears under the title and affiliation information,
%%% horizontally centered, and above the two columns of text. This is OPTIONAL.
%%% If you choose to have a ``teaser'' image, it needs to be placed between
%%% ``\begin{document}'' and ``\maketitle.''

\teaser{
\centerline{
\includegraphics[width=180mm]{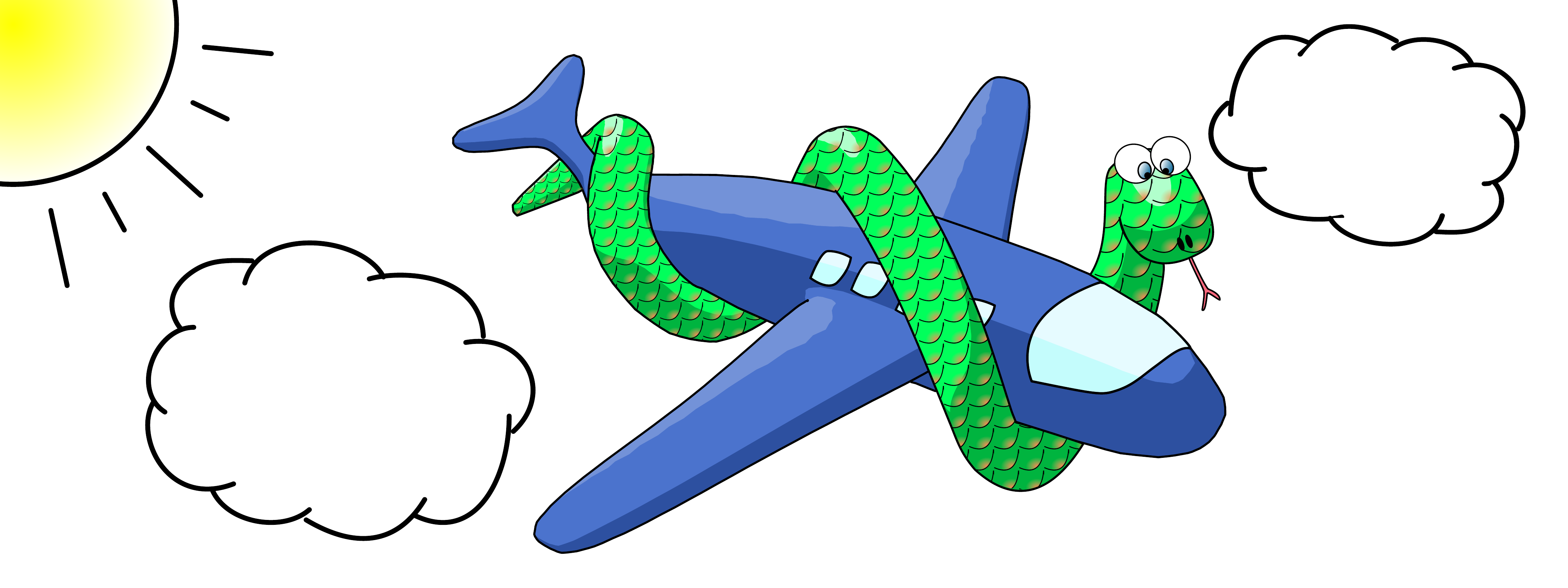}
}
\caption{This cartoon was generated by projecting snaxel fronts onto the view
plane. The snaxel fronts find and track the visual, shadow and shading
contours, and also form the planar map of the regions they bound, to simplify
vector art construction, stylization and animation.
%Tracking contours on meshes with other snake-based active contours can be
%problematic, leading some to complain, ``I've had it with these multi-front
%snakes on this meshed facet plane!''  
\vspace{-5mm}
}
\label{fig:teaser}
}

%%% The ``\maketitle'' command must appear after ``\begin{document}'' and,
%%% if you have one, after the definition of your ``teaser'' image, and
%%% before the first ``\section'' command.

\maketitle

%%% Your paper's abstract goes in its own section.

\begin{abstract} \vspace{-2mm}
While many algorithms exist for tracing various contours for illustrating a
meshed object, few algorithms organize these contours into region-bounding
closed loops. Tracing closed-loop boundaries on a mesh can be problematic due
to switchbacks caused by subtle surface variation, and the organization of
these regions into a planar map can lead to many small region components due
to imprecision and noise. This paper adapts ``snaxels,'' an energy minimizing
active contour method designed for robust mesh processing, and repurposes it
to generate visual, shadow and shading contours, and a simplified
visual-surface planar map, useful for stylized vector art illustration of the
mesh. The snaxel active contours can also track contours as the mesh animates,
and frame-to-frame correspondences between snaxels lead to a new method to
convert the moving contours on a 3-D animated mesh into 2-D SVG curve
animations for efficient embedding in Flash, PowerPoint and other dynamic
vector art platforms.
\end{abstract}

%%% ACM Computing Review (CR) categories.
%%% See <http://www.acm.org/class/1998/> for details.
%%% The ``\CRcat'' command takes four arguments.

% \begin{CRcatlist}
%   \CRcat{I.3.7}{Computer Graphics}{Three-Dimensional Graphics and Realism}{Animation};
% \end{CRcatlist}

%%% The ``\keywordlist'' command prints out the keywords.

\keywordlist

%%% The ``\TOGlinkslist'' command will insert hyperlinked icon(s) to your
%%% paper. This includes, at a minimum, hyperlinked icons to the ACM article
%%% page and the ACM Digital Library-held PDF. If you added URLs to
%%% ``\TOGprojectURL'' or the other, similar commands, they will be added to
%%% the list of icons.
%%% Note: this functionality only works for annual-conference papers.

\TOGlinkslist

%%% The ``\copyrightspace'' command 
%%% Do not remove this command.

\copyrightspace

\section{Introduction} \vspace{-2mm}

The automatic conversion of meshed surfaces into stylized vector art has become
a useful tool for generating illustrations, instructions and visual
accompaniments. Early work focused on simulating a hand-drawn sketched
appearance that could be generated from disjoint contour elements,
e.g. \cite{winkenbach94}, whereas
many more modern techniques rely on topologically correct closed-loop visual
contours whose planar map bounds image regions, e.g. \cite{Eisemann:2009}.
Such contours are useful for stylization, e.g. \cite{grabli04},
and diffusion gradients, e.g. \cite{Orzan:2008}.

While finding the elements of a visual contour from a meshed surface can be
straightforward, continuing them into region-bounding closed loops and forming
their planar map can require intricate geometric operations for correctness
and robustness, c.f.
\cite{Gangnet:1989,Asente:2007,stroila2008,Eisemann_cgf08,Eisemann:2009}.

We propose a significantly simpler approach that efficiently and robustly (1)
extracts closed-loop visual, shadow and shading contours and the regions they
define, (2) forms a visible-surface planar map of simple, mesh resolution
components and (3) provides correspondences between contours over time to
convert 3-D contour motion into 2-D curve motion, to convert illustrations of
dynamic 3-D objects into dynamic vector art.

This simpler approach is built on {\it snaxels}, a surface contour formulation
previously designed to support active contour propagation over a meshed surface
\cite{Jung:2004uf} instead of their original domain of the regular grid
structure of an image \cite{Kass:1988tk}. The original snaxel active contours
are formulated as an energy minimization over an irregular meshed domain
embedded in 3-D. Section~\ref{sec:snaxels} focuses this formulation on the
specific problem of extracting illustration contours, which simplifies its
construction and implementation. For
example, we replace the energy functional with an implicit contour function and
revise the integration appropriately. By definition, these active contours are
level sets (of regular values) and hence form and maintain closed loops. The
snaxel approach robustly tracks these closed loop contours, subdividing and
merging as necessary as the system tracks 
their shape through changes in the surface, view and lighting.

Section~\ref{sec:npr} defines implicit contour functions for the visual
contour generator as well as shadow and shading contours. It also shows how to
initialize the snaxels to capture these contours and the regions they define.
It further describes how to track these contours on a moving surface. The
snaxels detect and adapt visual contour topology across visual events, and
these events can be used to detect the presence of parabolic points on a
meshed surface.

Section~\ref{sec:planar} shows how the snaxel evolution rules can be modified
to generate a visible-surface planar map. While such a planar map can be
generated through computational geometry techniques, the snaxel implementation
is far simpler, and the mesh-constrained resolution of the snaxel formulation
naturally filters out the geometric noise of small components that
sometimes accompany precise arithmetic approaches.

Once the visual and other contours are defined, Section~\ref{sec:anim} shows
how the snaxel formulation can be used to convert the 3-D motion of
surface contours into the 2-D motion of image curves. The evolution of the
snaxel front provides the correspondences between the contours of each frame,
providing a natural framework for animating the contours and their stylization
into, e.g. animated SVG, in a form that can be compactly and
conveniently inserted into Flash, PowerPoint or other dynamic vector art
platforms.

Section~\ref{sec:conc} concludes with a summary, demonstration of our
interactive system for extracting and stylizing contours on animated meshes,
and a discussion of the directions of further research inspired by this
novel approach to contouring for illustrating meshed objects.

\section{Previous Work} \vspace{-2mm}
For a general view, the visual contours of a shape form closed loops. Numerous
techniques exist for extracting silhouettes and other contours from a
meshed object \cite{markosian97,elber98,kalnins03,decarlo2003,su05,burns05,%
olson2006,judd2007}. Many fewer methods exist that connect them into
region-bounding closed loops, which supports planar map construction and
various region stylization methods, ranging from hatching \cite{winkenbach94}
to diffusion gradients \cite{Orzan:2008}. A popular survey of silhouette
algorithms for polygonal models \cite{isenberg03} focuses on other issues than
forming closed loops and bounding regions.

The visual contour and other contours that follow edges of a mesh suffer from
well-known problems with switchbacks \cite{rusink08},
indicated by extraneous triangles
along the contour path. Past work has confronted this problem by subdividing
the mesh near the visual contour \cite{winkenbach96}, interpolating surface
normals across edges \cite{hertzmann2000} or extending a jagged shadow contour
outward into a smooth contour. As shown previously \cite{Eisemann_cgf08}, care
must be taken when tracing and integrating regions and contours extracted from
interpolated vectors, as the visual contour may not precisely surround
the projection of the surface mesh, leading to downstream region
classification problems.

The proposed snaxel method for
contour tracking likewise interpolates surface normals across the edges, but
could further smooth the contour by adding a smoothness term to the snaxel
energy functional. We extract a planar map through modified snaxel rules that
avoid the classification problems that could arise when post-processing a
planar map from the contours based on interpolated vertex normals.

Extraction of region-bounding contour loops and their planar map is complex
and has been fraught with robustness issues \cite{Gangnet:1989,Asente:2007}.
The region-robust contour loop and planar map
extraction of Eisemann et al. \cite{Eisemann_cgf08} relied on CGAL's
Arrangement package. The ``view-map'' layered planar map structure of
Eisemann et al. \cite{Eisemann:2009} is also numerically robust but only
through special tolerancing described by a separate appendix. The snaxel
approach simplifies these robustness issues by maintaining a planar map during
active contour evolution.

The snaxel visual contour approach builds on the idea of propagating contour
samples of Coherent Stylized Silhouettes \cite{kalnins03} and earlier work
\cite{Bourdev:1998wj}. That work used
chains of contour samples that adhered to the visual contour as it moved
but focused on maintaining consistent contour decoration throughout an animated
sequence. In particular it did not confront the changes in contour topology
caused by visual events, which the snaxel approach handles robustly and
simply. Because of their similarity, the methods used for coherent stylized
silhouettes apply directly to the evolving contours produced by snaxels.

\section{Snaxels Redux} \label{sec:snaxels} \vspace{-2mm}
Snaxels are active contours (a.k.a. ``snakes'' \cite{Kass:1988tk}) lifted from
the regular grids of a planar image to work on the irregular grid of a meshed
surfaces embedded in 3-D. In this form, snaxels have been useful for segmenting
mesh features~\cite{Jung:2004uf}, detecting relief on meshes~\cite{Liu:2006uu}
and a variety of other applications. Snakes are an attractive approach for
many applications because of their robustness and their behavior is
controlled primarily by the choice of which ``energy'' they should seek to
minimize. Here we review the relevant snaxel details and show how to adapt the
framework to find and track the contours commonly used for illustration.

\begin{figure}
\centerline{
\includegraphics[width=.45\columnwidth]{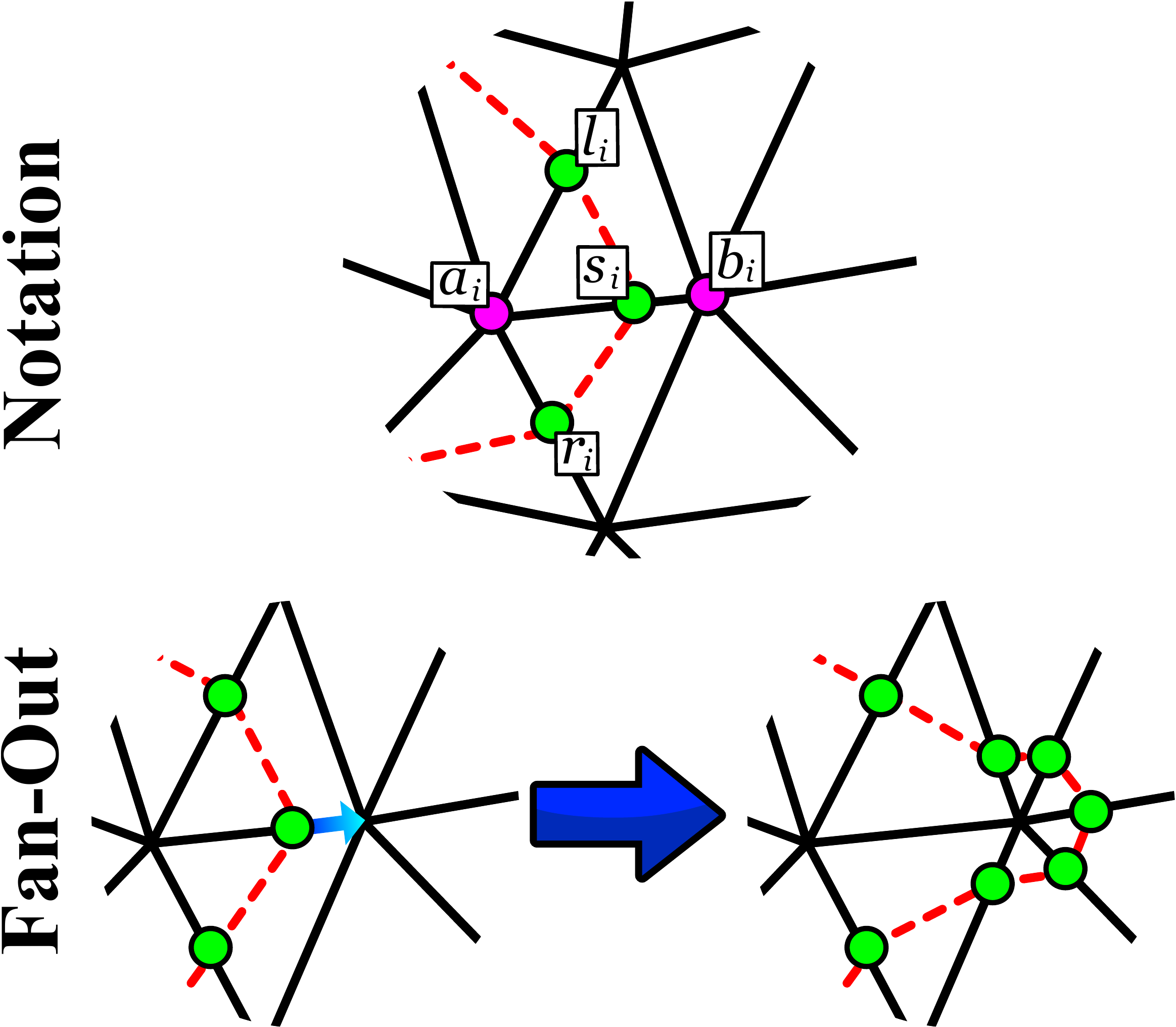} \hfill
\includegraphics[width=.45\columnwidth]{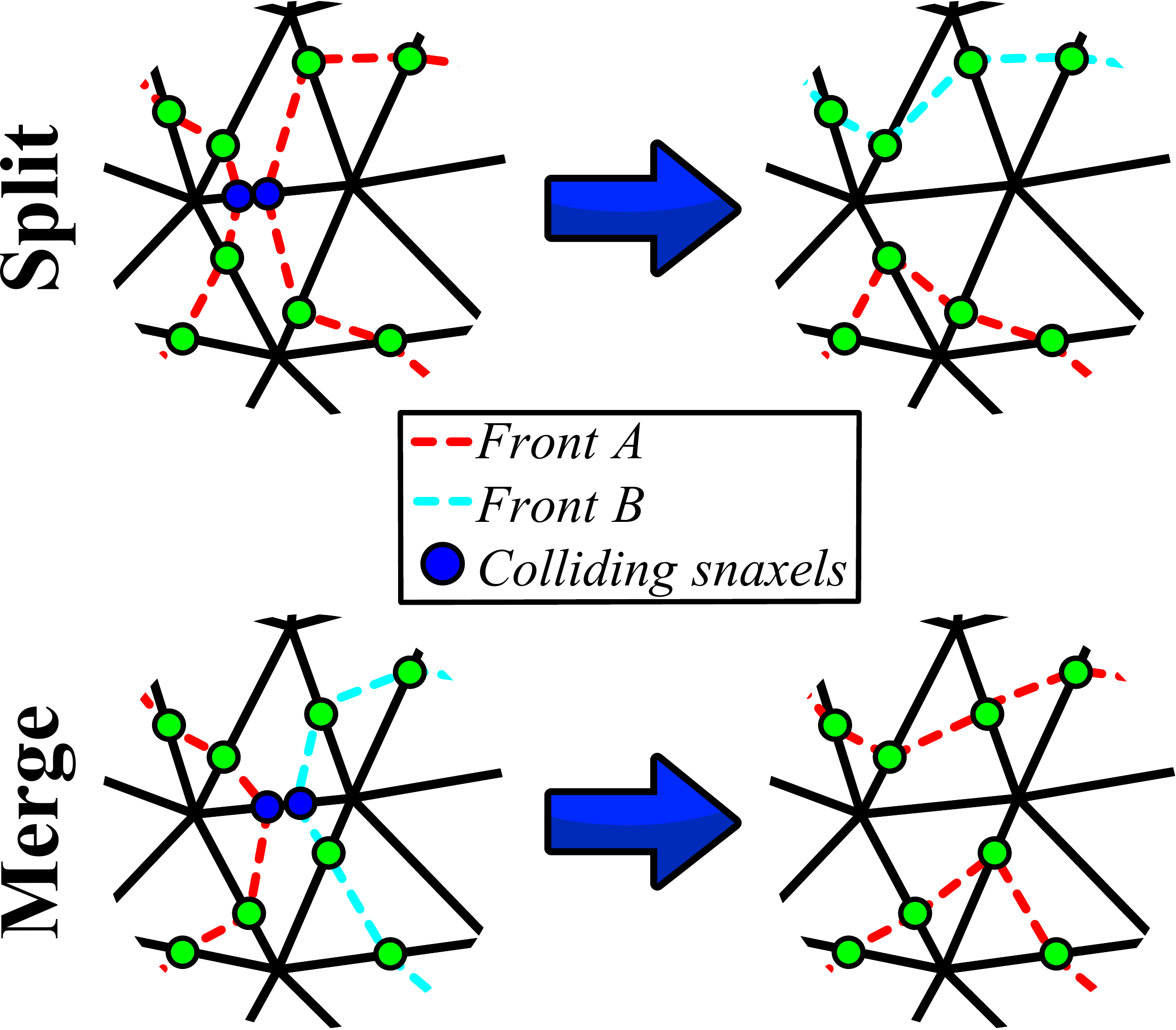}\vspace{-2mm}
}
\caption{A moving front evolving on a meshed surface represented as snaxels
traveling on the mesh edges until they reach a vertex where they fan out, or
encounter another snaxel on the same edge, changing the topology of the front.}
\label{fig:snaxelinfo}
\end{figure}

\begin{figure}
\begin{minipage}[b]{.2\columnwidth} \centering
\includegraphics[width=\textwidth]{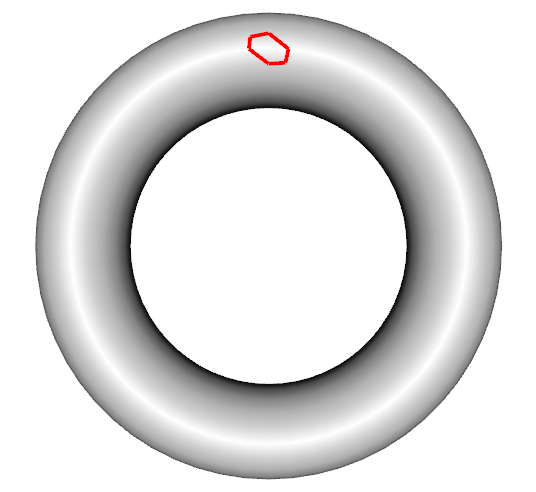} \\ (a)\vspace{-2mm}
\end{minipage}%
\begin{minipage}[b]{.2\columnwidth} \centering
\includegraphics[width=\textwidth]{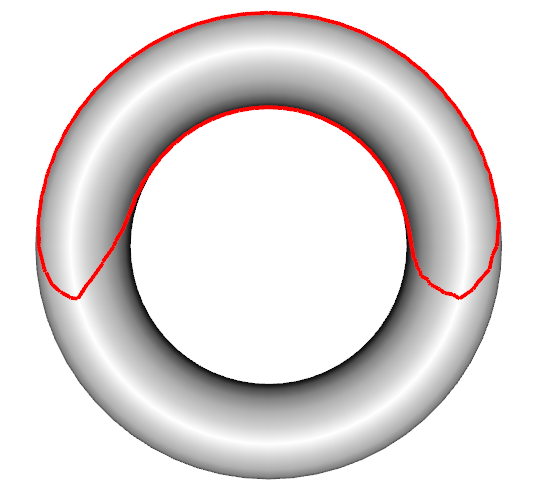} \\ (b)\vspace{-2mm}
\end{minipage}%
\begin{minipage}[b]{.2\columnwidth} \centering
\includegraphics[width=\textwidth]{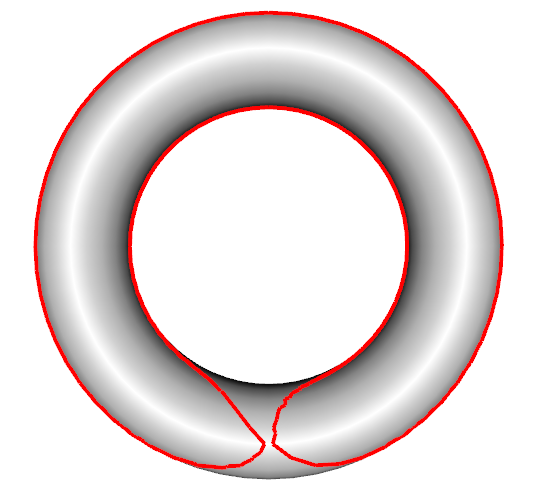} \\ (c)\vspace{-2mm}
\end{minipage}%
\begin{minipage}[b]{.2\columnwidth} \centering
\includegraphics[width=\textwidth]{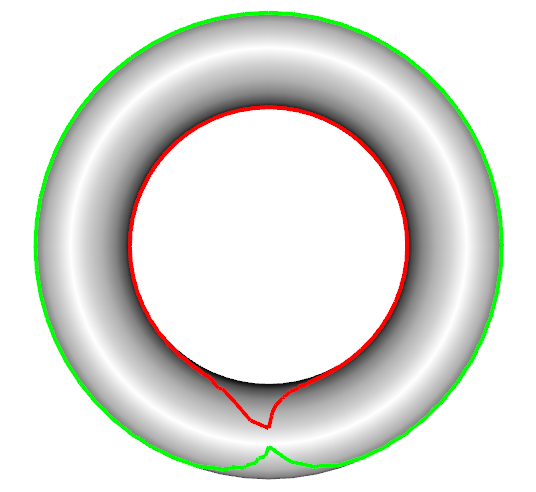} \\ (d)\vspace{-2mm}
\end{minipage}%
\begin{minipage}[b]{.2\columnwidth} \centering
\includegraphics[width=\textwidth]{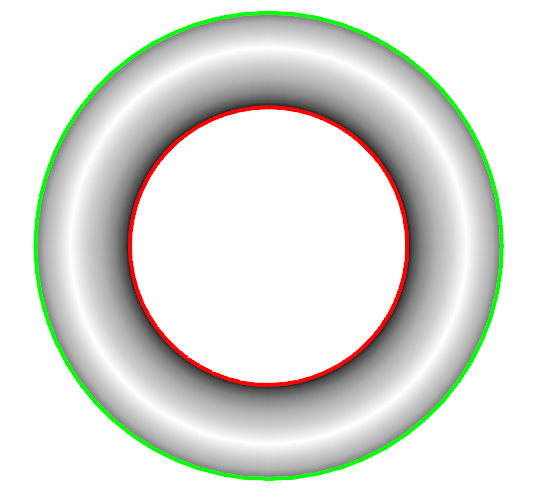} \\ (e) \vspace{-2mm}
\end{minipage} 
\caption{Demonstration of snaxel front evolution into a visual contour.
A snaxel front (red curve) is initialized (a) and evolves according to
the energy function in Eq.~\ref{eq:vcenergy}. A topological
event occurs between (c) and (d) that splits the single front into two fronts.
\vspace{-3mm}}
\label{fig:torusExample}
\end{figure}

Summarizing Bischoff et al.~\shortcite{Bischoff:2005va}, we represent an active
contour on a surface mesh as a collection of ``snaxels,'' points traveling
along the edges of the mesh, and each snaxel is connected to each of its two
neighboring snaxels by a segment lying on the face of the mesh.

\begin{enumerate}
\item {\bf Representation:} Each snaxel lies on a mesh edge
\begin{equation}
s_i = (1-t_i) a_i + t_i b_i,
\end{equation}
an edge-normalized distance $t_i$ from start vertex $a_i$ to end vertex $b_i,$
and is connected to neighboring snaxels referenced as $l_i$ and $r_i$ by line
segments across edge-neighboring faces.
\item {\bf Update:} The snaxels evolve the active contour as
\begin{equation}
t_i \gets t_i -\frac{\Delta t}{||b_i - a_i||} f(s_i),
\end{equation}
where $\Delta t$ is a time step (we use 0.1), and $f(s_i)$ is an implicit
contour function, evaluated at snaxel position $s_i,$ designed to lead the
snaxels over the meshed surface to the $f = 0$ contour. Active contours are
often designed using an energy functional $E(s_i)$ in which case
$f(s_i) = \nabla E(s_i).$
\item {\bf Fan-Out:} If $t_i > 1$ then create a new snaxel on every edge
adjacent to $b_i$ with parameter $t_i - 1,$ else if $t_i < 0$ then do the same
about $a_i$ with parameter $t_i + 1.$
\item {\bf Topology:} Delete any two snaxels sharing an edge and connect the
pair of dangling snaxels on the left face and again on the right face.
\end{enumerate}

Figure~\ref{fig:snaxelinfo} illustrates these rules.

We implement each active contour as a doubly-linked circular list to 
efficiently support the many insertions and deletions of snaxels as the
contours evolve over the meshed surface. The topological operations can
sometimes yield small contours consisting of only one or two snaxels, which
can be tested and deleted during a cleanup sweep through the snaxels.

\begin{figure}
\begin{minipage}[b]{.25\columnwidth} \centering
\includegraphics[width=\textwidth]{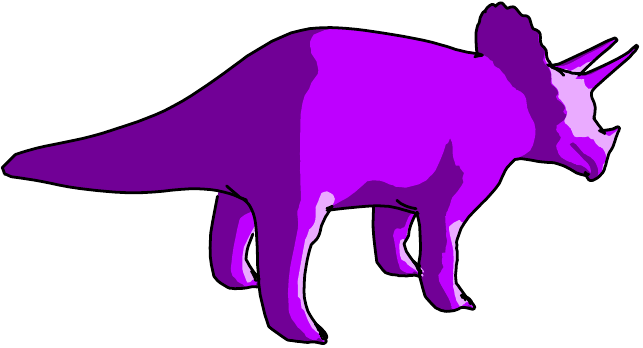} \\
\includegraphics[width=\textwidth]{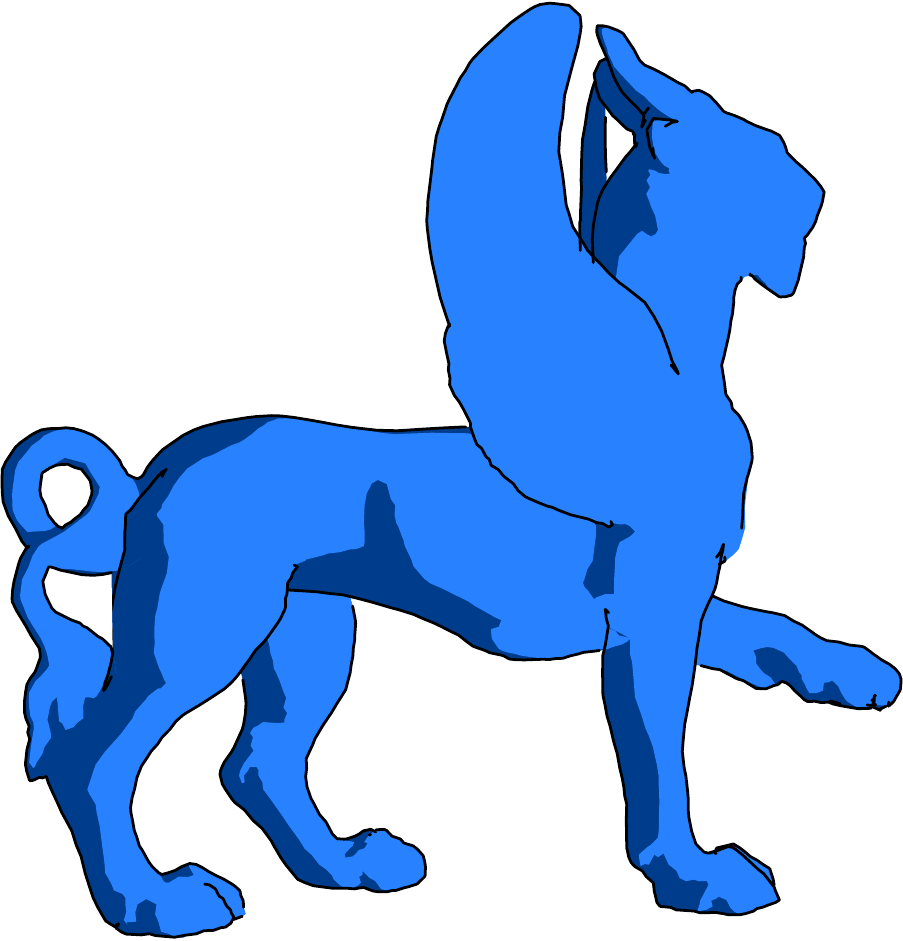}
\end{minipage} \hfill
\begin{minipage}[b]{.25\columnwidth} \centering
\includegraphics[width=\textwidth]{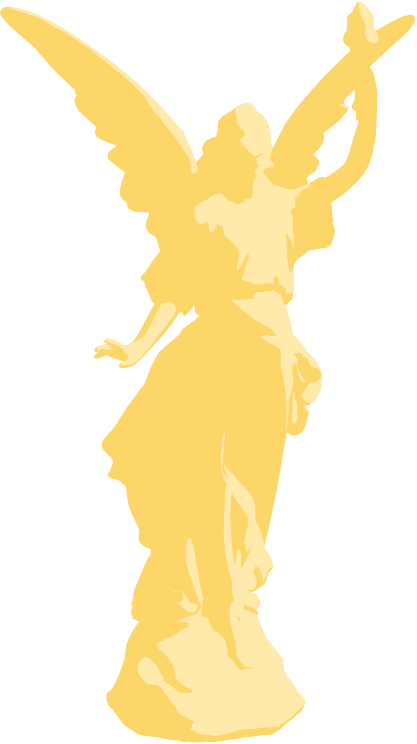}
\end{minipage} \hfill
\begin{minipage}[b]{.35\columnwidth} \centering
\includegraphics[width=\textwidth]{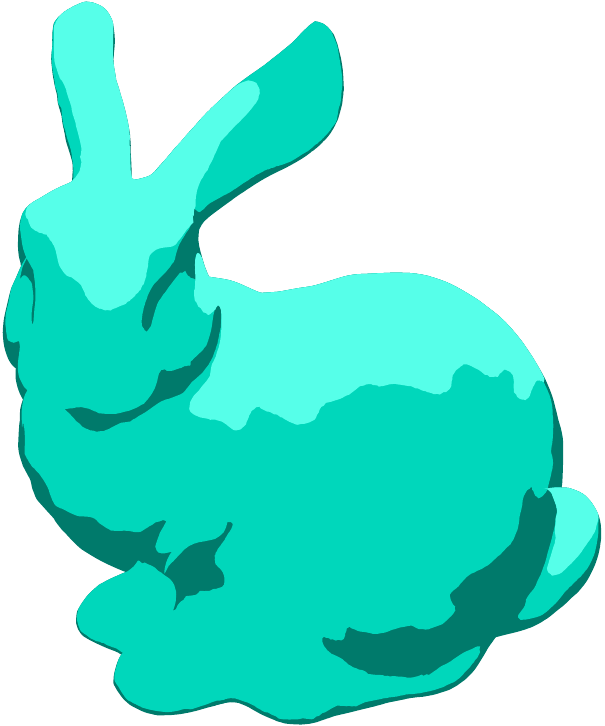}
\end{minipage}
\caption{Visual, shadow and diffuse isophote contours tracked by snaxels and
combined to illustrate four common meshed models. \vspace{-5mm}}
\label{fig:gleams}
\end{figure}

\section{NPR Contouring} \label{sec:npr}  \vspace{-2mm}
The snaxel framework described in the previous section is used to track active
contours that seek to minimize an energy functional. If we define the energy
functional appropriately, these active contours can track the contours
used for illustrative and expressive rendering algorithms.

Most importantly, snaxels can be used to extract the visual contour generator.
The visual contour generator forms the visible and hidden outlines of an object
embedded
in the object surface in 3-D, defined on a smooth surface by the points where
the view vector is tangent to the surface. Hence we define the implicit
contour function for the visual contour generator as
\begin{equation}
f(s_i) = N_i \cdot V_i
\label{eq:vcenergy}
\end{equation}
where $N_i$ is the surface normal at $s_i$ and $V_i$ is a direction vector at
$s_i$ pointing toward the viewer.

% A snaxel's normal vector ($s_N$) is calculated by interpolating the vertex normals of the edge that the snaxel is parameterized by. Given a snaxel $s$ parameterized by an edge consisting of vertices $v_{from}$ and $v_{to}$ (with normals $N_{from}$ and $N_{to}$, respectively), and $s_d \in [0,1]$ as the normalized position of $s$ along the oriented edge , then the snaxel's normal is defined as
% \begin{equation}
% s_N = \left\{ \begin{array}{cc}
% N_{from} & \text{if } \theta = 0 \\
% \frac{\sin( (1-s_d)\theta)}{\sin(\theta)} N_{from} + \frac{\sin( s_d\theta)}{\sin(\theta)} N_{to} & \text{if } \theta > 0\\
% \end{array} \right.
% \end{equation}
% where $\theta \in [0,\pi)$ is the angle between vertex normals, given by $\theta = \arccos (N_{from}\cdot N_{to})$.
% We use spherical linear interpolation to guarantee the resulting interpolated vector is unit, and mesh vertex normals are computed using a weighted average of incident face normals (weights are assigned based on surface area of the face).

For a meshed surface, we define the normal vector $N_i$ at snaxel $i$ along
edge $(a_i,b_i)$ as the spherical linear interpolation of the normals
$N_a, N_b$ defined at the vertices $a_i, b_i,$
\begin{equation}
N_i = \frac{\sin((1-t_i) \theta)}{\sin(\theta)} N_a +
	\frac{\sin(t_i \theta)}{\sin(\theta)} N_b,
\end{equation}
where $\theta = \arccos(N_a \cdot N_b)$ is the angle between the normals, and
assume the limit $N_i \rightarrow N_a$ as $\theta \rightarrow 0.$ 
Figure~\ref{fig:torusExample} demonstrates a snaxel front propagating across a 
torus while adhering to the implicit contour function as in Eq.~\ref{eq:vcenergy}.

As shown previously \cite{Eisemann_cgf08}, using normals interpolated from
vertex normals to extract a smooth visual contour generator over mesh faces
instead of edges \cite{hertzmann2000}, can lead to a silhouette that might
not strictly contain the projection of the mesh. When a smoothed silhouette
curve segment traverses a face, the entire face is either visible or occluded
and when visible includes the portion outside the silhouette curve. This can
be problematic when classifying regions on the projection or intersecting
multiple contours (e.g. visual contour with shadow contour to segment visible
illuminated portions from visible shadowed portions). As shown in the next
section, the planar map produced by revised snaxel rules avoids such
misclassifications.

We can likewise generate a shadow contour and isophotes of diffuse illumination
by setting the implicit contour function to 
\begin{equation}
f(s_i) = \min( N_i \cdot L_i - k, \ N_i \cdot V_i)
\label{eq:isophote}
\end{equation}
where $L$ is a direction vector from vertex $i$ toward a light source, and $k$
is the isovalue of the isophote. For example, if $k = 0,$ then
Eq. \ref{eq:isophote} generates a shadow contour whereas setting $k > 0$
generates a contour of constant diffuse reflection. The minimum between the
two terms is taken so that snaxels on isophote contours do not lie on backfacing
polygons, which allows for stylization as illustrated in Fig.~\ref{fig:gleams}.

The interaction between snaxels of different types (e.g. visual contour v.
shadow contour) depends on the form of planar map used, as described in the
next section. For example, if a planar map is post-processed, then visual
contour snaxels would not interact with shadow contour snaxels. But if the
snaxel method for planar map generation is used, then the two kinds of snaxels
would indeed interact with each other.

\begin{figure}
\includegraphics[width=40mm]{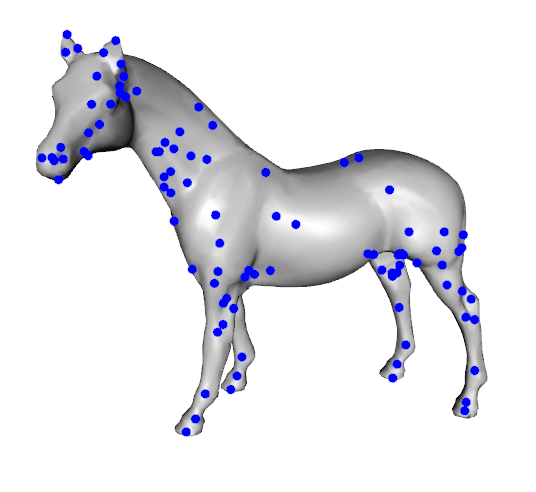}
\includegraphics[width=40mm]{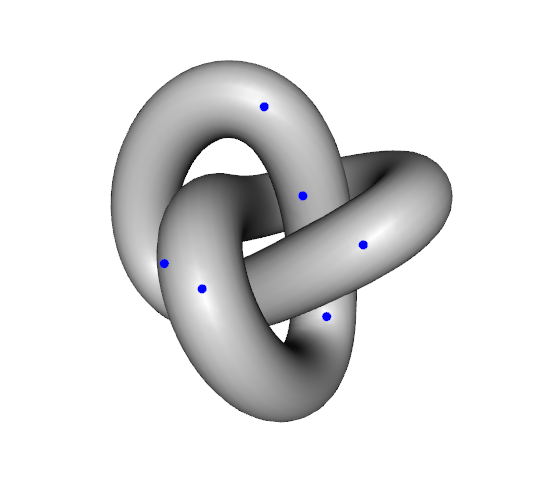} \vspace{-5mm}
\caption{Snaxel fronts for finding the visual contour generator are initialized
to surround vertices whose normal points more toward the viewer than do its
neighboring vertices.\vspace{-4mm}}
\label{fig:snaxel_initialization}
\end{figure}
 
\subsection{Initialization} \vspace{-2mm}
Snaxels must be created properly to ensure that every contour-bounded region is
tracked and generated. Since we seek contours where $f = 0,$ we initialize
contours at the (local) extremes of $f$ as shown in
Figure~\ref{fig:snaxel_initialization}. As this figure shows, smooth surfaces
benefit from fewer initialization points and need less merging, whereas
noisier meshes, such as those reconstructed from scanned points, generate many
initial contours that require significant merging into the visual contour
generator. We have found the snaxel framework is robust enough to nevertheless
handle such cases.

Since the snaxel front expands through all vertices as it descends (splitting
and merging as necessary) from local maxima to a zeroset contour, it can label
vertices as its sweep defines the region bounded by the active contour. Hence
the snaxel approach not only generates region bounding contours but also
labels all vertices (and edges and faces if needed) within the region. These
region delineations can be convenient for various region-based stylization
methods. 

Some contours could be more problematic, such as small specular highlights.
Phong reflectance can be used as an implicit contour function
\begin{equation}
f(s_i) = V_i \cdot (2(N_i \cdot L_i)N_i - L_i),
\label{eq:gleams}
\end{equation}
which generates contours surrounding specular gleams. However, as Phong showed,
these contoured gleam regions can occur between vertices, and so contain no
vertex, and these specular regions might be missed by a per-vertex snaxel
initialization.

\begin{figure}
\vspace{-5mm}
\hspace{-5mm}
\begin{minipage}[b]{.37\columnwidth} \centering
\includegraphics[width=\textwidth]{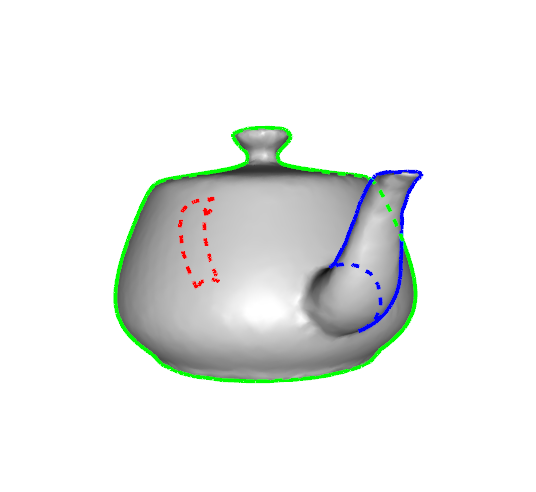} \\[-2ex] (a)
\end{minipage}%
\begin{minipage}[b]{.37\columnwidth} \centering
\includegraphics[width=\textwidth]{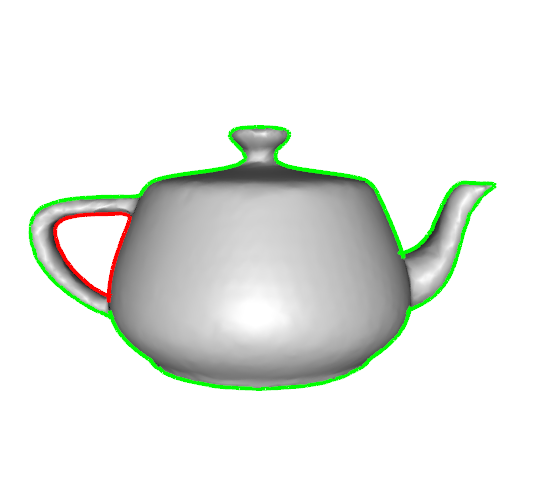} \\[-2ex] (b)
\end{minipage}%
\begin{minipage}[b]{.37\columnwidth} \centering
\includegraphics[width=\textwidth]{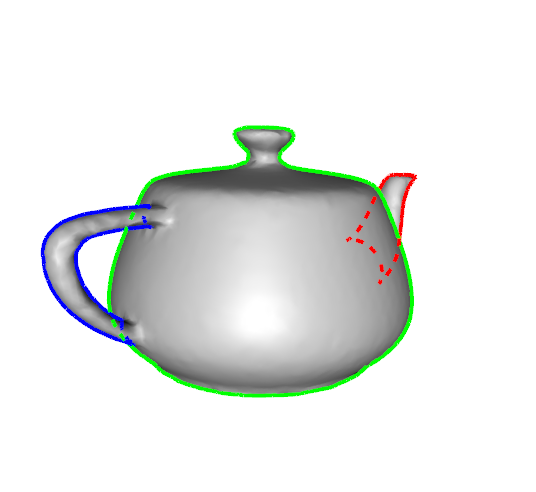} \\[-2ex] (c)
\end{minipage}
\caption{As the view changes, the snaxels move accordingly to track the
new visual contour positions. Our method is robust to meshes with complex
visual events, and captures the entire visual contour generator (occluded
portions dashed).\vspace{3mm}}
\label{fig:teapotExample}
% \end{figure}

% \begin{figure}
\begin{centering}
\begin{minipage}[b]{.33\columnwidth} \centering
\includegraphics[width=\textwidth]{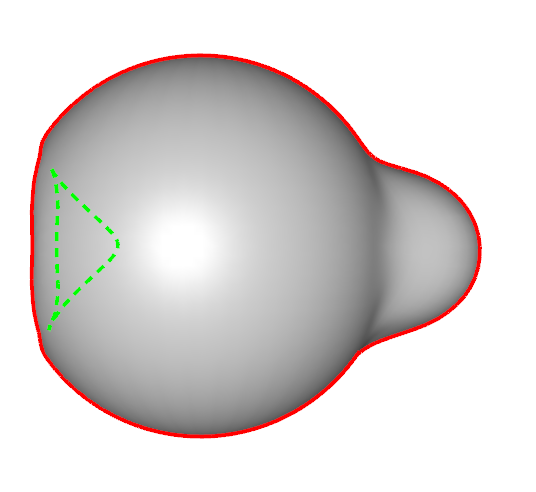} \\ \vspace{-1.5mm} (a)\vspace{0mm}
\end{minipage}\hfill%
\begin{minipage}[b]{.33\columnwidth} \centering
\includegraphics[width=\textwidth]{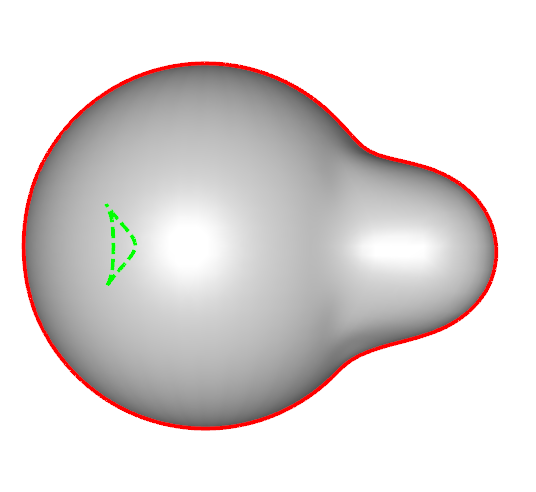} \\  \vspace{-1.5mm}(b)\vspace{0mm}
\end{minipage}\hfill%
\begin{minipage}[b]{.33\columnwidth} \centering
\includegraphics[width=\textwidth]{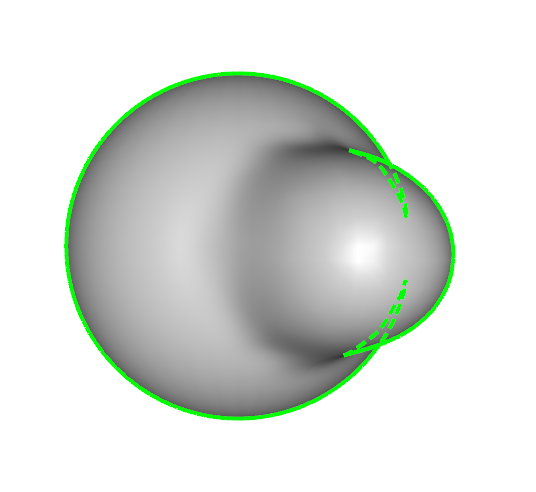} \\ \vspace{-1.5mm} (c)\vspace{0mm}
\end{minipage}\\
\begin{minipage}[b]{.33\columnwidth} \centering
\includegraphics[width=\textwidth]{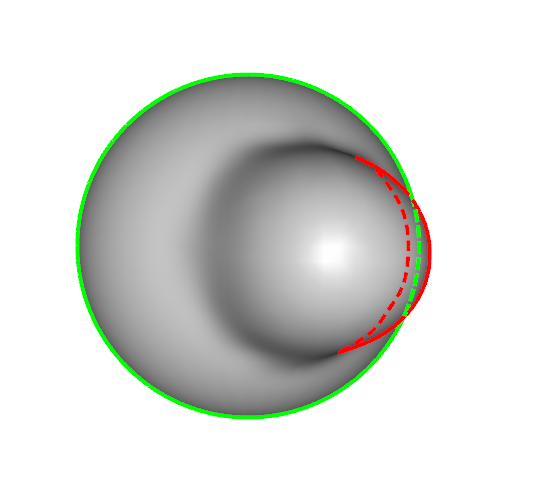} \\ \vspace{-1.5mm} (d)\vspace{0mm}
\end{minipage}\hfill%
\begin{minipage}[b]{.33\columnwidth} \centering
\includegraphics[width=\textwidth]{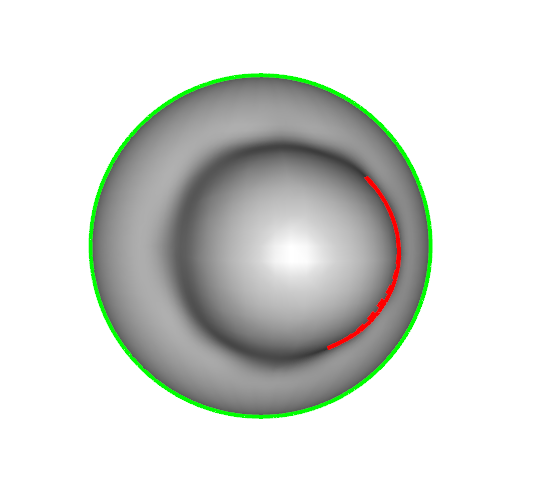} \\ \vspace{-1.5mm} (e)\vspace{0mm}
\end{minipage}\hfill%
\begin{minipage}[b]{.33\columnwidth} \centering
\includegraphics[width=\textwidth]{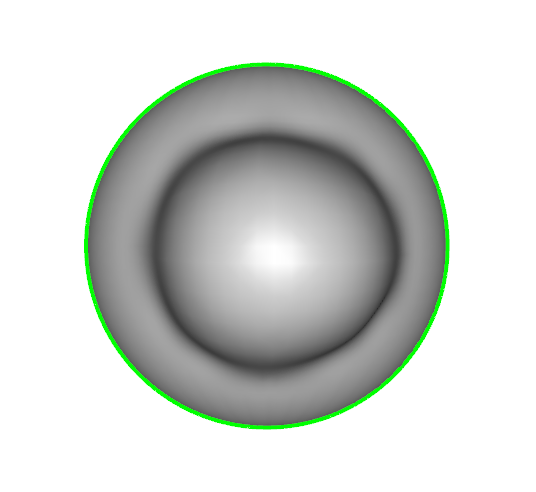} \\ \vspace{-1.5mm} (f)\vspace{0mm}
\end{minipage}\\
\begin{minipage}[b]{.33\columnwidth} \centering
\includegraphics[width=\textwidth]{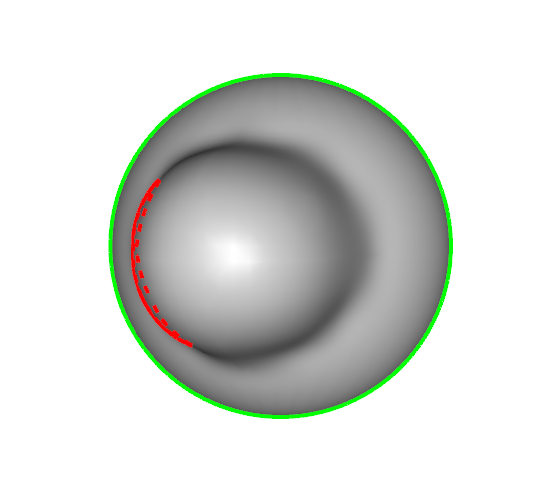} \\ \vspace{-1.5mm} (g)\vspace{0mm}
\end{minipage}\hfill%
\begin{minipage}[b]{.33\columnwidth} \centering
\includegraphics[width=\textwidth]{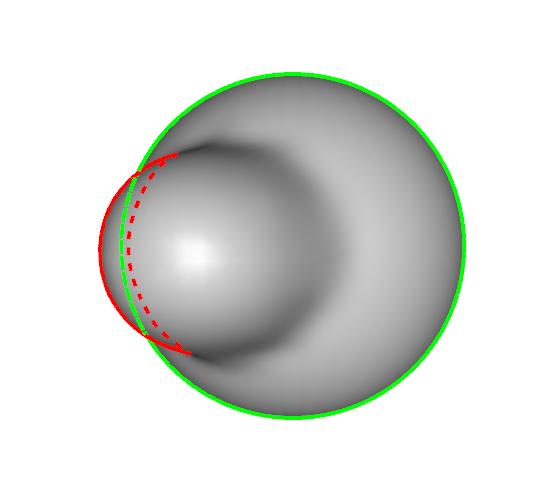} \\ \vspace{-1.5mm} (h)\vspace{0mm}
\end{minipage}\hfill%
\begin{minipage}[b]{.33\columnwidth} \centering
\includegraphics[width=\textwidth]{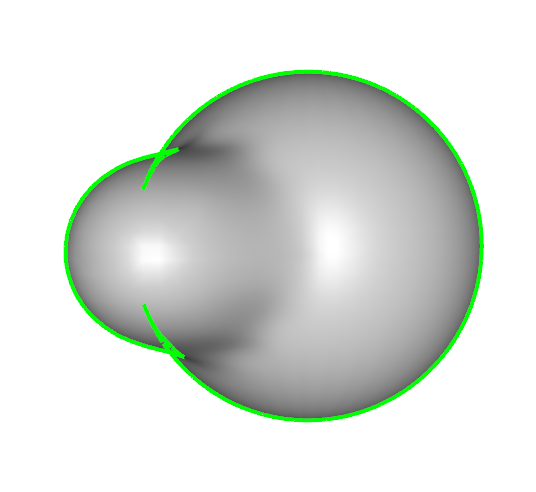} \\ \vspace{-1.5mm} (i)\vspace{0mm}
\end{minipage}\\
\begin{minipage}[b]{.33\columnwidth} \centering
\includegraphics[width=\textwidth]{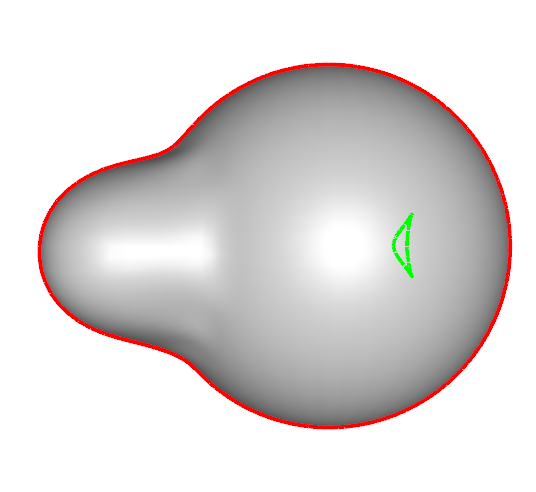} \\ \vspace{-1.5mm} (j)\vspace{0mm}
\end{minipage}\hfill%
\begin{minipage}[b]{.33\columnwidth} \centering
\includegraphics[width=\textwidth]{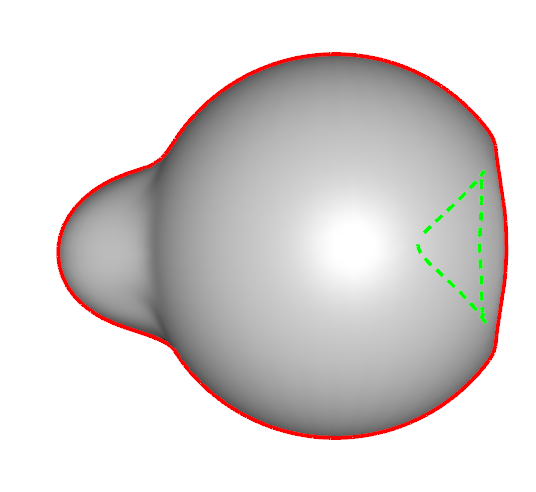} \\  \vspace{-1.5mm}(k)\vspace{0mm}
\end{minipage}\hfill%
\begin{minipage}[b]{.33\columnwidth} \centering
\includegraphics[width=\textwidth]{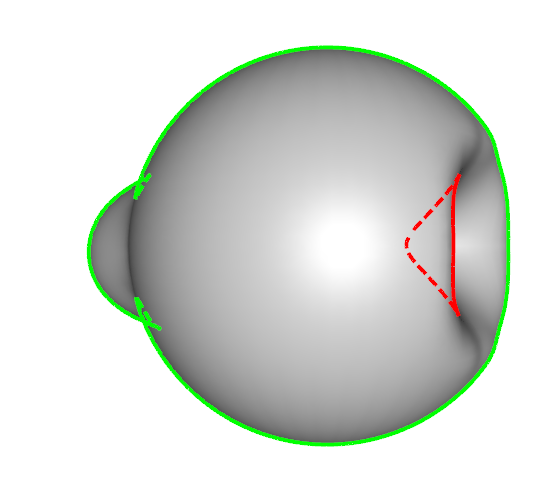} \\  \vspace{-1.5mm} (l)\vspace{0mm}
\end{minipage}\\
\begin{minipage}[b]{.33\columnwidth} \centering
\includegraphics[width=\textwidth]{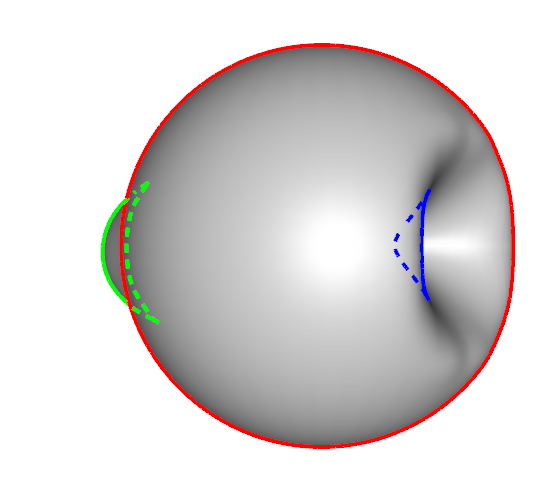} \\ \vspace{-1.5mm} (m) \vspace{0mm}
\end{minipage}\hfill%
\begin{minipage}[b]{.33\columnwidth} \centering
\includegraphics[width=\textwidth]{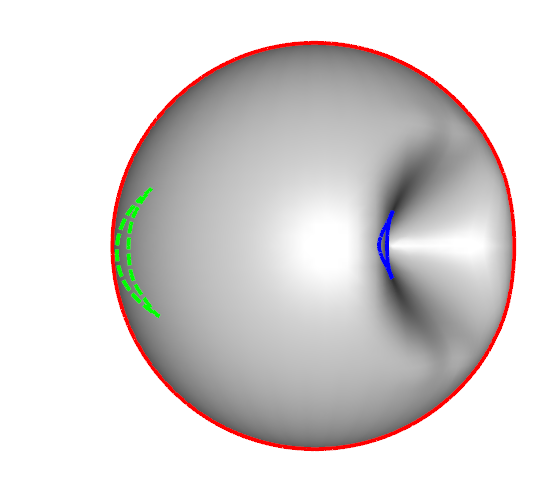} \\\vspace{-1.5mm} (n)\vspace{0mm}
\end{minipage}\hfill%
\begin{minipage}[b]{.33\columnwidth} \centering
\includegraphics[width=\textwidth]{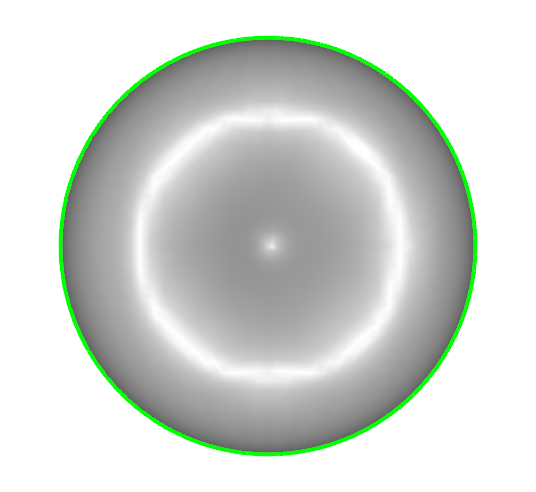} \\ \vspace{-1.5mm} (o)\vspace{0mm}
\end{minipage}\\
\begin{minipage}[b]{.33\columnwidth} \centering
\includegraphics[width=\textwidth]{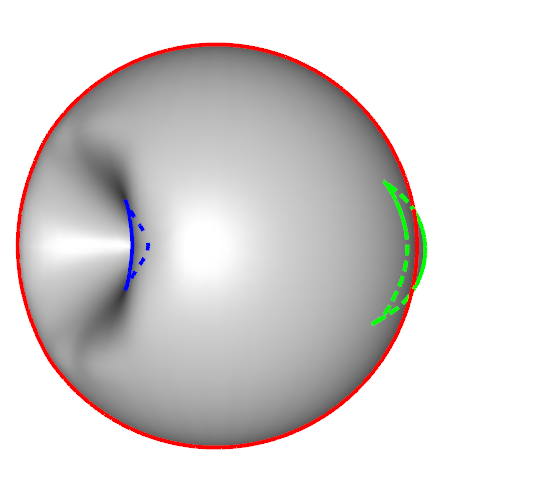} \\  \vspace{-1.5mm}(p)\vspace{0mm}
\end{minipage}\hfill%
\begin{minipage}[b]{.33\columnwidth} \centering
\includegraphics[width=\textwidth]{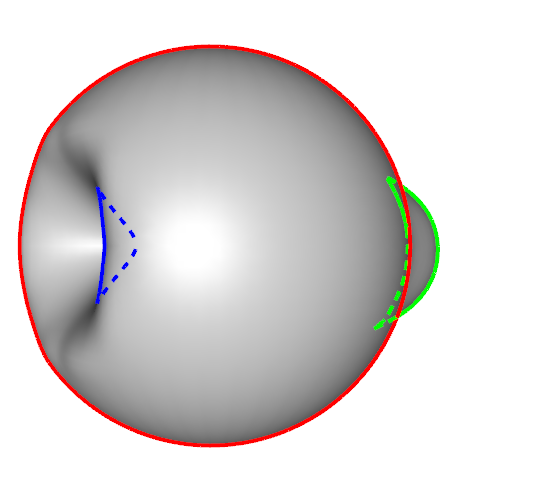} \\  \vspace{-1.5mm}(q)\vspace{0mm}
\end{minipage}\hfill%
\begin{minipage}[b]{.33\columnwidth} \centering
\includegraphics[width=\textwidth]{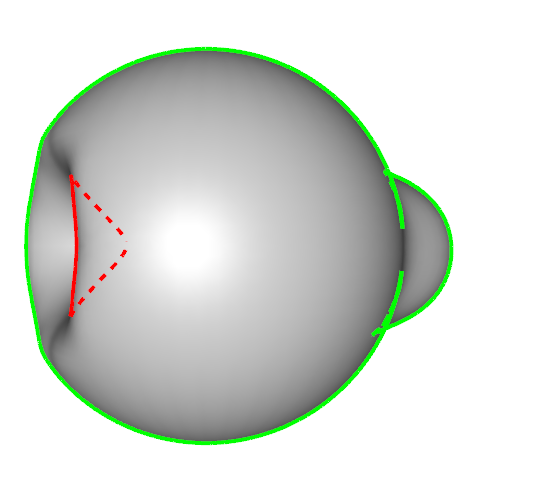} \\  \vspace{-1.5mm}(r)\vspace{0mm}
\end{minipage}
\end{centering}
\caption{Visual events managed by the snaxel fronts. Fronts are split after
(c), (k) and (l), merged after (h), (k) and (q), annihilated
after (b), (e) and (n) and created after (f), (i) and (o). Note that between
(k) and (l) the snaxel fronts robustly handle the combination of a merge and a
split (which flipped the contour color labels). \vspace{-1mm}}
\label{fig:nose}
\end{figure}

\begin{figure}
\centerline{
\includegraphics[width=40mm]{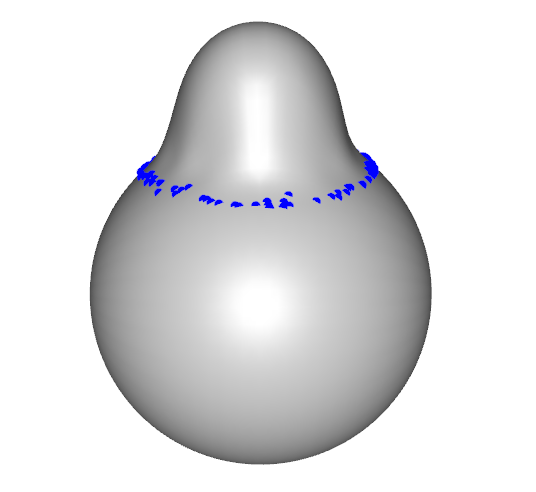}
\includegraphics[width=40mm]{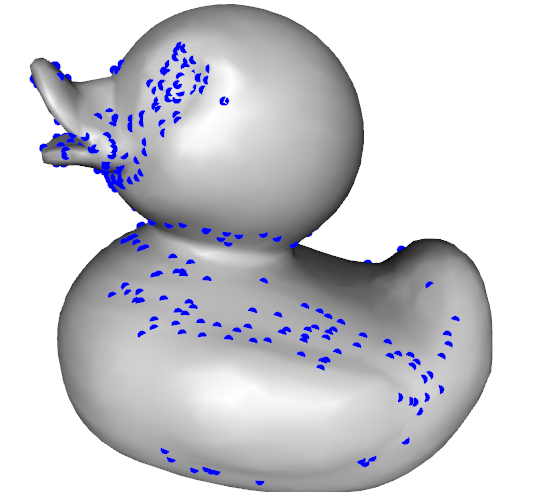}
}
\caption{The snaxel framework can identify parabolic points on a mesh as
locations where the visual contour generator changes topology.\vspace{-3mm}}
\label{fig:parabolic_points}
\end{figure}

\subsection{Animation} \vspace{-2mm}
In addition to growing visual and other contours on meshed surfaces, the snaxel
approach is particularly good at tracking existing contours as the surface
changes. These contours change as the view changes, the light position moves
or the surface rotates, and also as a dynamic mesh changes shape. In these
cases, the snaxels need not be reinitialized from scratch but can simply
evolve to correct the contour to the new situation; figure~\ref{fig:teapotExample}
shows an example of this. 

As the contours move to accommodate changing views, lighting and shape, we
maintain the contour region labeling. Each vertex keeps track of whether or
not a front has visited it, and such visits only occur during the ``fan-out''
stage of snaxel propagation. Since merging and splitting events happen only on
mesh edges, this visited flag is unaffected by topology changes.

When the view of a surface changes, the visual contour generator can undergo
topological changes, which are denoted as visual events. These visual events
can cause contours to divide or merge, or to be created or destroyed. The
flexible topology of snaxel evolution handles the divide and merge cases, and
the snaxel cleaning phase manages the annihilation of entire contour loops.
The remaining case is the creation of a contour loop, which is difficult
because there may not be any snaxels near the point where a new contour
appears.

At each frame in the animation, snaxels are initialized at extrema of $f$; however,
we ensure that snaxel fronts are not created around any vertex that has already
been visited by a snaxel front. Since we search for all extrema of $f$ (rather than
only minima or maxima), we detect backfacing and occluded visual contours as
well as frontfacing visual contours. Several such contours are properly detected 
and shown in three frames of the rotating ``nose'' example shown in 
Fig.~\ref{fig:nose}.

%A new loop element in the visual contour generator is created when a
%portion of the viewed surface that faced the viewer now faces away from the
%viewer. The visual contour generator of a faceted surface lies along the edges
%between frontfacing and backfacing facets, but we are using interpolated vertex
%normals. Hence we detect the creation of new contour loops in the
%normal-interpolated visual contour generator by an iterative test of
%all mesh vertices, searching for any whose normal switched from frontfacing to
%backfacing. If any such vertices are found that have not been traversed by
%an existing snaxel front, then a new snaxel front is initiated on each of
%them. Such contour creates happens and is properly detected in three cases in
%the rotating nose example shown in Fig.~\ref{fig:nose}.
%
%We likewise initialize new shadow contour fronts on is olated vertices
%that now face away from the light source. For isophotes, we retest the dot
%product for any changes. In general, for any contour defined by an implicit
%contour function $f,$ we create a new front around any isolated vertex where
%$f$'s sign changed from positive to negative (or zero).

\subsection{Mesh Parabolic Points} \vspace{-2mm}
On smooth surfaces, the location where a visual event causes a change in the
topology of the visual contour generator occurs at a parabolic point. These are
points where the Gaussian curvature vanishes because one of the principle
curvatures is flat. Such positions have been difficult to precisely locate on
meshed surfaces. Curvature measures on meshed surfaces are notoriously
susceptible to noise and other problems, yielding some to resort to a full
global fitting of an implicit surface to a mesh just to evaluate curvature
over the mesh \cite{ohtake04}. The Gaussian curvature of a mesh is zero
everywhere except its vertices, but typically all curvatures on a mesh are
evaluated with respect to its vertices, which further confounds the detection
of parabolic points.

We can use the snaxel framework to detect parabolic points on a mesh. By
sampling the sphere of (orthographic) view directions, we can evolve the
visual contour generator over the mesh. When the snaxels defining this
contour alter topology, we can mark that location as a parabolic point.
Figure~\ref{fig:parabolic_points} demonstrates some examples of these
markings.

\section{Planar Map Generation} \label{sec:planar} \vspace{-2mm}
Given several sets of overlapping contours (e.g. the visible and occluded
visual contour generator, shadow contour and other shading contours) and the
regions they delineate, the planar map decomposes them into a planar graph of
homogeneous regions. The planar map can be constructed of just visible
portions of a surface, or can consist of all surfaces, visible and occluded,
in which case regions in the planar map share depth complexity in addition to
other region attributes.

\begin{figure} \centering
\begin{minipage}[b]{.45\columnwidth} \centering
\includegraphics[width=\textwidth]{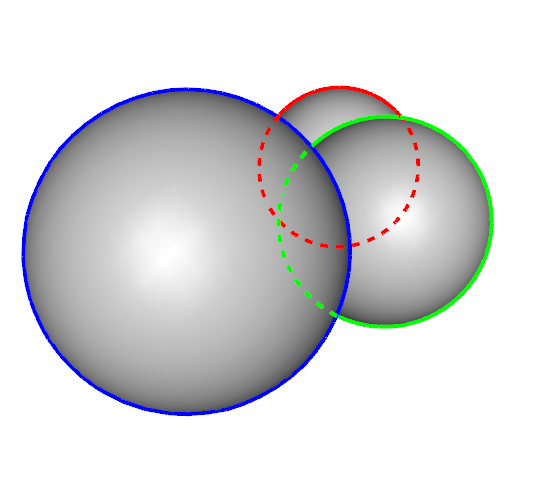}\vspace{-9mm}
\end{minipage}%
\begin{minipage}[b]{.45\columnwidth} \centering
\includegraphics[width=\textwidth]{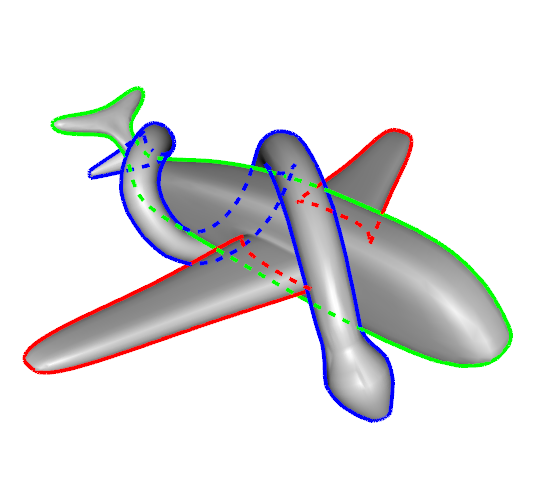}\vspace{-9mm}
\end{minipage}\\
Meshed Geometry \\[2ex] 
\begin{minipage}[b]{.45\columnwidth} \centering
\includegraphics[width=\textwidth]{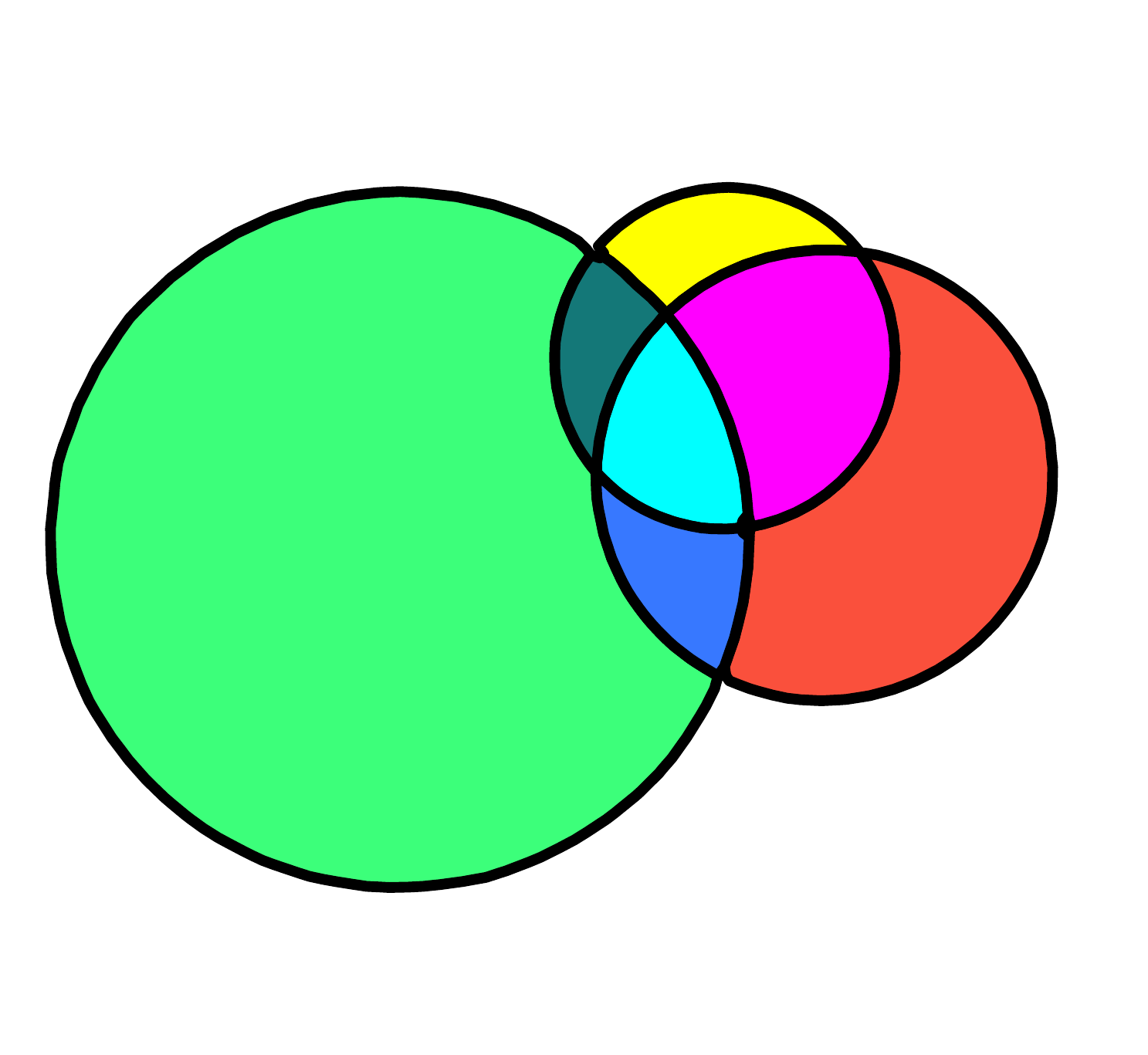}\vspace{-7mm}
\end{minipage}%
\begin{minipage}[b]{.45\columnwidth} \centering
\includegraphics[width=\textwidth]{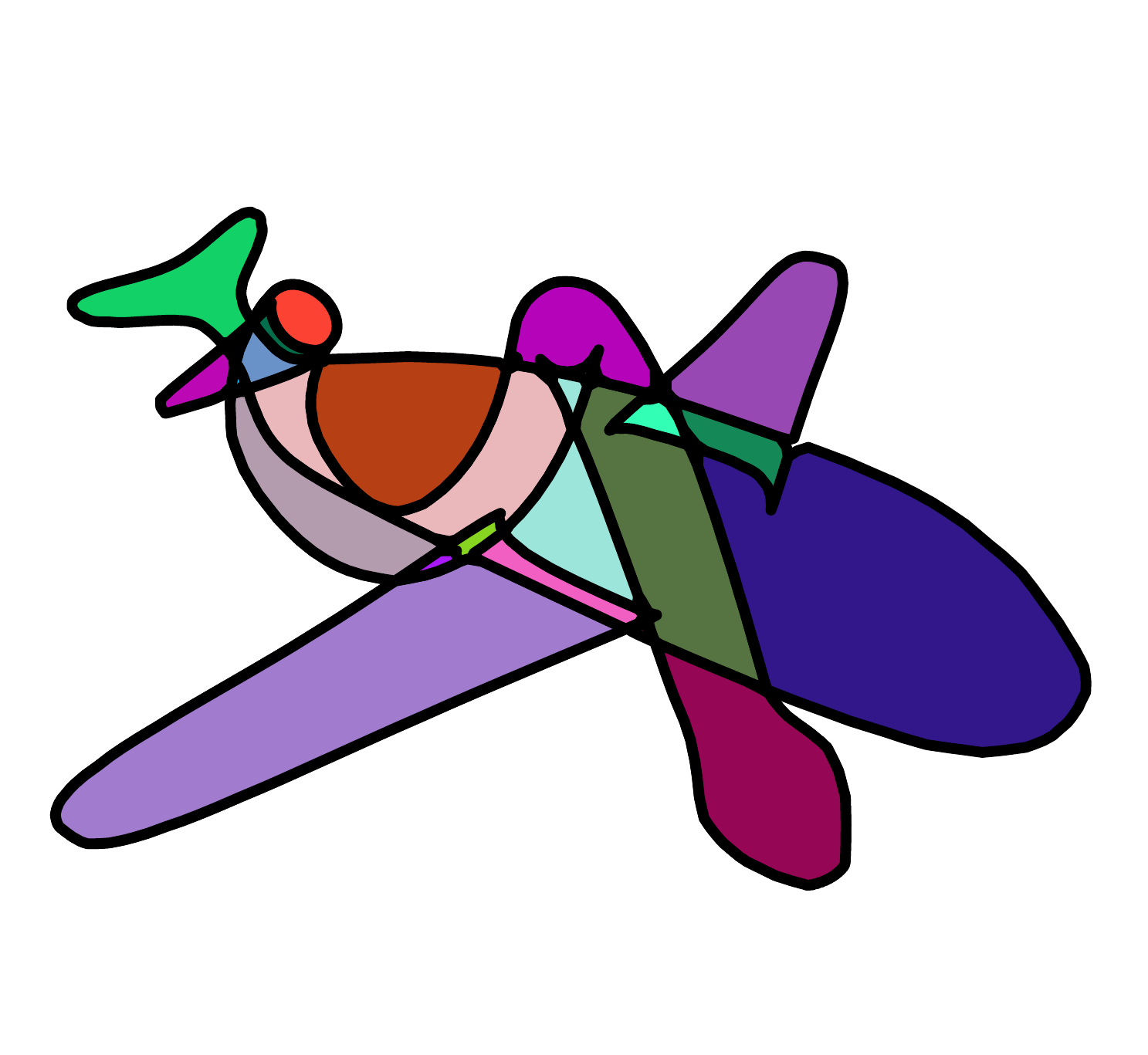}\vspace{-7mm}
\end{minipage}\\
Post-Processed \\[2ex]
\begin{minipage}[b]{.45\columnwidth} \centering
\includegraphics[width=\textwidth]{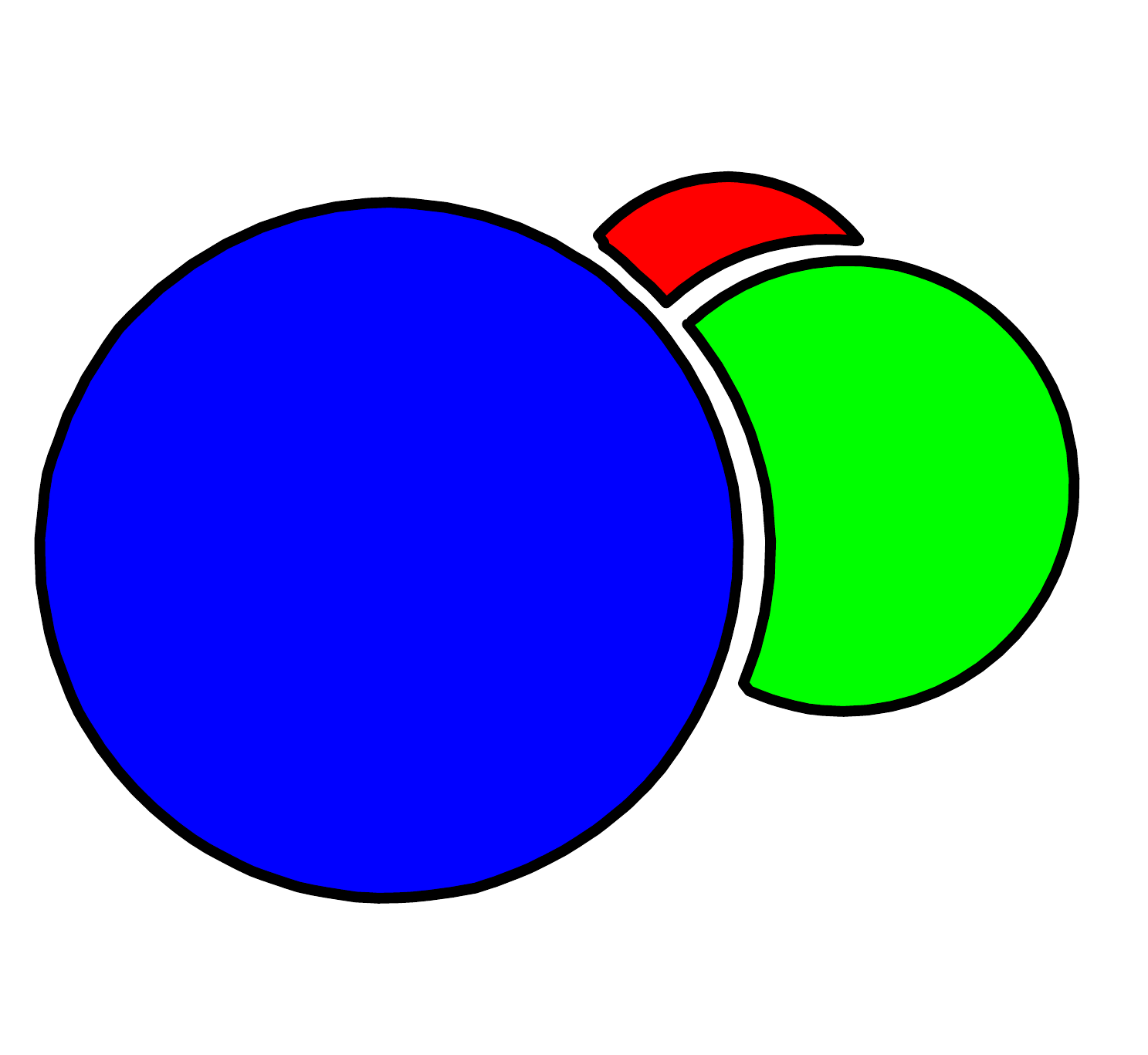} \vspace{-11mm}
\end{minipage}
\begin{minipage}[b]{.45\columnwidth} \centering
\includegraphics[width=\textwidth]{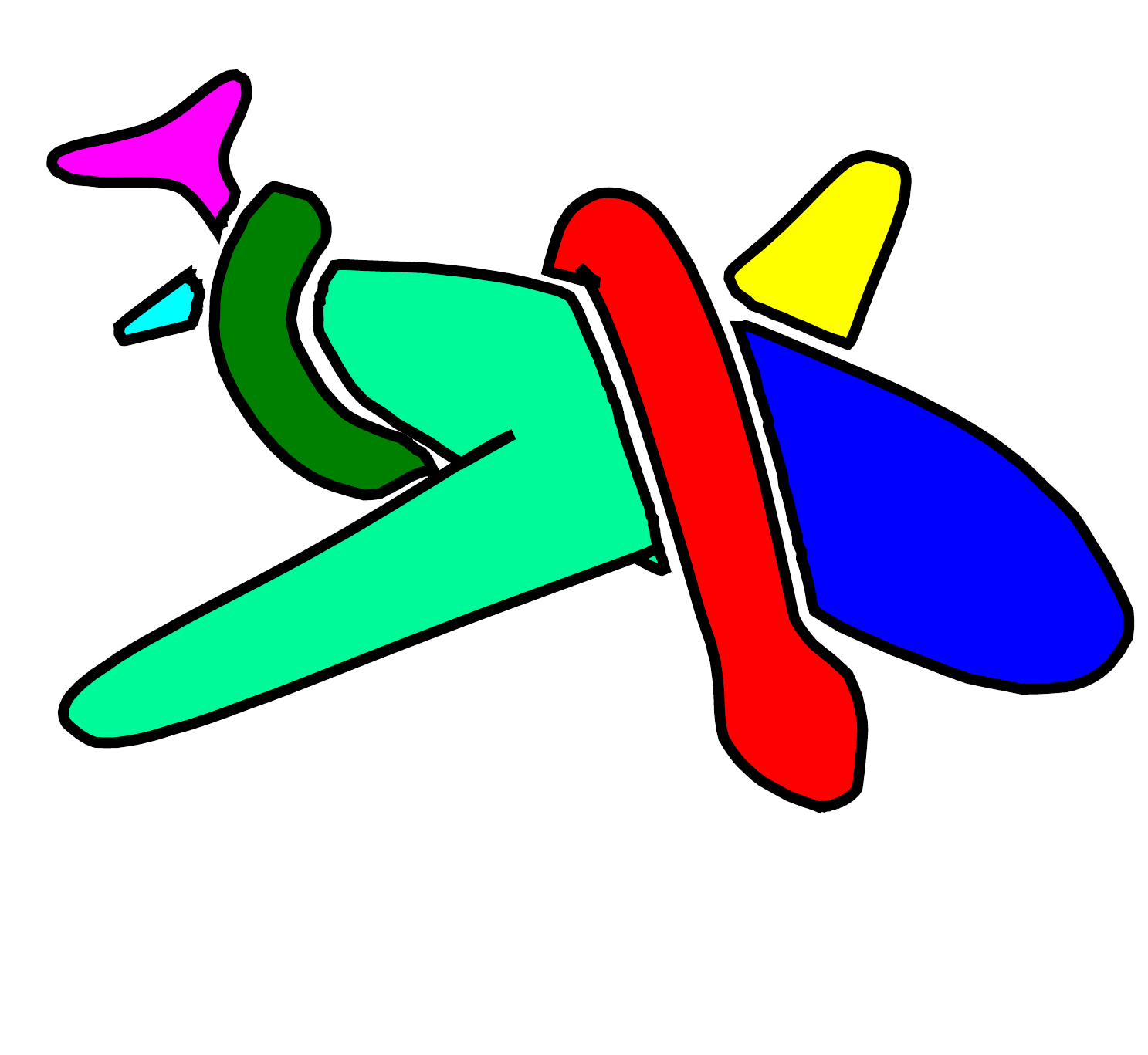} \vspace{-11mm}
\end{minipage}\\
Snaxel-Defined
\caption{Planar maps computed for two scenes. A full planar map is computed as a post-process by polygon clipping the snaxel contours. Alternatively, a visible
surface planar map is constructed using alternate snaxel propagation rules. \vspace{-3mm}} 
\label{fig:planarmap}
\end{figure}

Given the overlapping visual, shadow and shading
contours generated from the previous section, a planar map can be generated as
a post-process, e.g. by the CGAL Arrangement package
\cite{fhhn-dipmc-00,wfzh-aptac-05} or the
General Polygon
Clipper library \cite{Murta} (based on \cite{Vatti92}), as shown in the center
row of Figure~\ref{fig:planarmap}. Such
processing requires integration of a rather heavyweight library, and can yield
numerous small regions (see e.g. \cite{biermann01}) that can hinder stylization.

\begin{figure*}
\centerline{
\includegraphics[width=0.30\columnwidth]{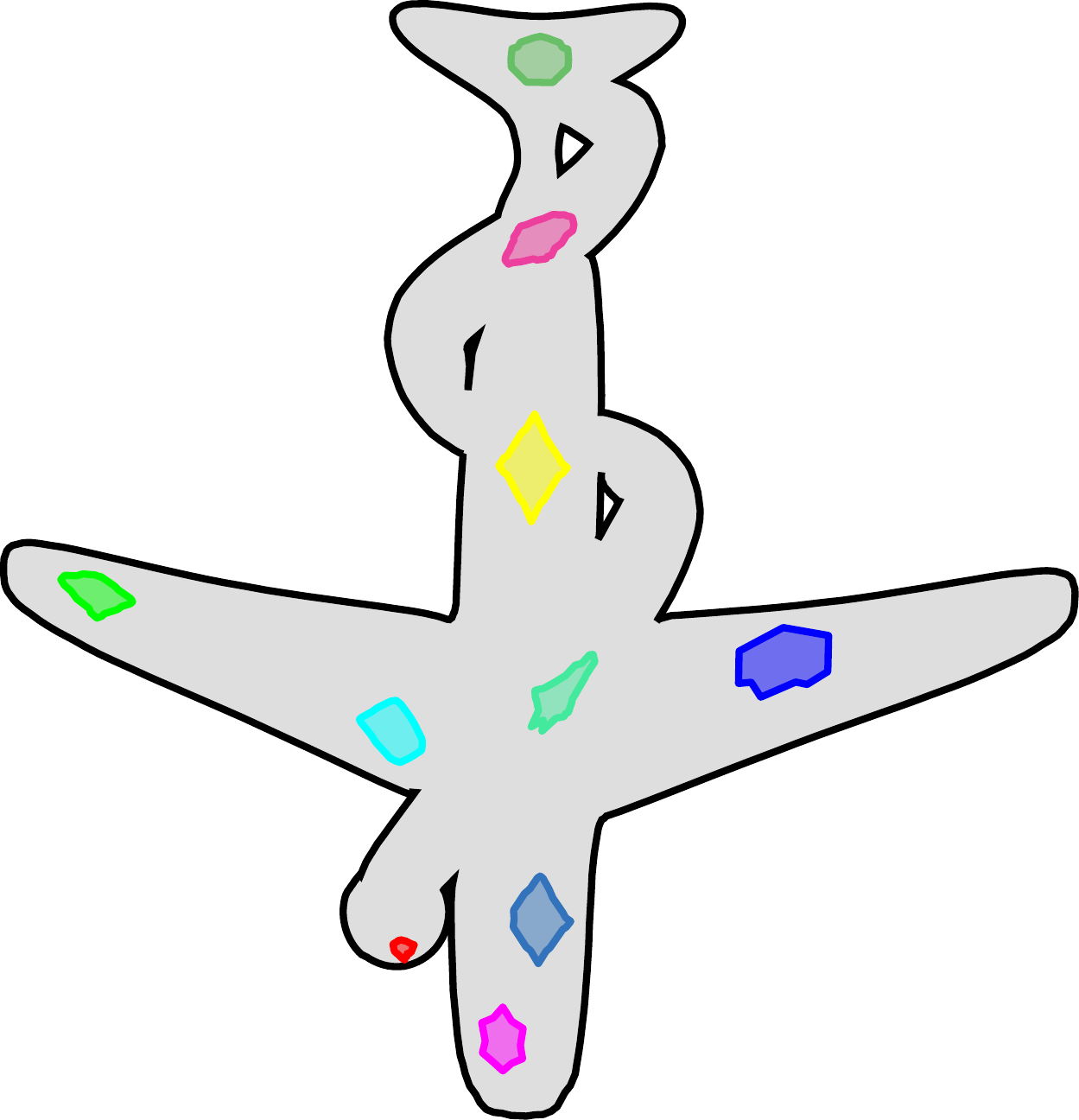} \vspace{2mm} \hspace{7.7mm}
\includegraphics[width=0.30\columnwidth]{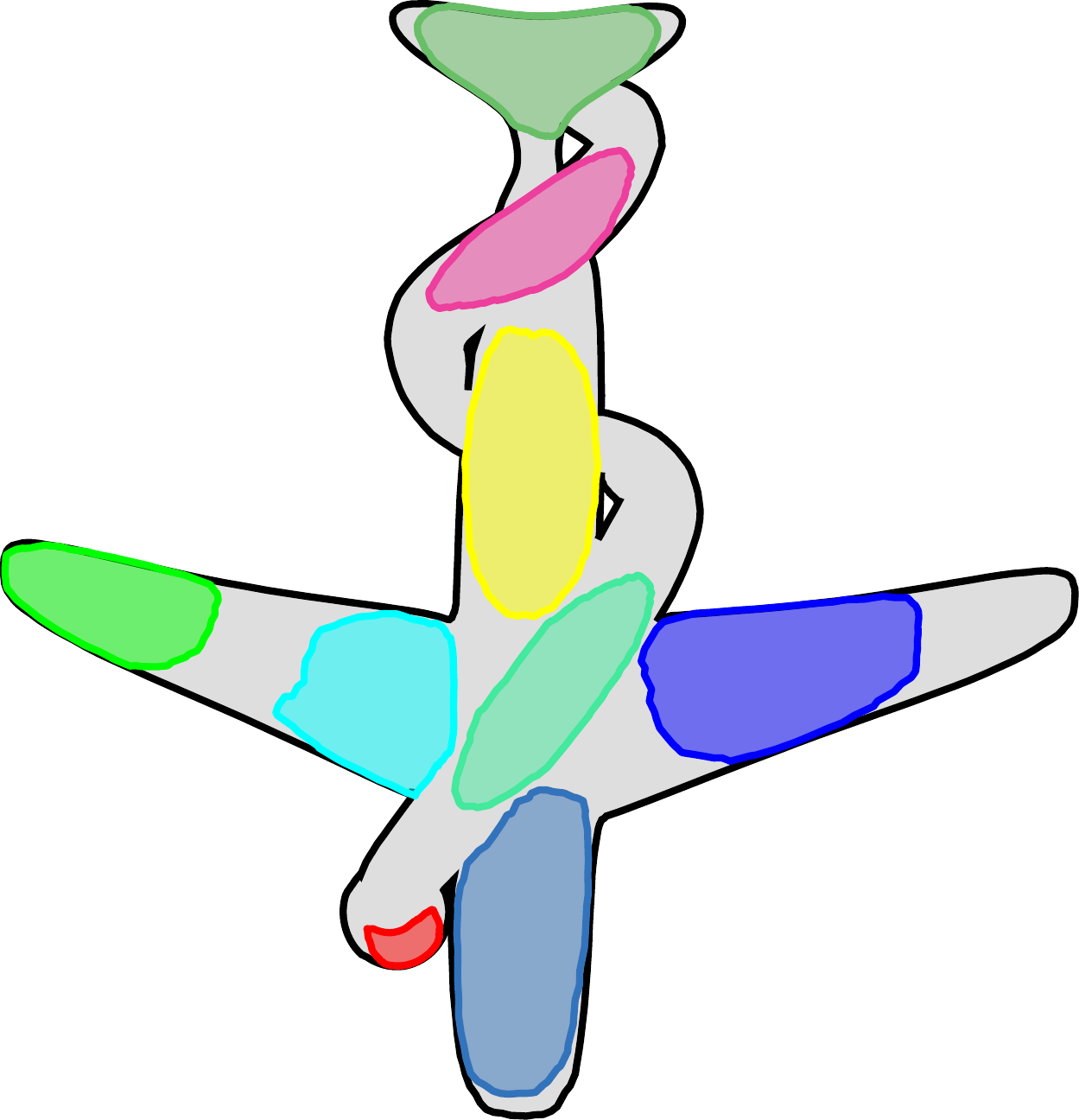} \vspace{2mm} \hspace{7.7mm}
\includegraphics[width=0.30\columnwidth]{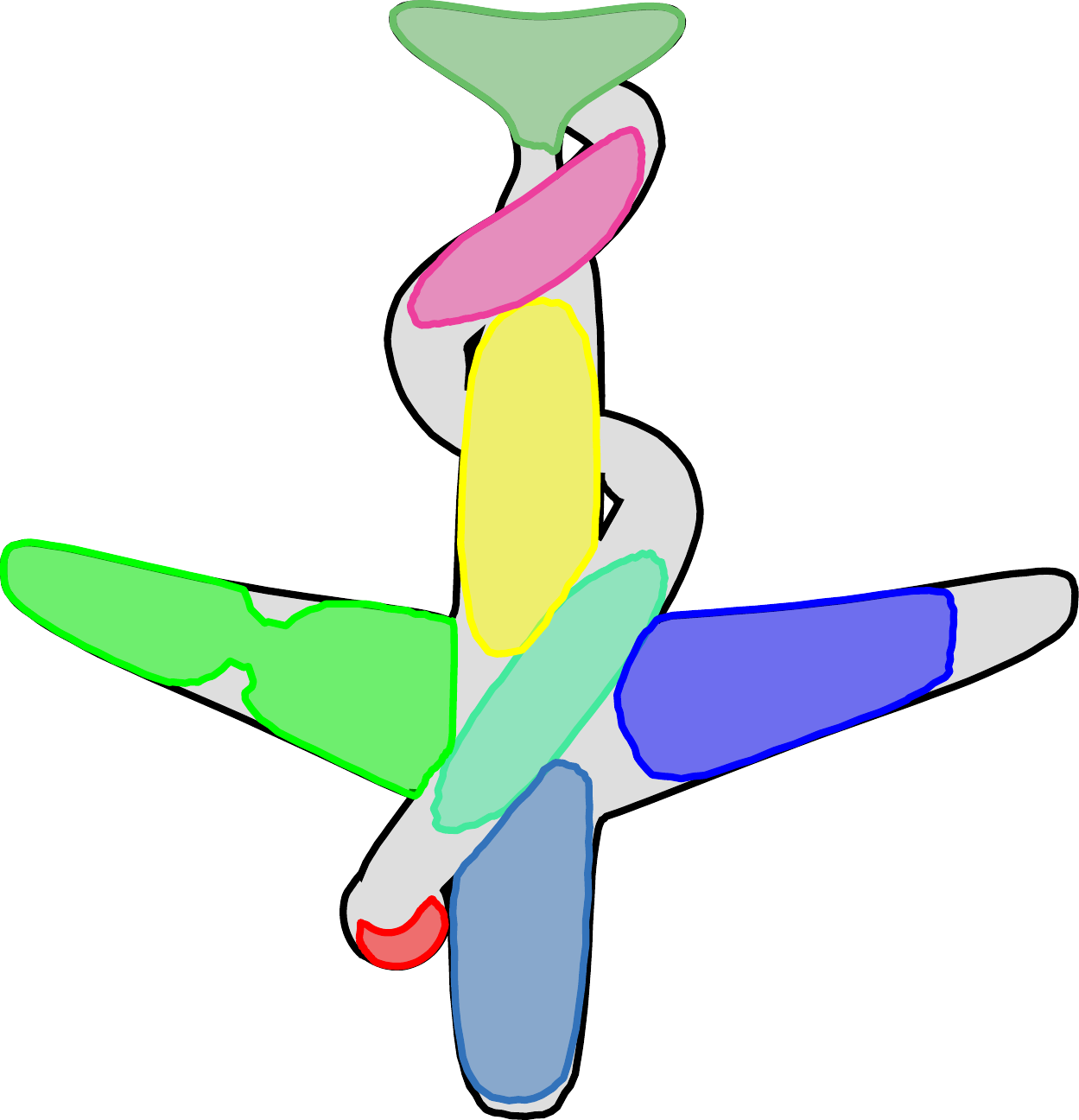} \vspace{2mm} \hspace{7.7mm}
\includegraphics[width=0.30\columnwidth]{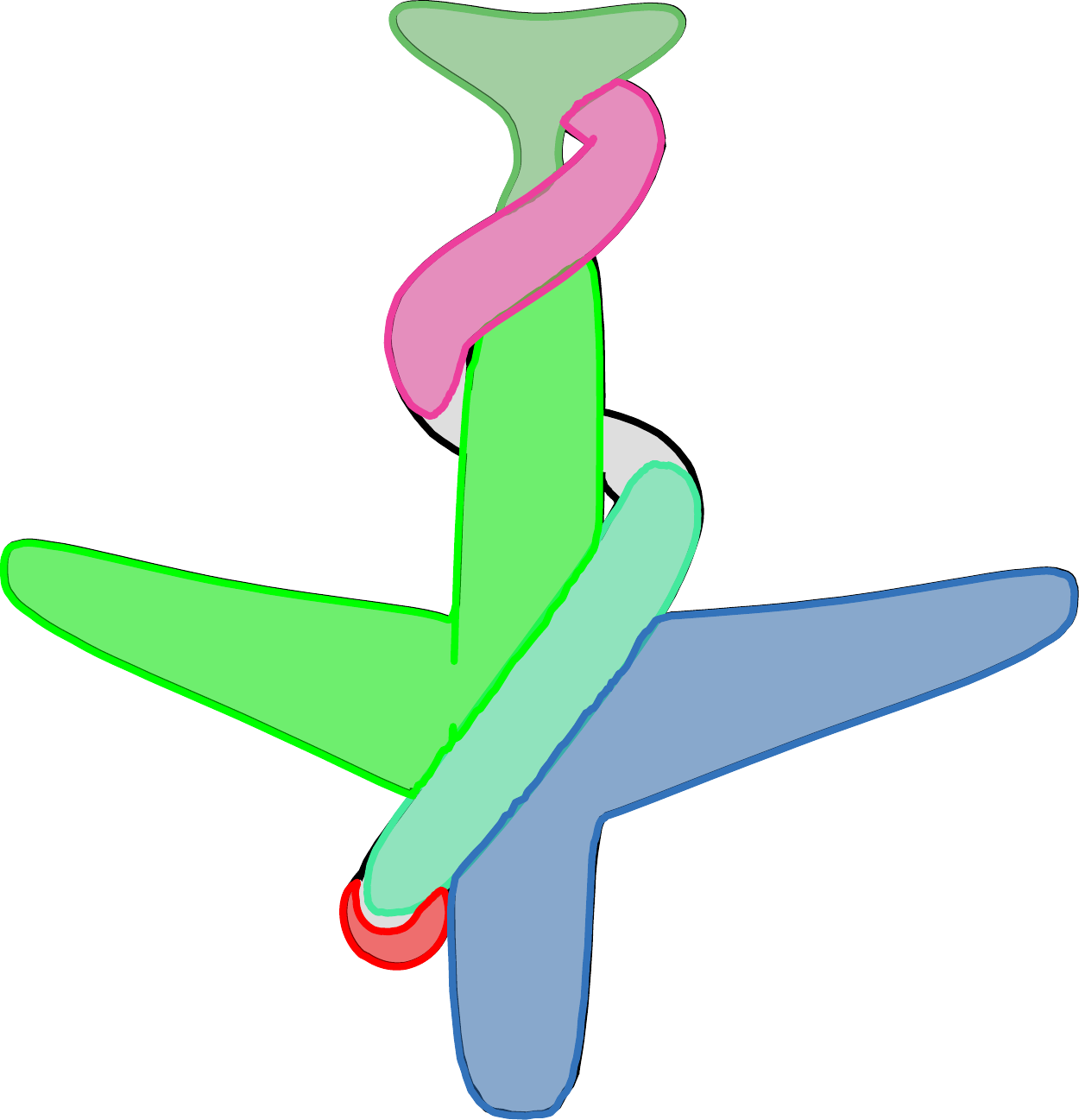} \vspace{2mm} \hspace{7.7mm}
\includegraphics[width=0.30\columnwidth]{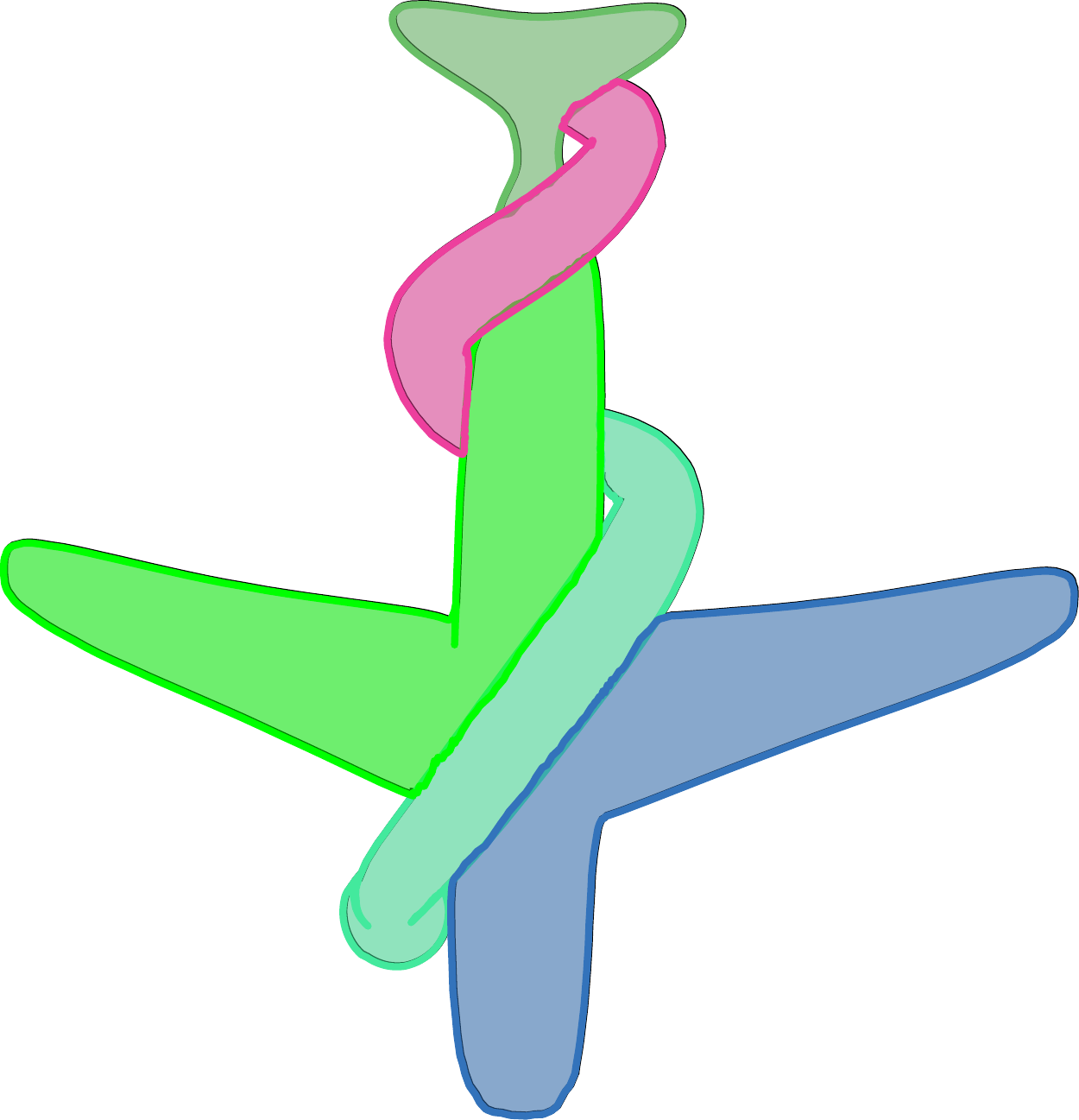}
}
\vspace{-7mm}
\centerline{ \Large \hspace{15mm} (a) \hfill (b) \hfill (c) \hfill (d) \hfill (e)  \hspace{15mm} }
\vspace{1mm}
\centerline{
\includegraphics[width=0.30\columnwidth]{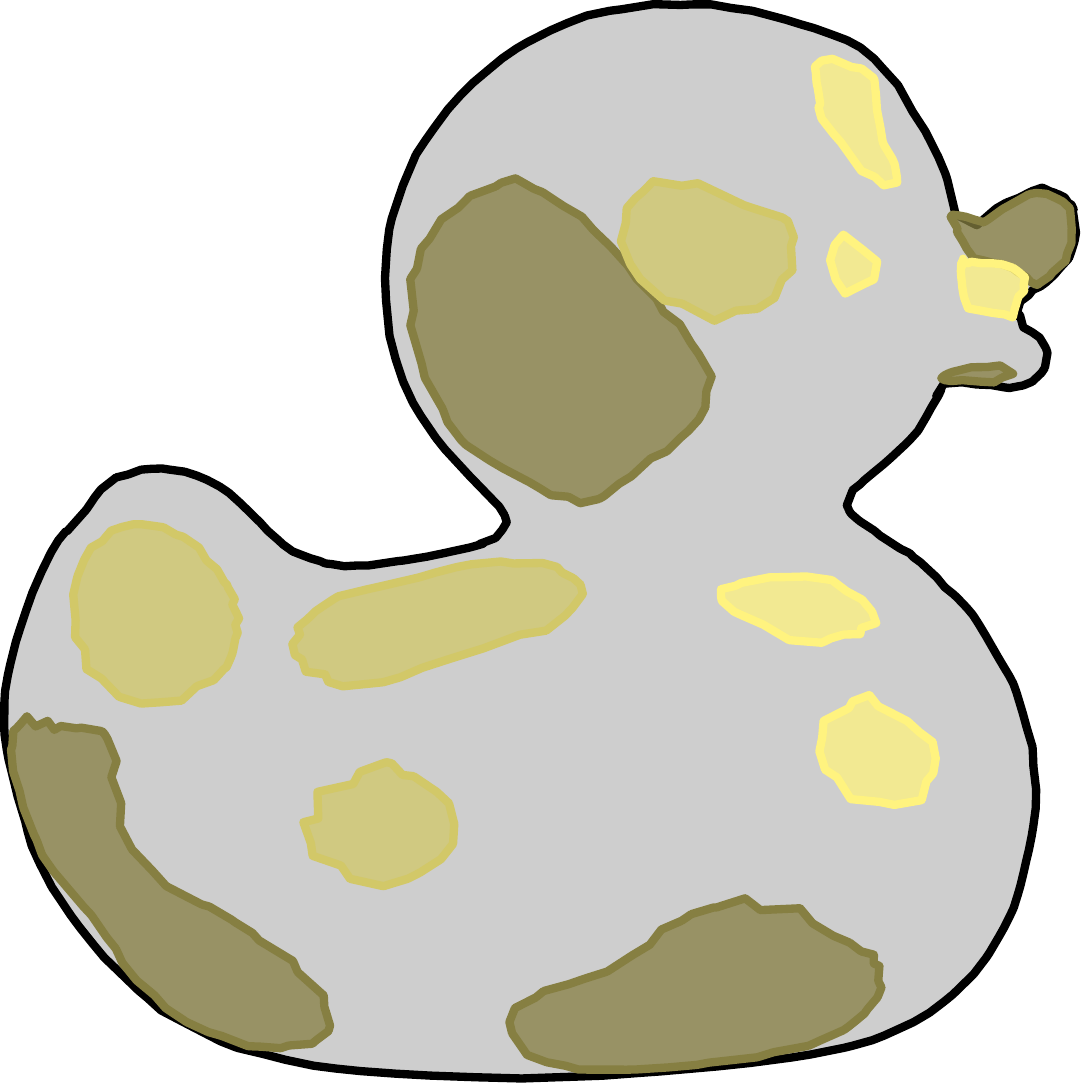} \vspace{2mm} \hspace{7.7mm}
\includegraphics[width=0.30\columnwidth]{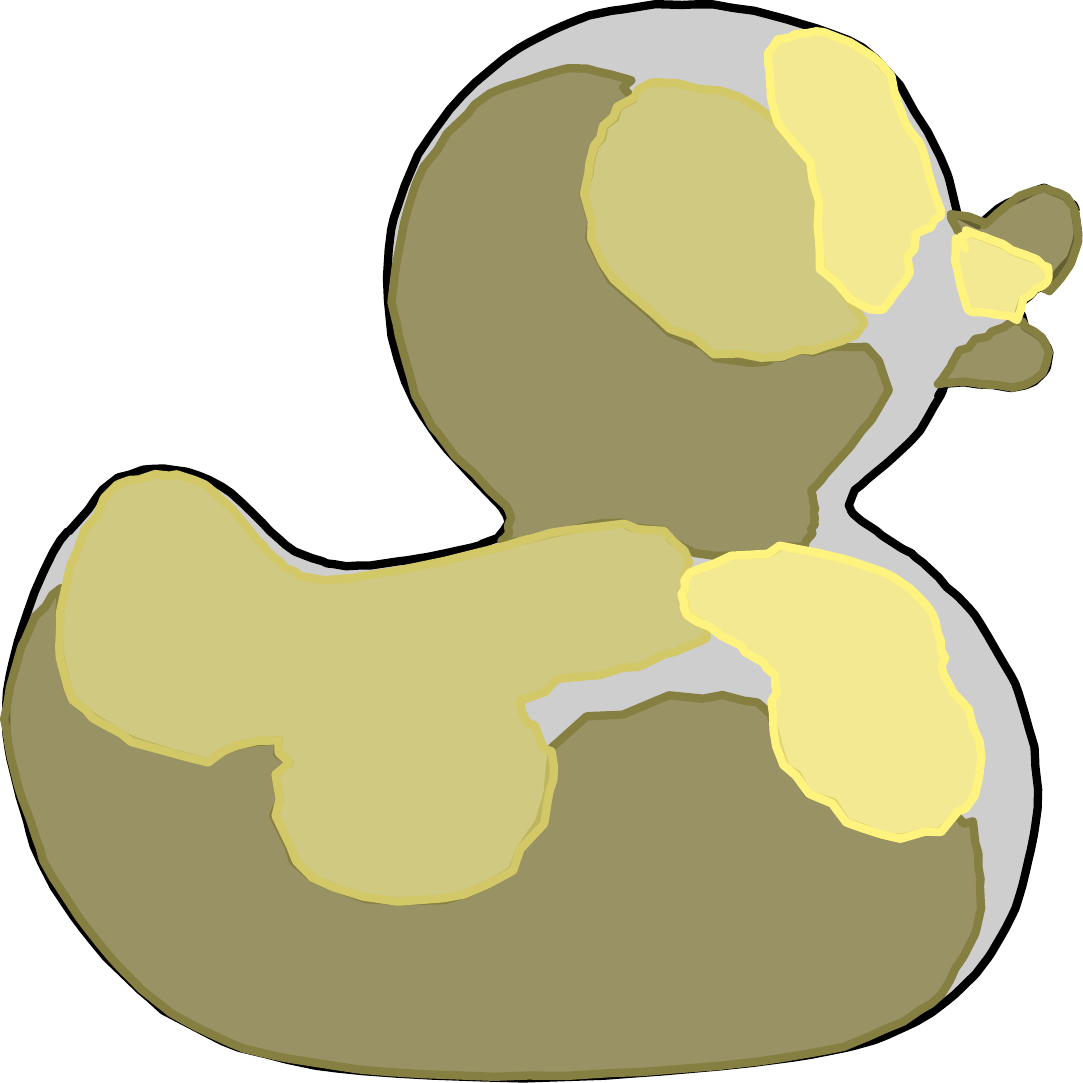} \vspace{2mm} \hspace{7.7mm}
\includegraphics[width=0.30\columnwidth]{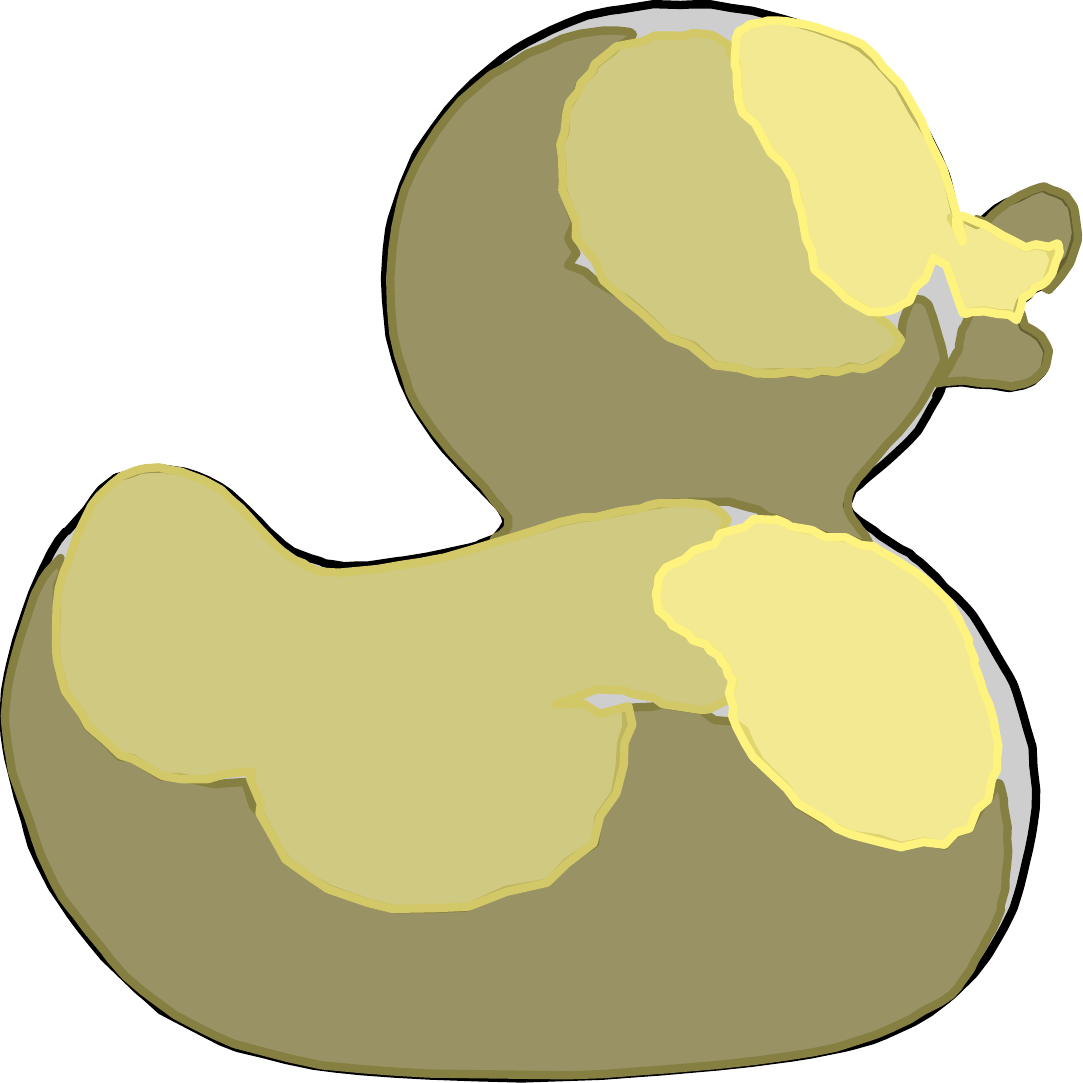} \vspace{2mm} \hspace{7.7mm}
\includegraphics[width=0.30\columnwidth]{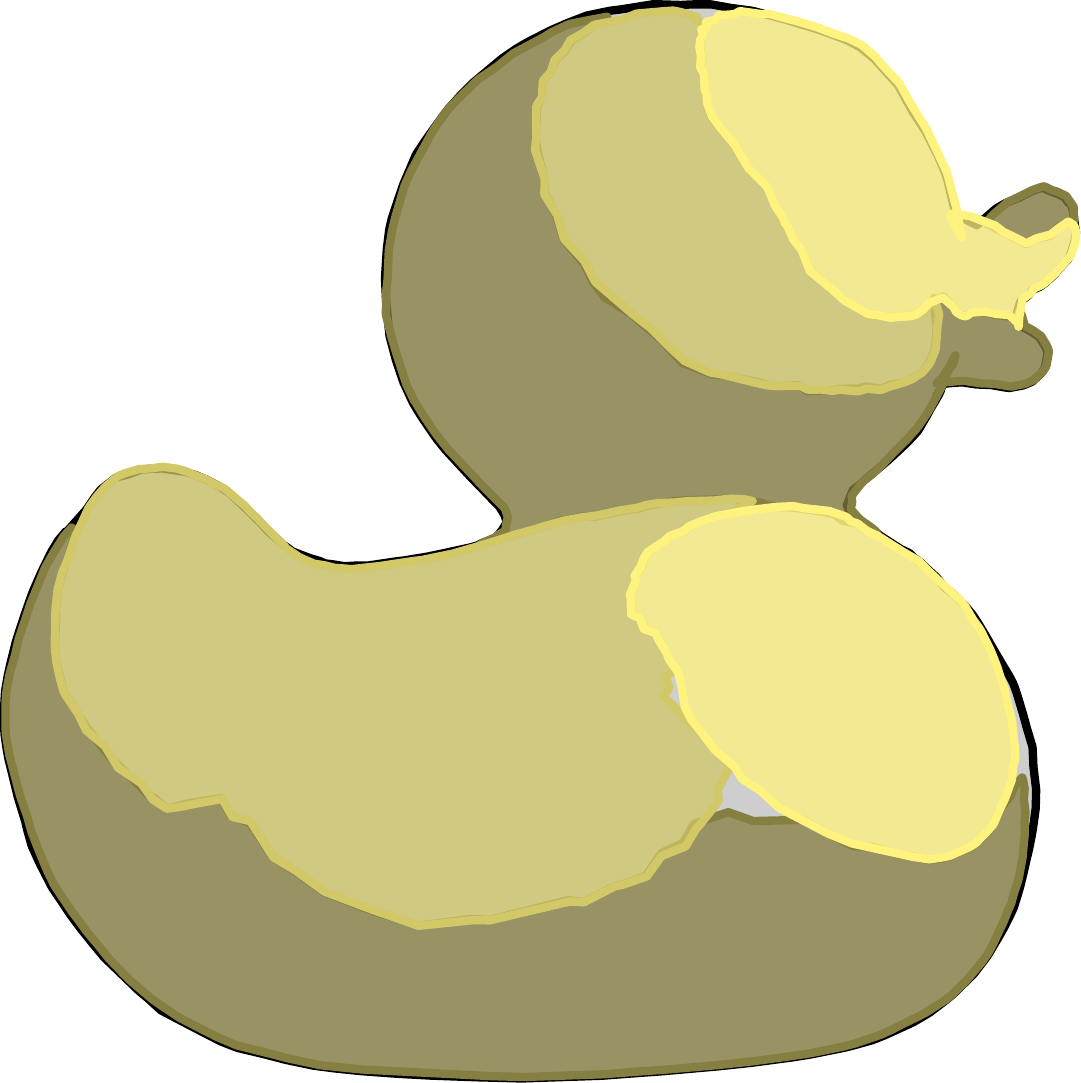} \vspace{2mm} \hspace{7.7mm}
\includegraphics[width=0.30\columnwidth]{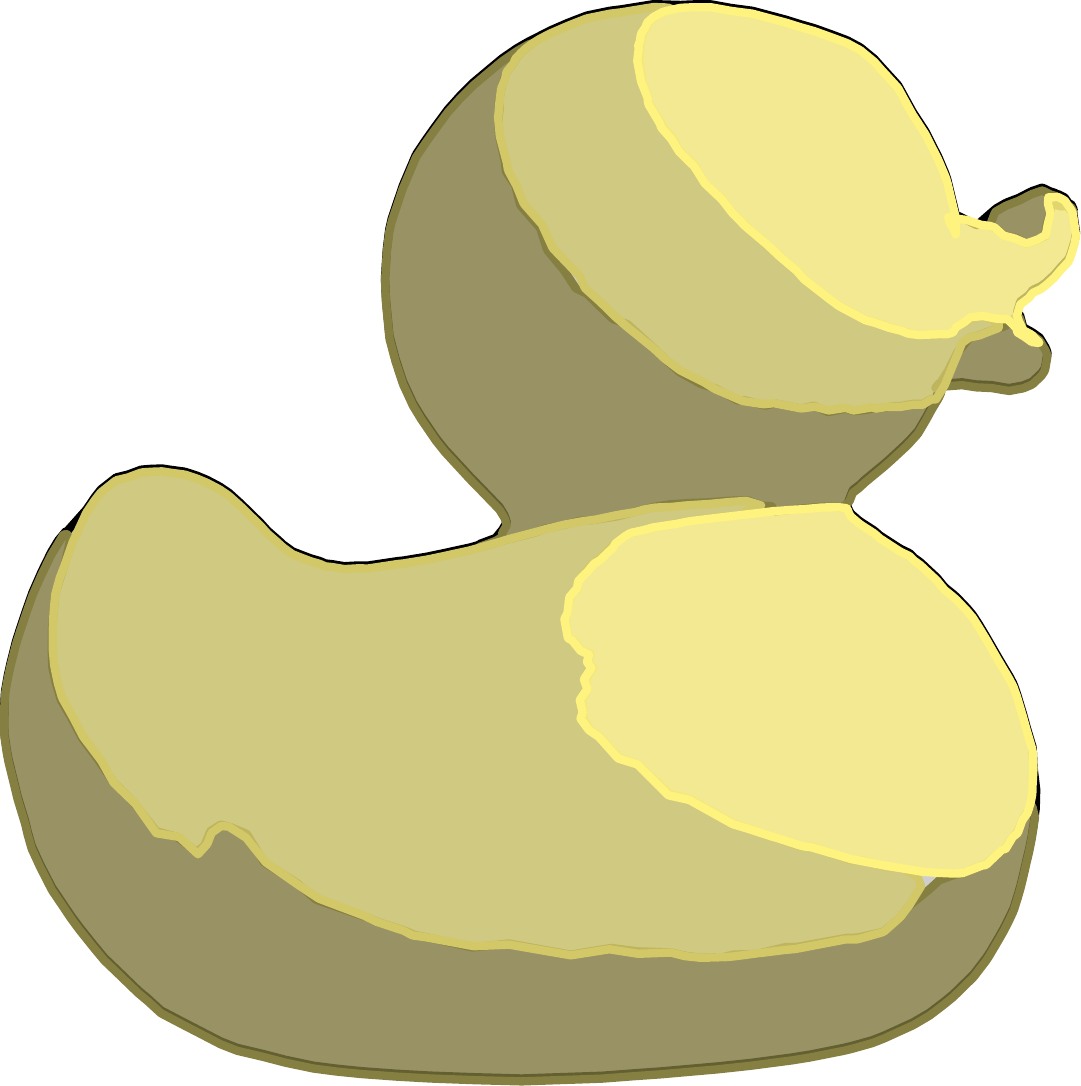}
}
\vspace{-8mm}
\centerline{ \Large \hspace{15mm} (f) \hfill (g) \hfill (h) \hfill (i) \hfill (j)  \hspace{15mm} } \vspace{-3mm}
\caption{Demonstration of planar map generation using our snaxel propagation technique. Our method works not only for visual contours \emph{(top)}, but also for shadow and diffuse isophote contours \emph{(bottom)}. Snaxel fronts are initialized automatically (left images), and these fronts evolve based on rules described in Section~\ref{sec:planar}. Rule 3 is illustrated as fronts collide and merge on the plane's wings (b-d), and ``darker'' isophote contours yield to ``brighter'' isophote contours based on rule 4 (h-j). The gray background denotes the constant-shaded mesh. \vspace{-3mm}}
\label{fig:planarmap_demo}
\end{figure*}

A specific problem that can occur when post-processing a planar map from
extracted contours is that the contours may not precisely bound the
projections of the regions they represent \cite{Eisemann_cgf08}. The silhouette
formed by the visual contour extracted from interpolated vertex normals does
not strictly contain the projection of the mesh, and other contours on the
mesh may escape its bounds. A post-processed planar map in these cases would
yield small regions near these imprecise locations caused by the approximation
of normal interpolation.

The snaxel formulation can be modified to generate a planar map that fixes
these problems, providing further control over precision and stylization as
well as simplifying
implementation. To generate a planar map, we amend the
snaxel topology rules as follows:
\begin{enumerate}
\item We detect collisions between snaxels of any type (visual, shadow,
shading).
\item We detect snaxel collisions in the image plane as well as on mesh
edges. We hence maintain the projection of each snaxel on the image plane in
addition to its position on the mesh edge.
\item Snaxel fronts of the same type that collide on the 3-D surface are allowed
to merge and split, but snaxel fronts of different types whose image
projections collide are not allowed to merge, and instead push against each
other.
\item Occluded fronts must yield to visible fronts when their projections
collide. Similarly, ``darker'' isophote contours must yield to ``brighter'' isophote
 contours (e.g. a contour with $k=0.5$ will have precedence over another contour with $k=0$).
\end{enumerate}
These rules generate a visible-surface planar map consisting of regions of
different combinations of shading (e.g. illuminated v. shadow). These rules are easily expressed in terms of a new energy functional
 \begin{equation}
 \begin{split}
f_{\text{pmap}}(s_i) &= -\delta(s_i) f(s_i) , \text{ where}\\
 \delta(s_i) &= \left\{ \begin{array}{rl}  1 & \text{ if $s_i$ is within a closer front } \\ -1 & \text{ otherwise} \end{array} \right.
 \end{split}
 \label{eq:planarmapenergy}
 \end{equation}
where $f$ is the chosen energy function (e.g. Eq~\ref{eq:vcenergy}, \ref{eq:isophote}, or \ref{eq:gleams}). The statement ``$s_i$ is within a closer front'' means that the snaxel $s_i$ is inside of the 2D polygon defined by another front's image-space projection, and that the given front is closer to the camera than $s$. Figure~\ref{fig:planarmap_demo}  demonstrates snaxels propagating over a mesh while adhering to these rules; equivalently, deforming based on Eq.~\ref{eq:planarmapenergy}. 

We keep track of snaxel position on the mesh edges as well as their projected
positions in the image plane. To detect when snaxel fronts overlap in image-space, 
we use the even-odd-rule algorithm, which also ensures that no contours
are self-intersecting. Such tests can be expensive for large meshes, and collisions
may also be detected at raster resolution using other, less expensive techniques
(e.g. occlusion queries). 

Initialization can be tricky for planar map evolution. Ideally one would want
a snaxel front initialized for every planar map region, but these regions are
not known a priori. Hence we test the mesh vertices after a planar map has
been generated to ensure the snaxel-defined regions match their actual
classification (e.g. a vertex in the shadowed region is actually shadowed).
Any mis-classified vertices are reinitialized with corrected fronts and the
planar map is refined.

Snaxels define an active contour front based on snaxel positions along
edges, and when they form a visible-surface planar map they overcome the
problems caused by an approximate silhouette generated by interpolated vertex
normals. The snaxels live on mesh edges, and by definition form loops that
enclose regions, to a resolution defined by the surface mesh tessellation.
Hence small mesh corners exposed beyond an imperfect interpolated-normal
smoothed silhouette are not tessellated enough to support an isolated snaxel
front (including any discovered by the aforementioned re-initialization pass)
and so are conveniently filtered away by the snaxel visible-surface planar map.

\section{3-D SVG Animation} \label{sec:anim} \vspace{-2mm}
Our snaxel framework allows us to convert frame-based NPR animations into 2-D,
keyframed animations of projected outline curves and regions. In other words,
we can represent animated vector art not as a sequence of individual vector
art frames, but as a smaller subset of keyframes, interpolating the outline
curves between them by moving their vertices. The advantage of such a
representation for animated vector art are its compactness and intuitive
representation, as well as its support for temporally coherent stylization
\cite{kalnins03}.

The main challenge of converting the motion of a 3-D surface contour into the
motion of its 2-D projection is the correspondence of the curve representation
across frames of the animation. Given an animation sequence, such as rotating
or deforming a meshed object, the snaxel fronts will evolve to produce contours
for each frame of the sequence.
 The snaxels defining these contours provide the
necessary support to generate correspondences from frame to frame.
%, and we 
%can process the
%correspondences of each of these contours separately, independently of each
%other, so long as we consider the entire spacetime image of a contour. (E.g.
%two disjoint contours may merge into a single contour and then divide again
%later, which we would consider a single spacetime contour.)

\begin{figure} 
\includegraphics[width=\columnwidth]{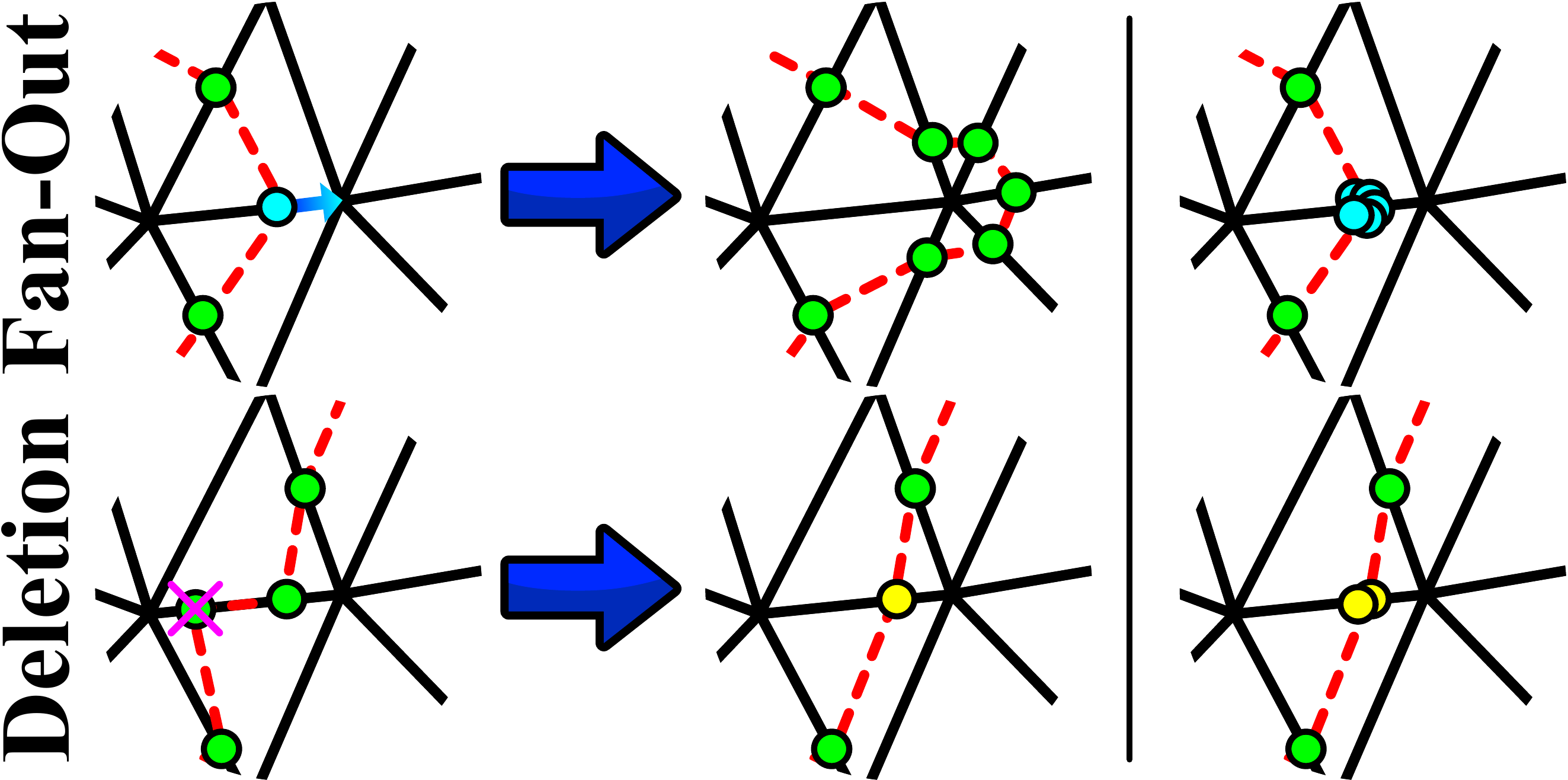} 
 \centerline{ \bf \large \hspace{7mm} Frame $k$ \hspace{15mm} Frame $k+1$ \hspace{8mm} Colocated \hfill }\vspace{-2mm}
\caption{Snaxels are tracked over time and added prior to fan-out and after deletion operations so that contours maintain the same number of vertices.  The colocated column shows the contours after the colocation process; teal and yellow colored circles represent time-corresponding snaxels before and after colocation.
\vspace{-3mm}}
\label{fig:coherence}
\end{figure}

SVG polyline animation does not support changes in topology, so we insert
keyframes at every contour topology change event, breaking the spacetime
contour into contour sequences that do not change topology, and process the
snaxels of each contour sequence into the vertex motion of an animated
polyline. Such animation also requires each contour within a given
sequence to have the same number of vertices, which we achieve by
detecting temporal correspondences between vertices, and then appropriately  
adding colocated vertices based on these correspondences (see Fig~\ref{fig:coherence}).

\begin{figure} \vspace{-3mm}
\centerline{
\includegraphics[width=0.45\columnwidth]{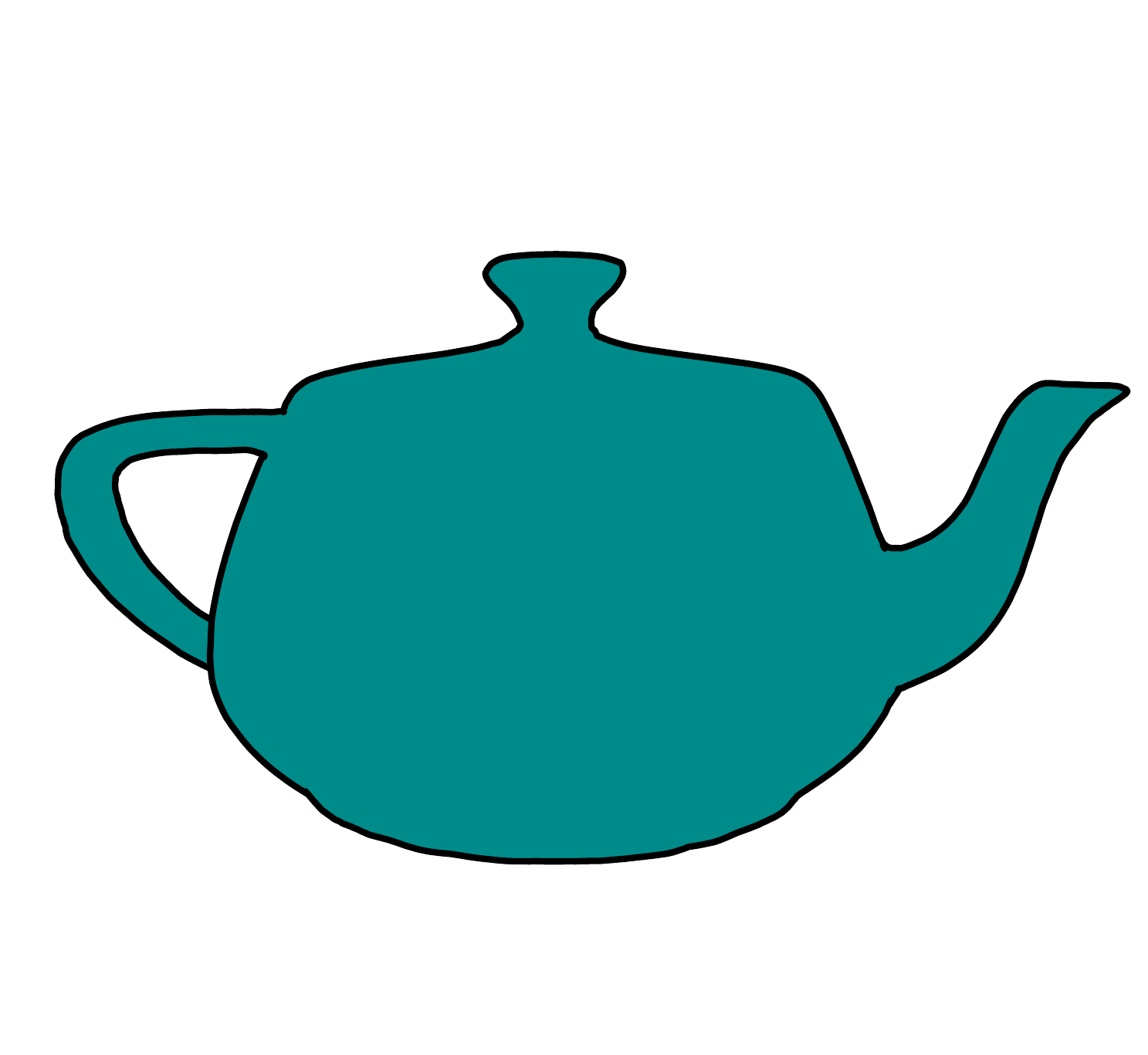}  \hfill
\includegraphics[width=0.45\columnwidth]{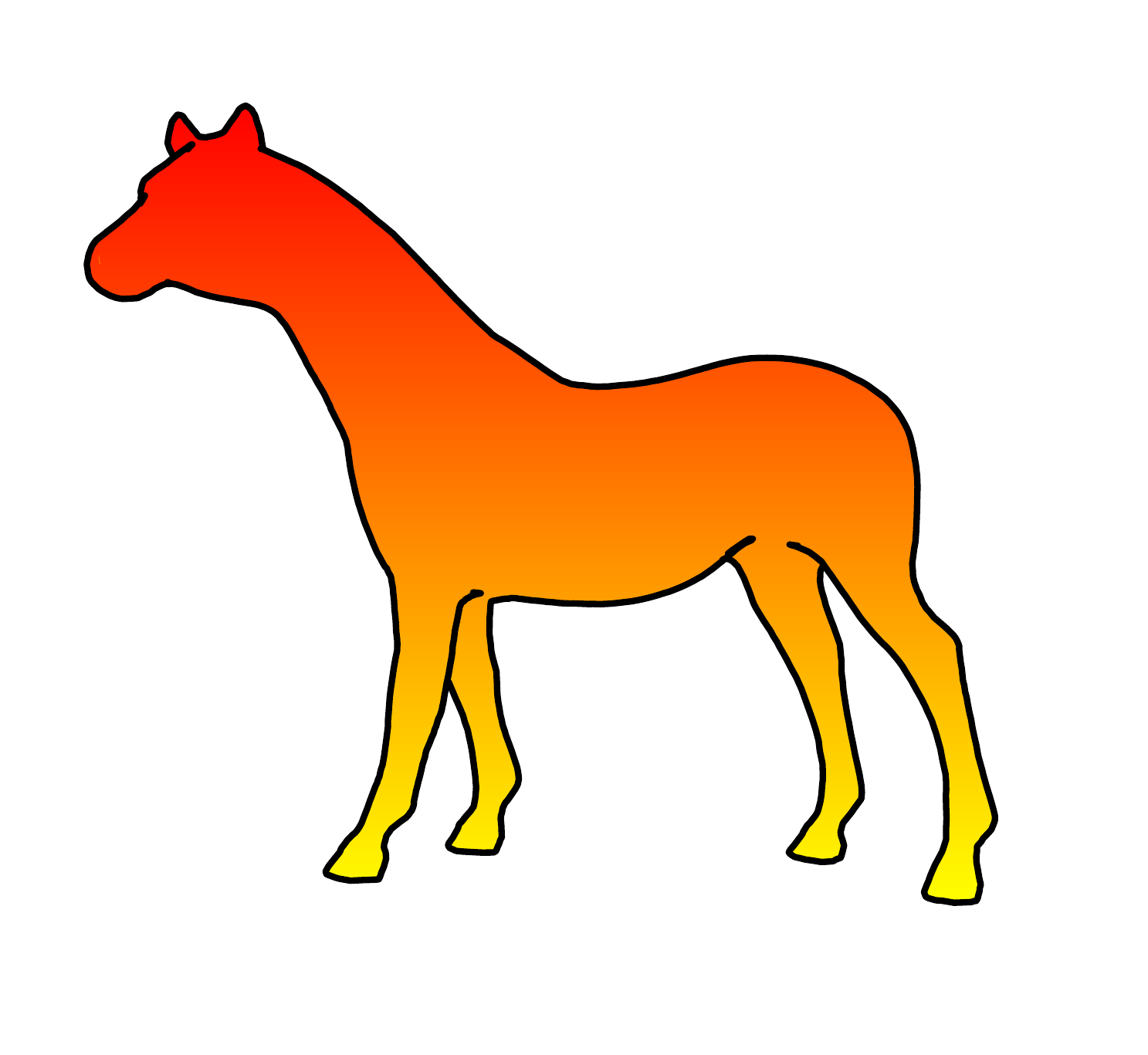}\\ \vspace{-2mm}
}
\centerline{
\includegraphics[width=0.45\columnwidth]{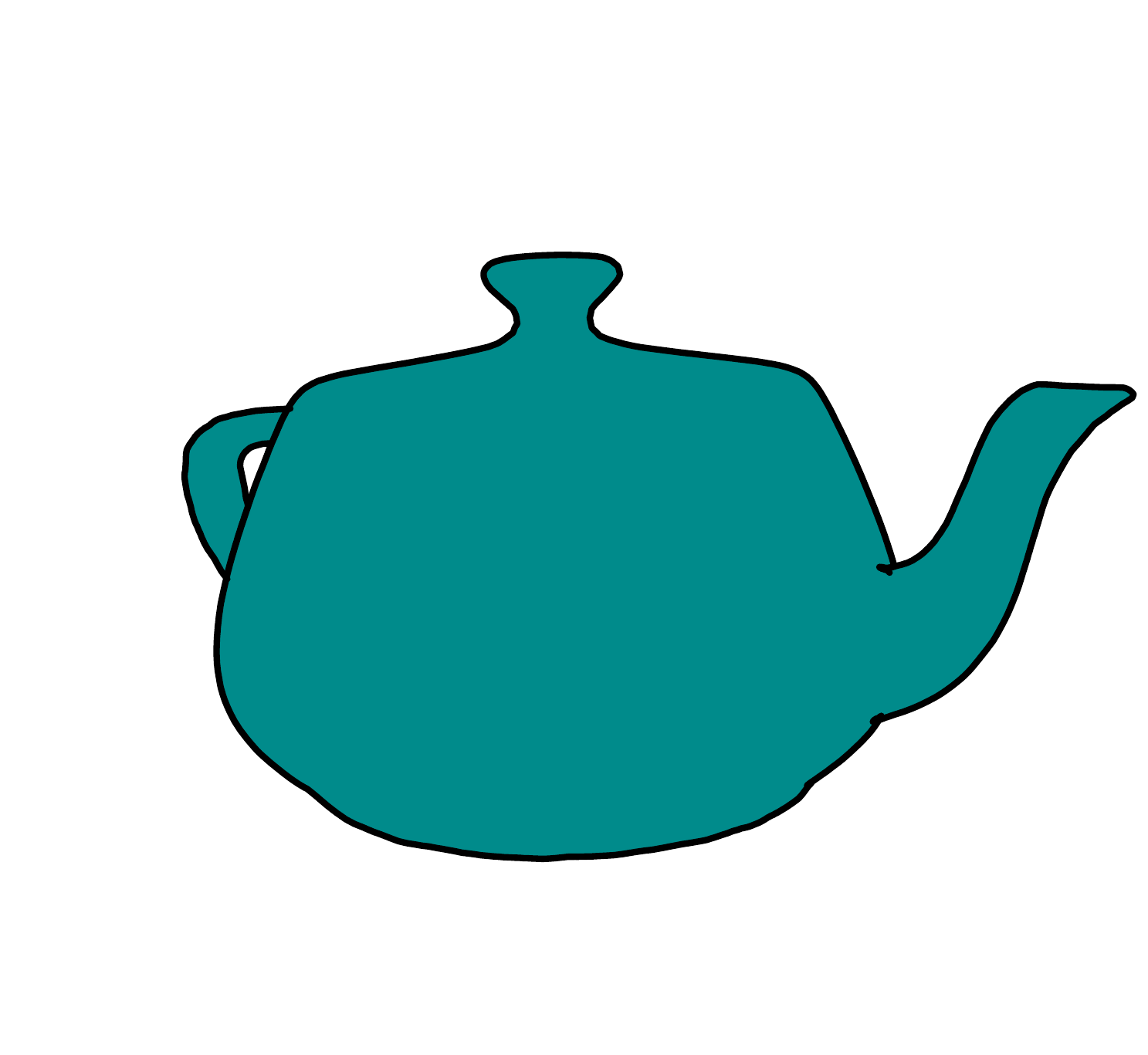} \hfill
\includegraphics[width=0.45\columnwidth]{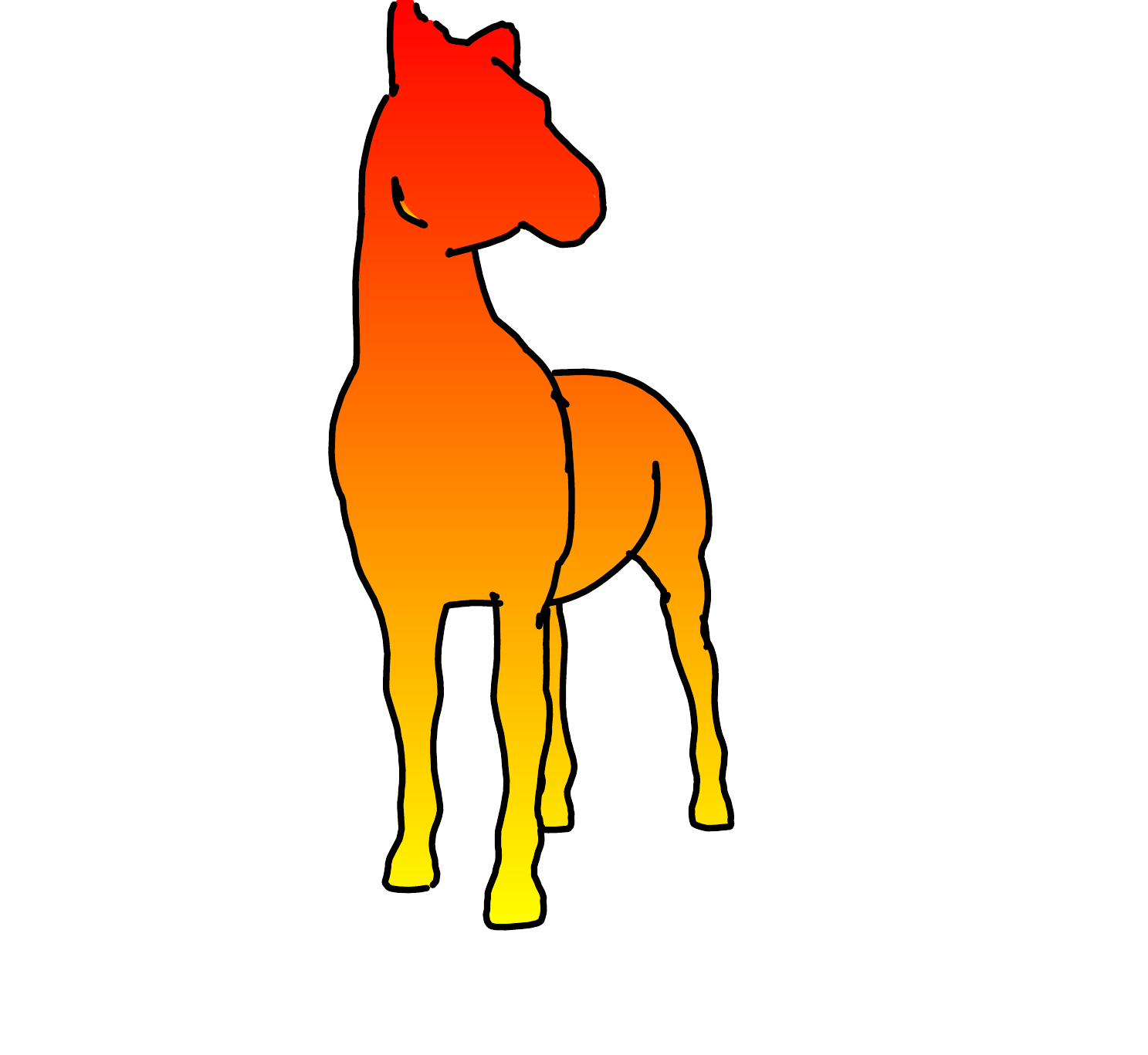}\\ \vspace{-6mm}
}
\centerline{
\includegraphics[width=0.45\columnwidth]{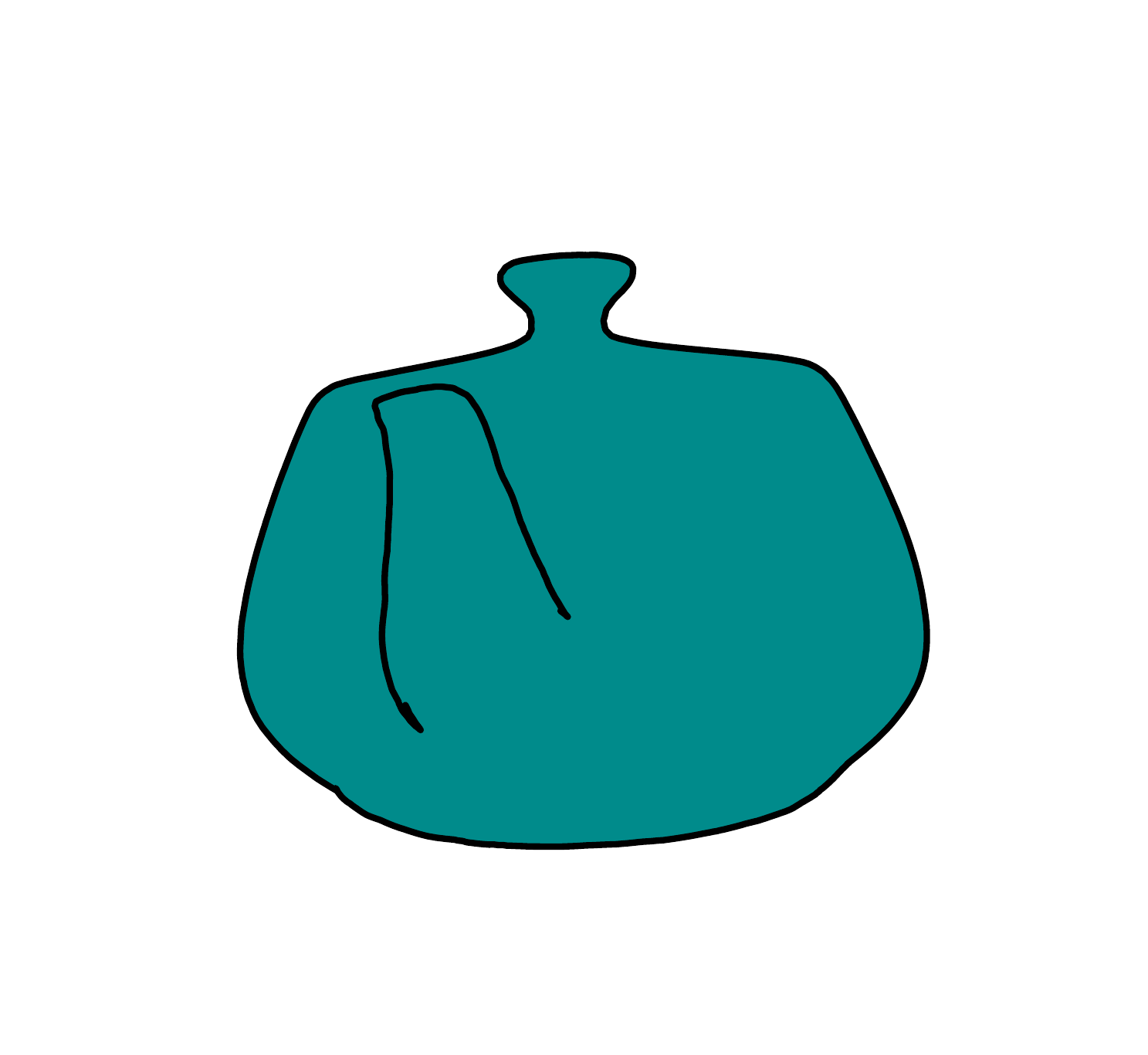} \hfill
\includegraphics[width=0.45\columnwidth]{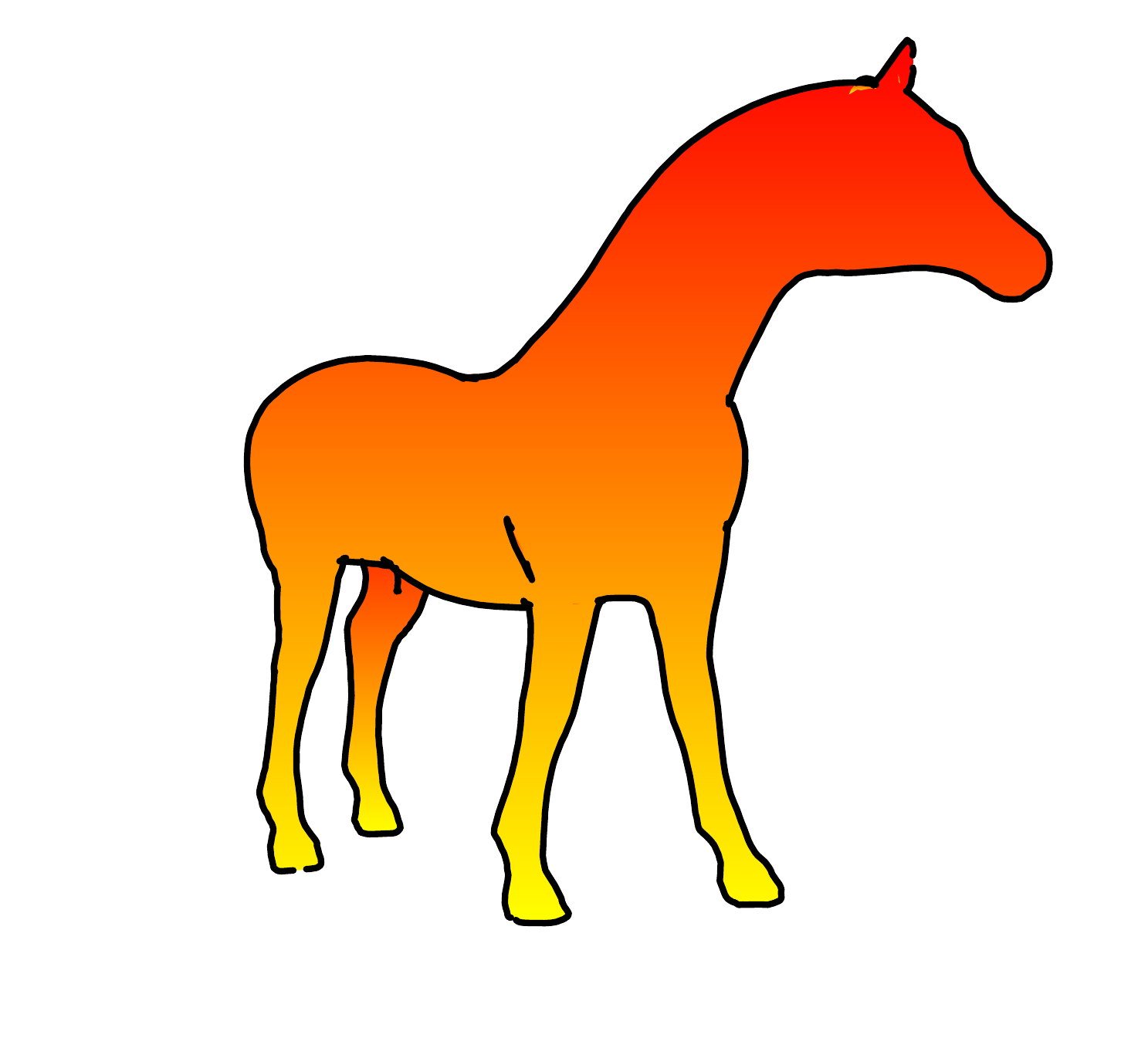}\\ \vspace{-6mm}
}
\centerline{
\includegraphics[width=0.45\columnwidth]{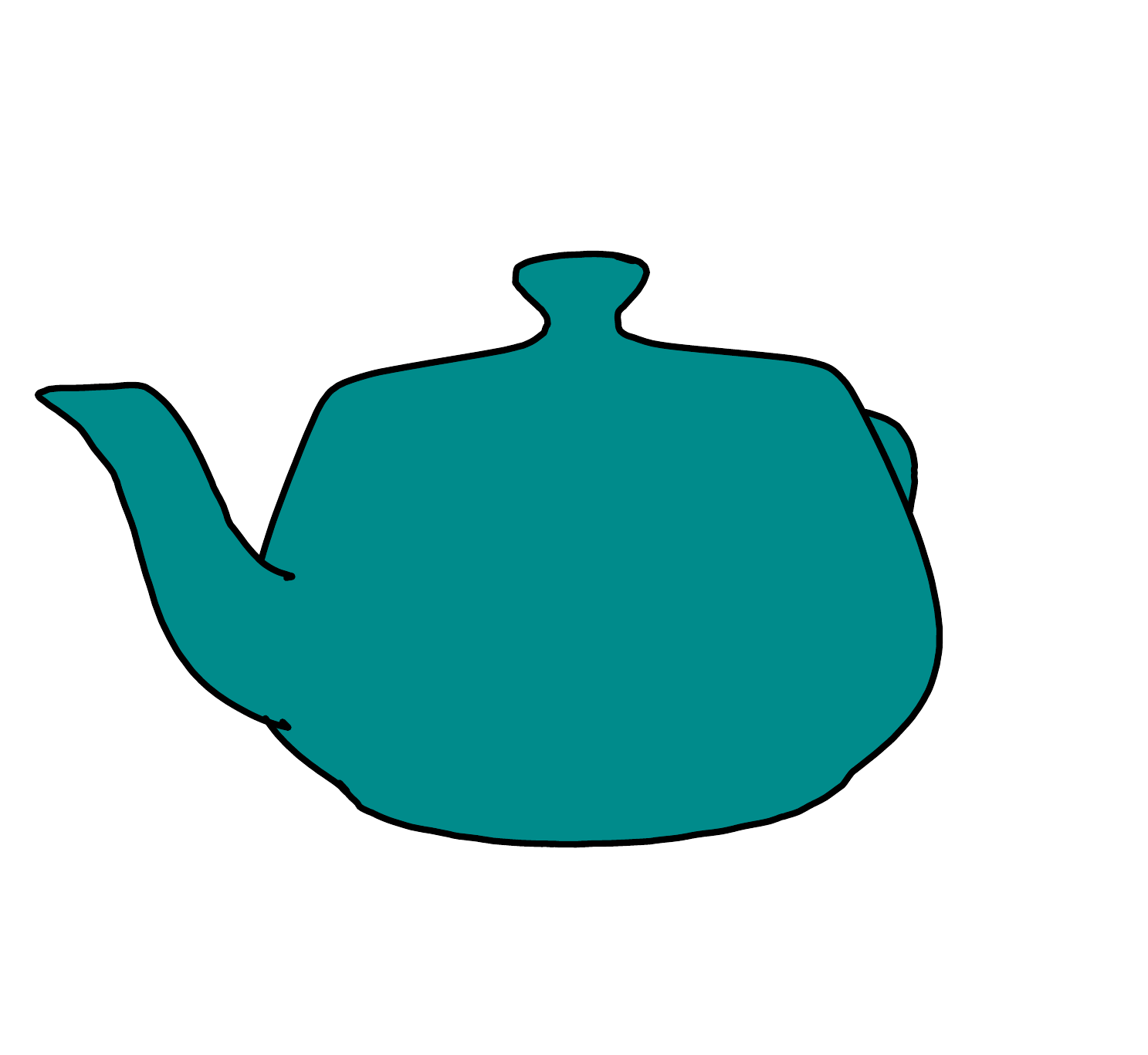} \hfill
\includegraphics[width=0.45\columnwidth]{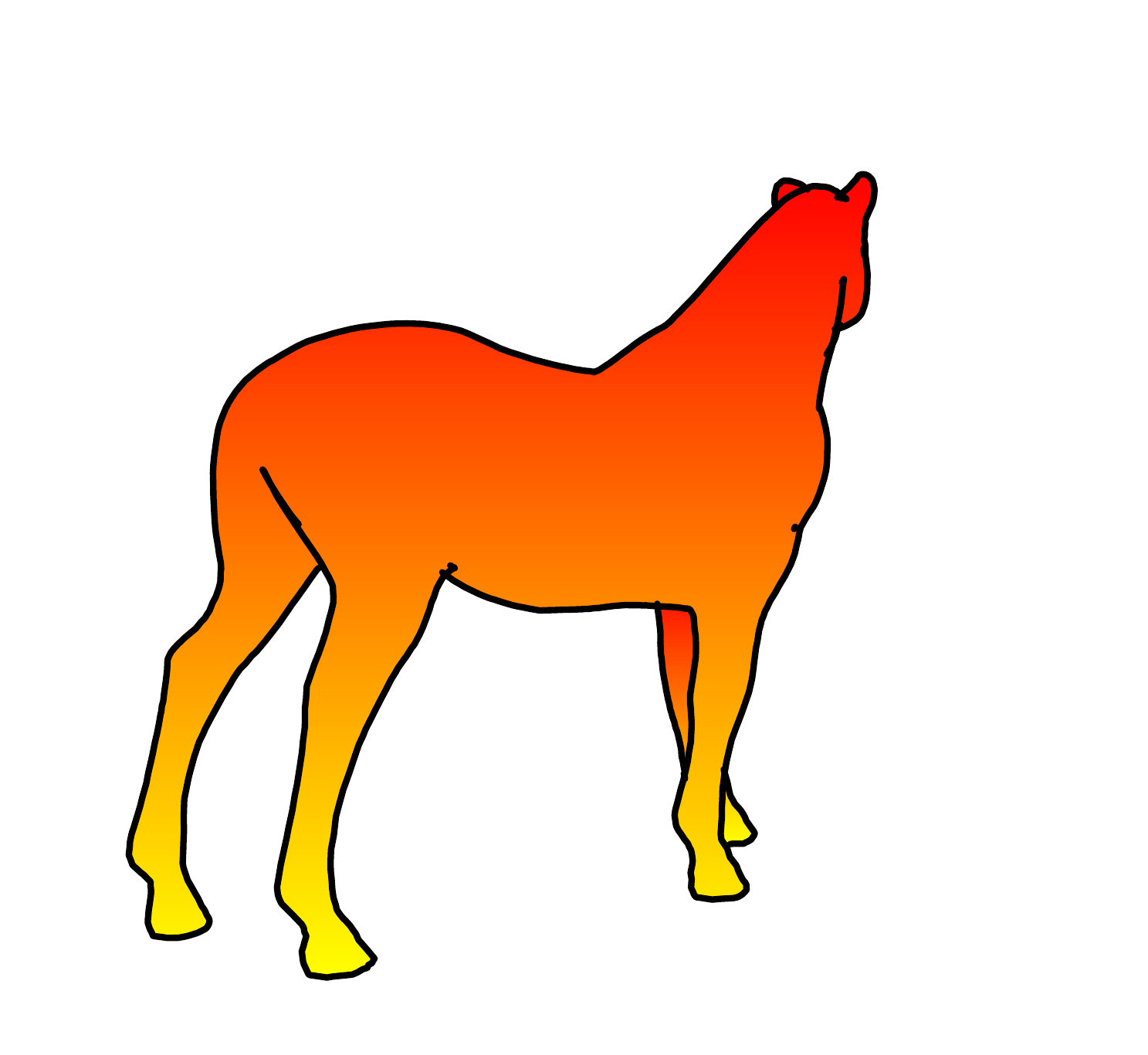}\\ \vspace{-6mm}
}
\centerline{
\includegraphics[width=0.45\columnwidth]{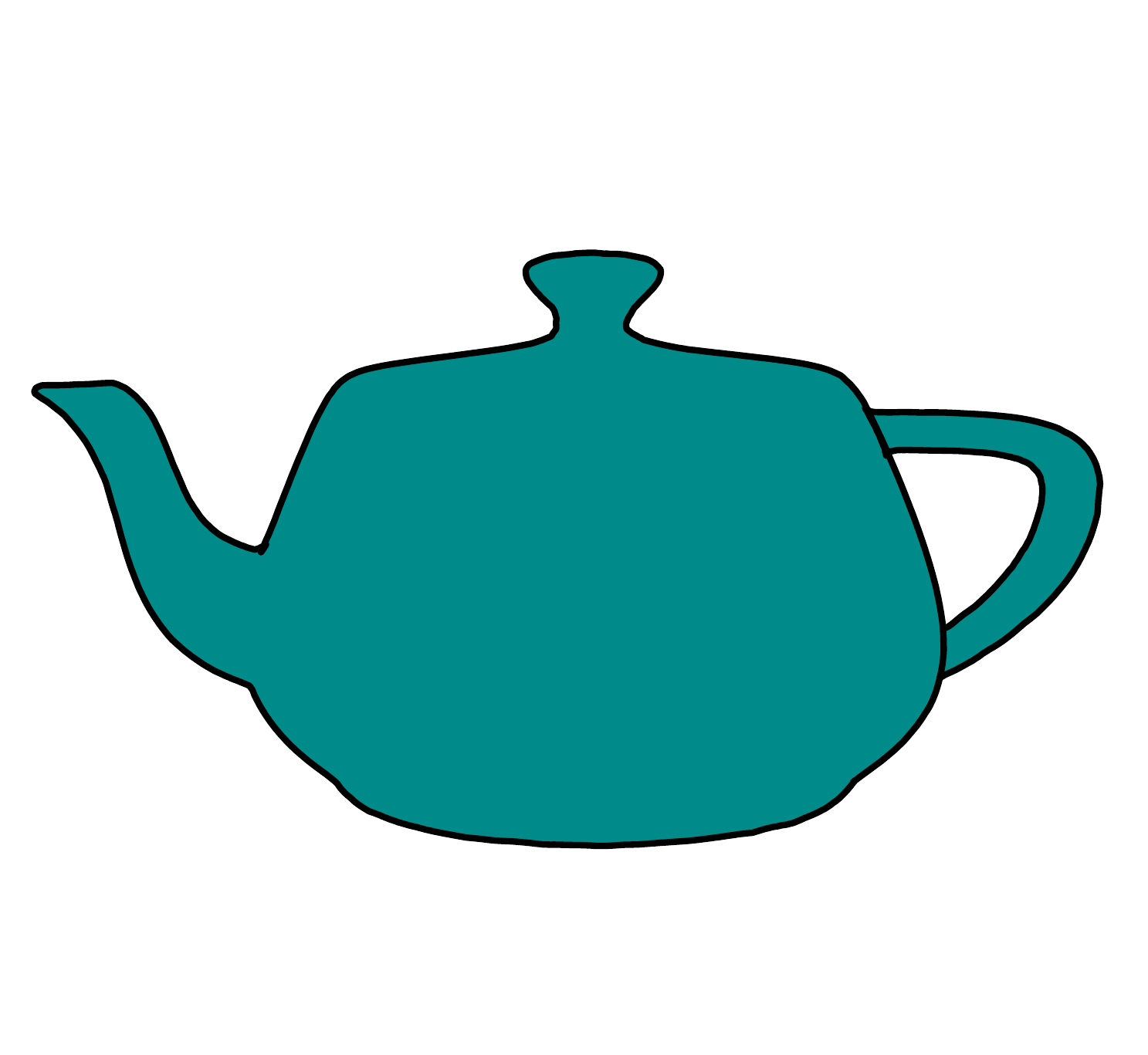} \hfill
\includegraphics[width=0.45\columnwidth]{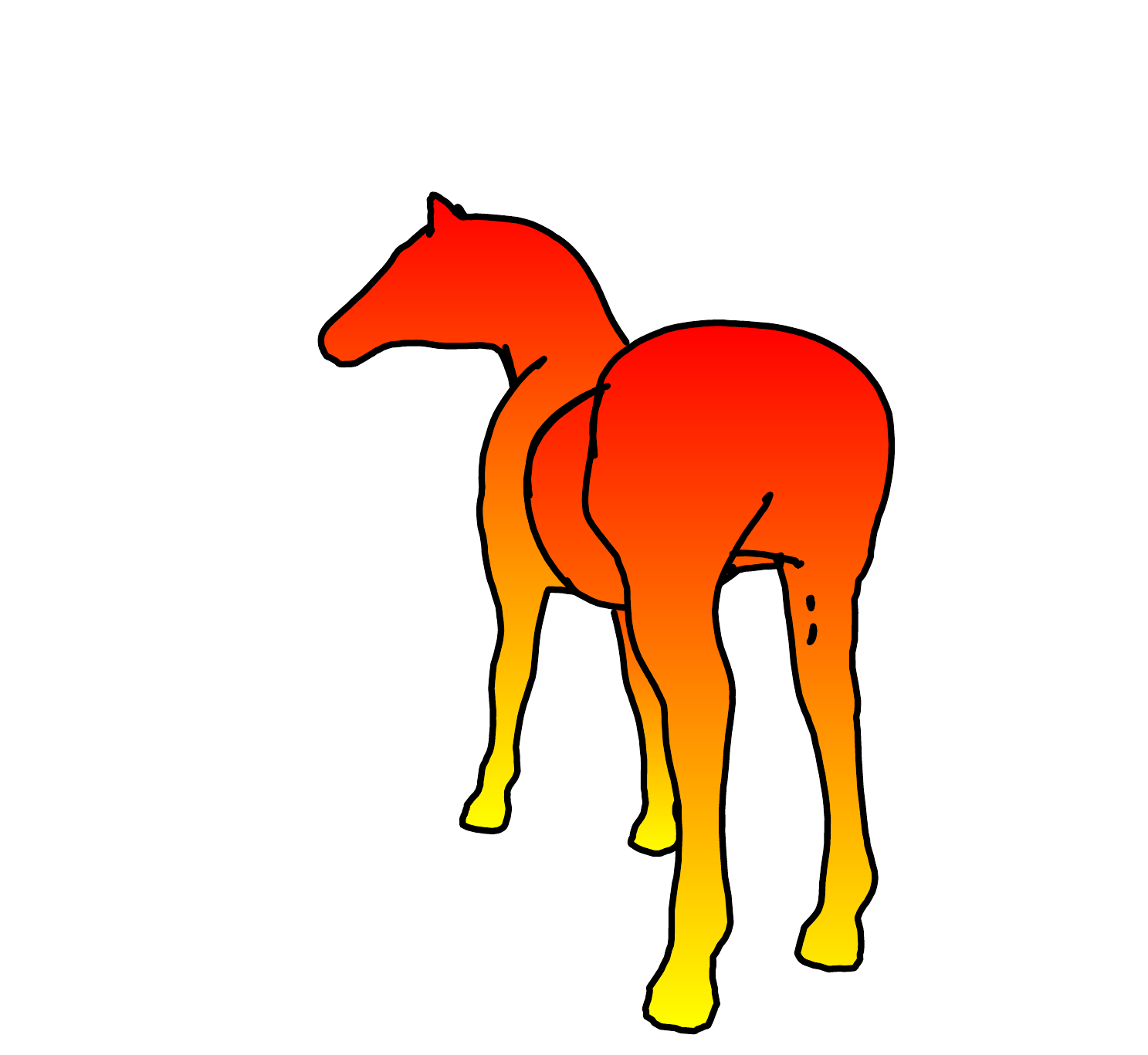} \vspace{-2mm}
}
\caption{Frames from animated SVG files produced by our system. The horse and teapot rotate 
about a vertical axis in 3-D, and the motion of their visual
contours is reproduced as 2-D polyline animation, by interpolating the
projected contour polyline vertices between keypoints. The resulting animation
can be replayed at any frame rate as the contours are interpolated continuously
across time. \vspace{1mm}}
\label{fig:2dcurveanim}
\end{figure}

\newcommand{\comment}[1]{}

\comment{
\begin{figure*}
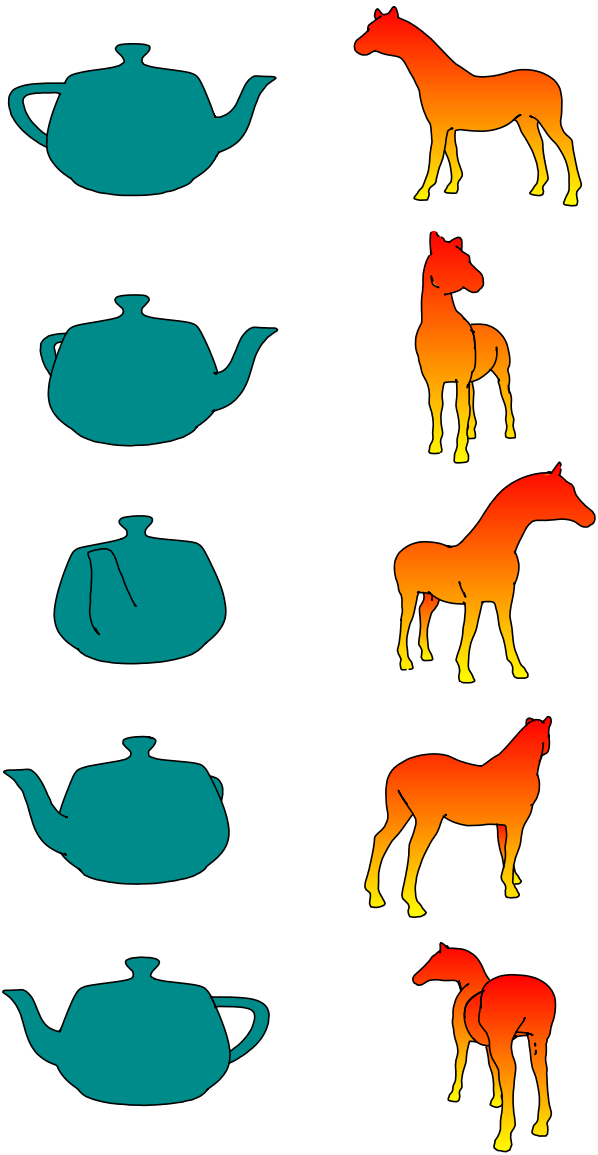

\begin{minipage}[t]{.2\textwidth} \centering
\includegraphics[width=\textwidth]{fig/2d_curve_anim/teapot_frame1}%
\end{minipage}%
\begin{minipage}[t]{.2\textwidth} \centering
\includegraphics[width=\textwidth]{fig/2d_curve_anim/teapot_frame2}%
\end{minipage}%
\begin{minipage}[t]{.2\textwidth} \centering
\includegraphics[width=\textwidth]{fig/2d_curve_anim/teapot_frame3}%
\end{minipage}%
\begin{minipage}[t]{.2\textwidth} \centering
\includegraphics[width=\textwidth]{fig/2d_curve_anim/teapot_frame4}%
\end{minipage}%
\begin{minipage}[t]{.2\textwidth} \centering
\includegraphics[width=\textwidth]{fig/2d_curve_anim/teapot_frame5}%
\end{minipage}\\
\begin{minipage}[t]{.2\textwidth} \centering
\includegraphics[width=\textwidth]{fig/2d_curve_anim/horse_frame1}%
\end{minipage}%
\begin{minipage}[t]{.2\textwidth} \centering
\includegraphics[width=\textwidth]{fig/2d_curve_anim/horse_frame2}%
\end{minipage}%
\begin{minipage}[t]{.2\textwidth} \centering
\includegraphics[width=\textwidth]{fig/2d_curve_anim/horse_frame3}%
\end{minipage}%
\begin{minipage}[t]{.2\textwidth} \centering
\includegraphics[width=\textwidth]{fig/2d_curve_anim/horse_frame4}%
\end{minipage}%
\begin{minipage}[t]{.2\textwidth} \centering
\includegraphics[width=\textwidth]{fig/2d_curve_anim/horse_frame5}%
\end{minipage}
\caption{Frames from animated SVG files produced by our system. The horse and teapot rotate 
about a vertical axis in 3-D, and the motion of their visual
contours is reproduced as 2-D polyline animation by interpolating the
projected contour polyline vertices between keypoints. The resulting animation
can be replayed at any frame rate as the contours are interpolated continuously
across time. \vspace{0mm}}
\label{fig:2dcurveanim}
\end{figure*}
}

\begin{figure}
\centerline{
\includegraphics[width=0.3\columnwidth]{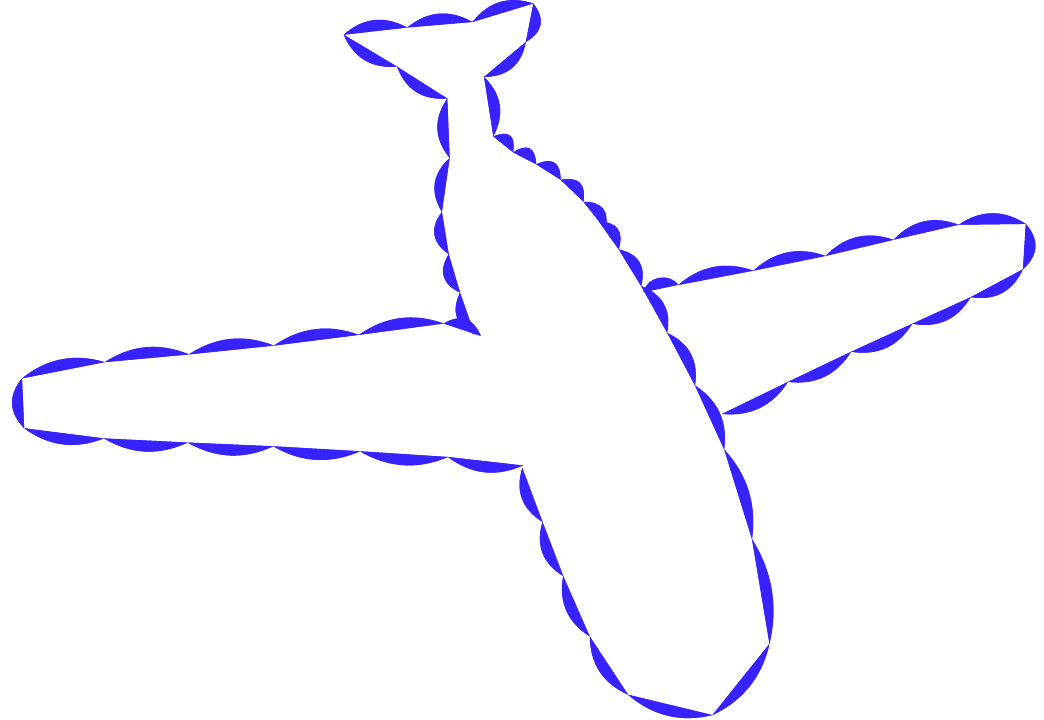}
\includegraphics[width=0.3\columnwidth]{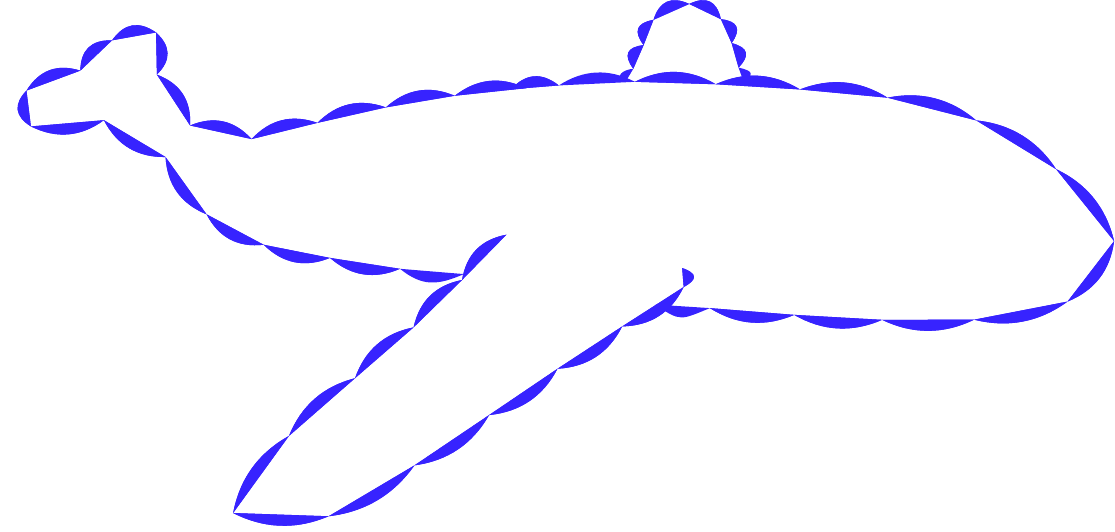}
\includegraphics[width=0.3\columnwidth]{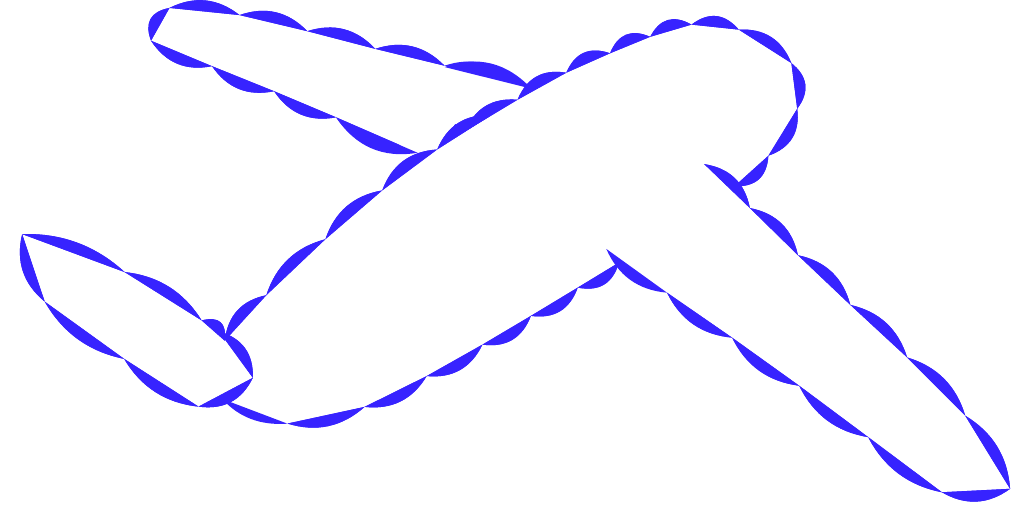}
}
%\centerline{
%\includegraphics[width=0.3\columnwidth]{fig/comparison/torus-fluffy1}
%\includegraphics[width=0.3\columnwidth]{fig/comparison/torus-fluffy2}
%\includegraphics[width=0.3\columnwidth]{fig/comparison/torus-fluffy3}
%}
%\centerline{
%\includegraphics[width=0.3\columnwidth]{fig/comparison/torus2}
%\includegraphics[width=0.3\columnwidth]{fig/comparison/torus3}
%\includegraphics[width=0.3\columnwidth]{fig/comparison/torus6}
%}
\centerline{
\includegraphics[width=0.3\columnwidth]{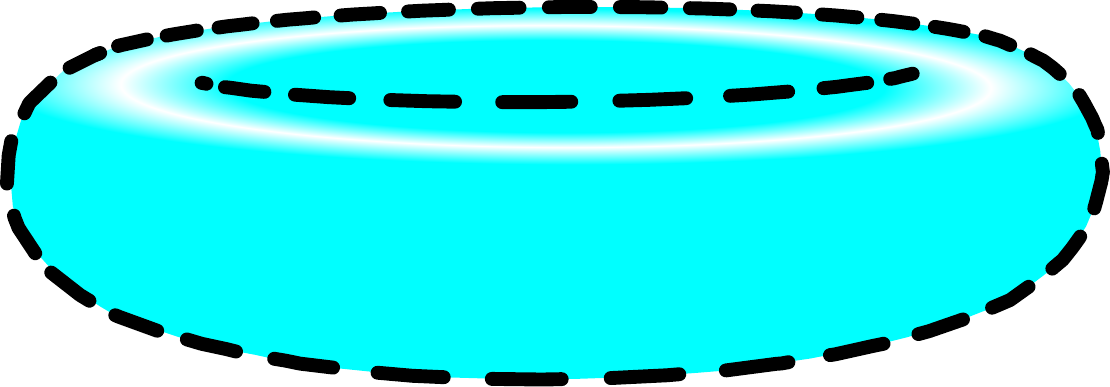}
\includegraphics[width=0.3\columnwidth]{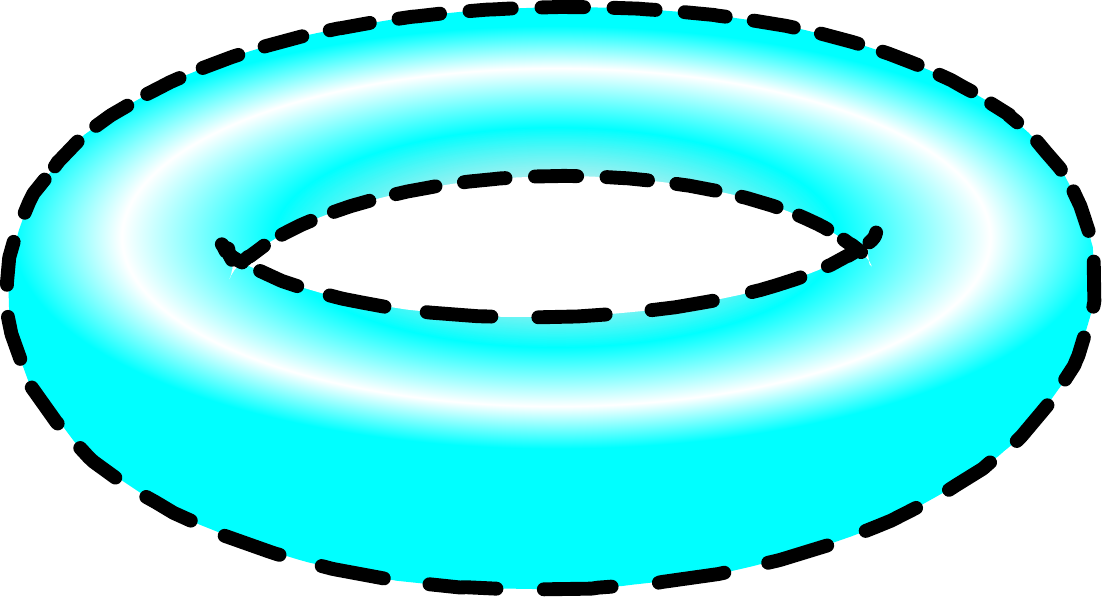}
\includegraphics[width=0.3\columnwidth]{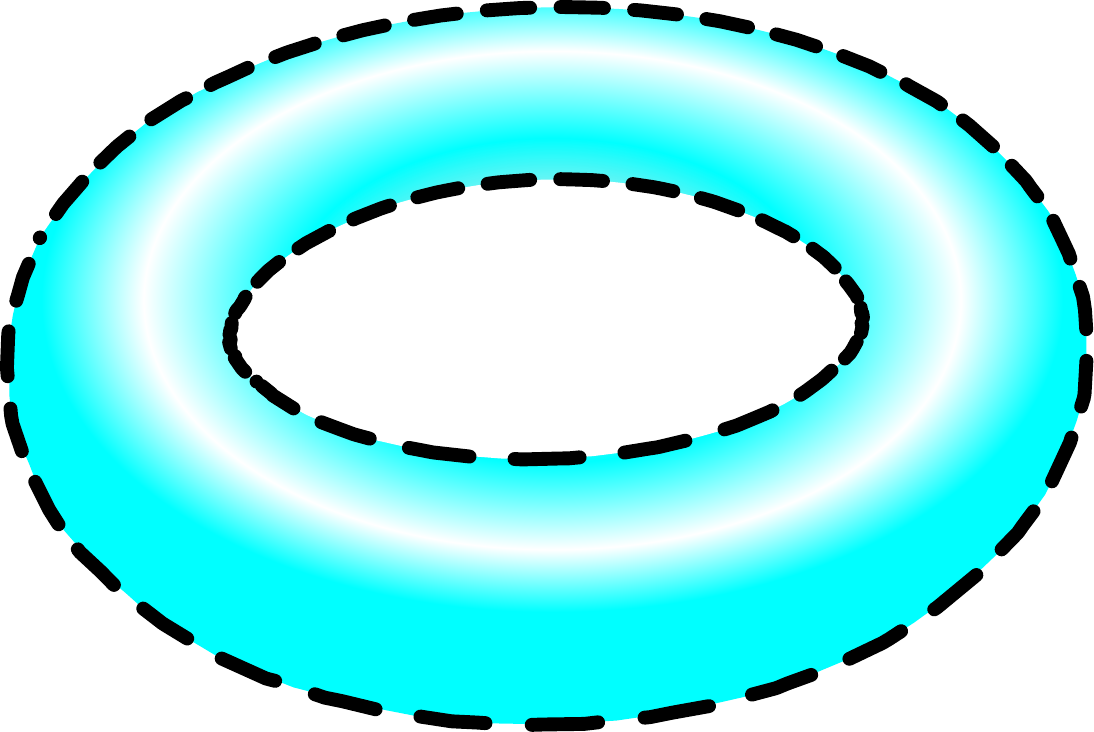}
}
\caption[]{Snaxels can be used to maintain temporally coherent silhouettes,
implementing Kalnins et al.~\shortcite{kalnins03} using the snaxels as contour
particles, parameterizing the dashed silhouettes in 3-D so they appear
fastened to the surface.}
\label{fig:comparison}
\end{figure}

If a snaxel simply moves along its mesh edge, then a correspondence can be
easily drawn between its start position and end position. When a snaxel is
deleted, we do not actually delete it but instead colocate it with the closer
of its two neighbors in the contour. When a snaxel undergoes a fan-out
operation, we replace the snaxel with multiple colocated copies of the snaxel,
each copy corresponding to one of the snaxels generated by the fan-out 
(as in Fig~\ref{fig:coherence}). Often a
fan-out generates several subsequent deletion events. We detect this case to
avoid excessive duplication and colocation of snaxels. These snaxels, along
with their colocated copies, become the vertices of animated
polylines representing the 2-D curve motion of the 3-D contour on the animated
mesh surface. Figure~\ref{fig:2dcurveanim} shows two animation sequences converted into 2-D keyframed animation using our technique.

With the tracking method presented here, we can use snaxels to reproduce the
coherent stylized silhouettes of Kalnins et al.~\shortcite{kalnins03}, as
demonstrated in Figure~\ref{fig:comparison}. Because
we know the correspondences and the snaxels' 3-D position as well as their 2-D
positions in the image space, we can sample the polylines defined by the
snaxels (in 2D, 3D, or a mixture of both) to achieve temporally coherent
silhouettes. Furthermore, when snaxel contours change topology, the
snaxel positions provide a basis for reparameterizing the contours with a
reasonable level of coherence across a topology change or visual event,
avoiding the ``pop'' discussed by Kalnins et al. \shortcite{kalnins03}.

\section{Conclusion} \label{sec:conc} \vspace{-2mm}
We have shown that snaxels, a formulation for active contours on unstructured
meshes, supports a variety of useful operations for constructing stylized
vector art illustrations of meshed surfaces. The flexible topology of snaxel
fronts allows them to evolve into visual and other contours, and to track
these contours across various changes in shape, view and lighting. Adjusting
the snaxel rules allows them to produce a useful visible surface planar map,
and by tracking the snaxel positions themselves as a shape is animated, the
3-D motion of the surface contours can be converted into 2-D motion of
animated SVG curves and polylines.

\begin{figure*}[t]
\centerline{
\includegraphics[height=33mm]{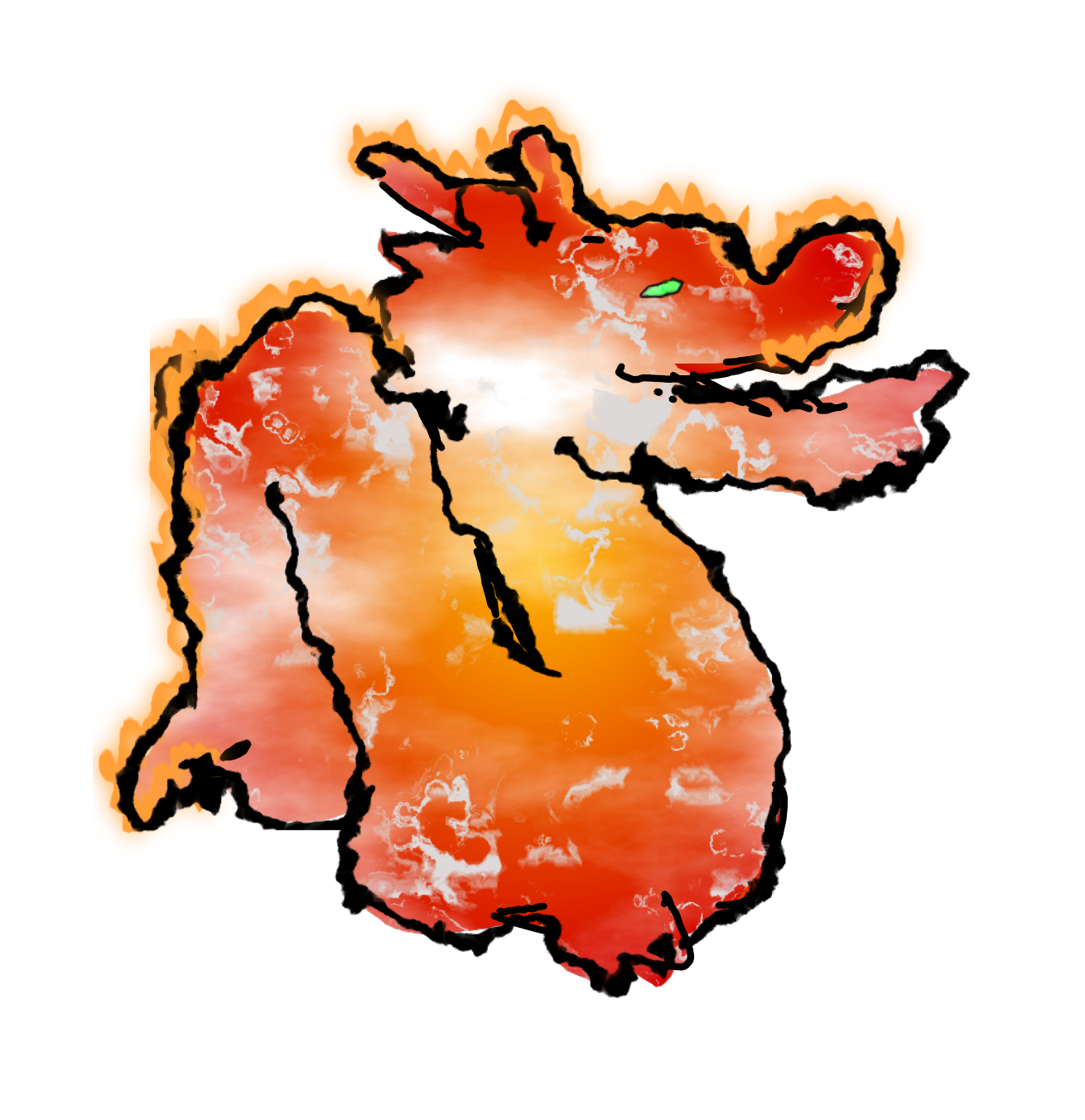} 
\includegraphics[height=33mm]{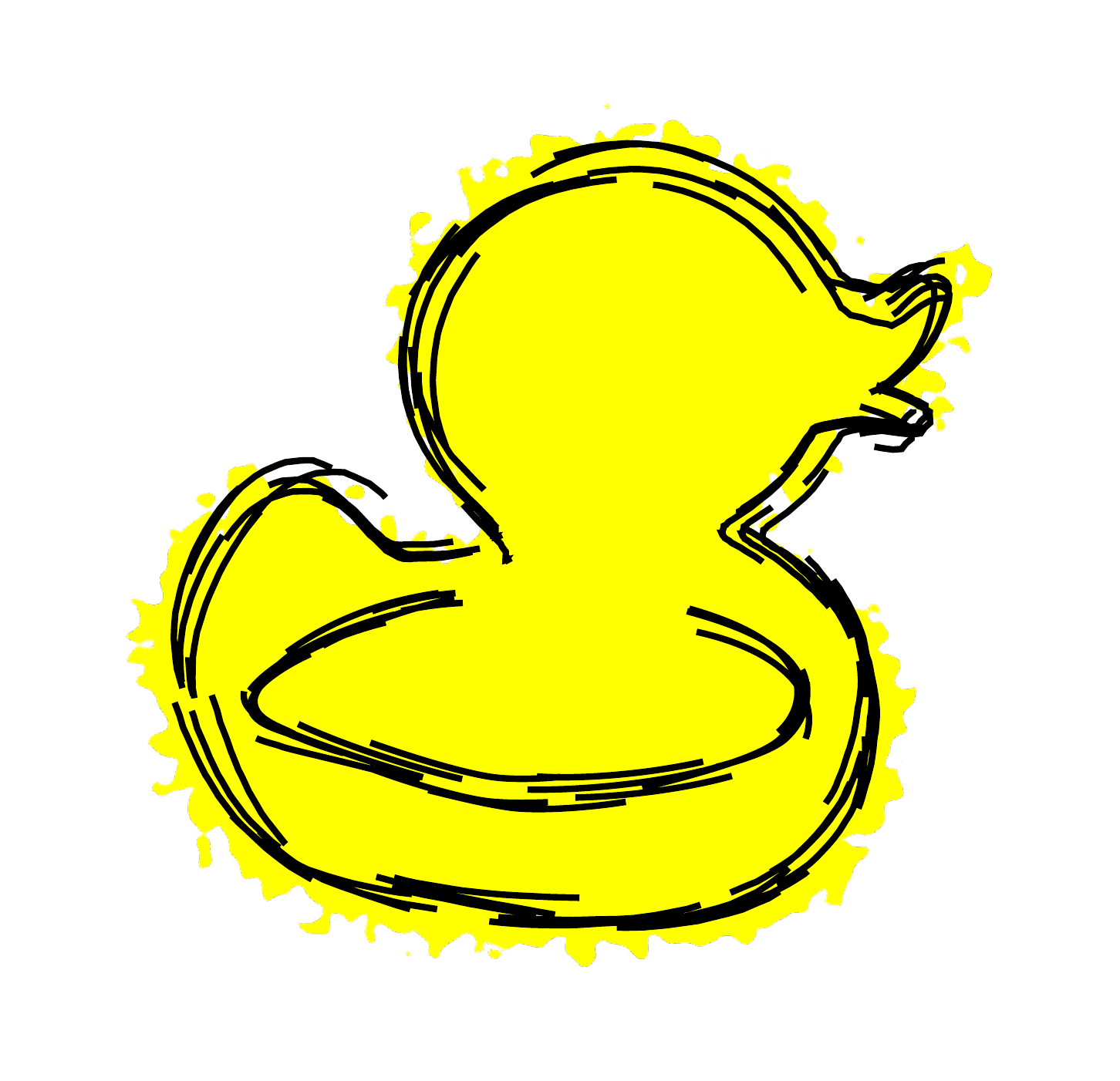}
\includegraphics[height=33mm]{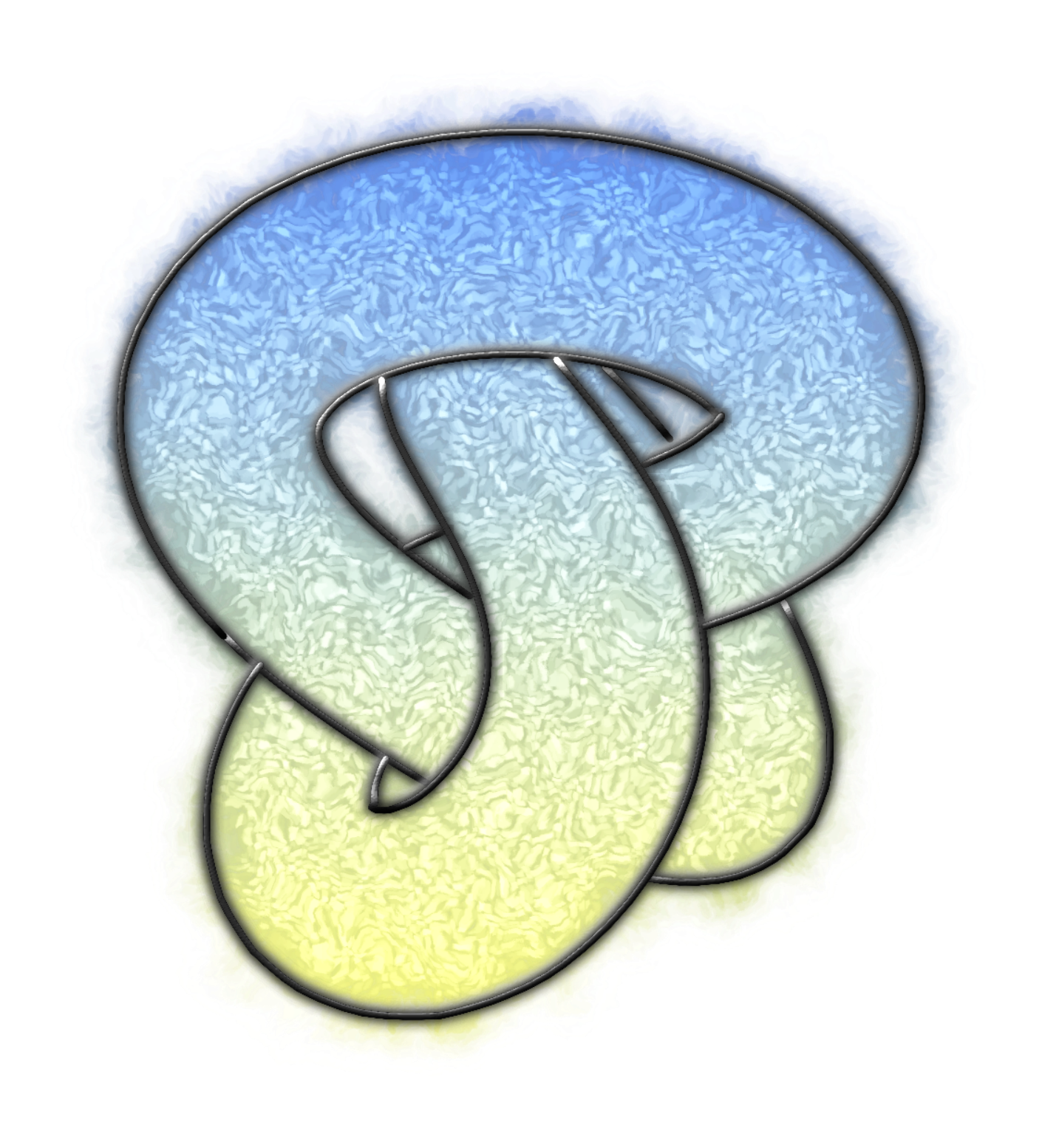}\hspace{2mm}
\includegraphics[height=33mm]{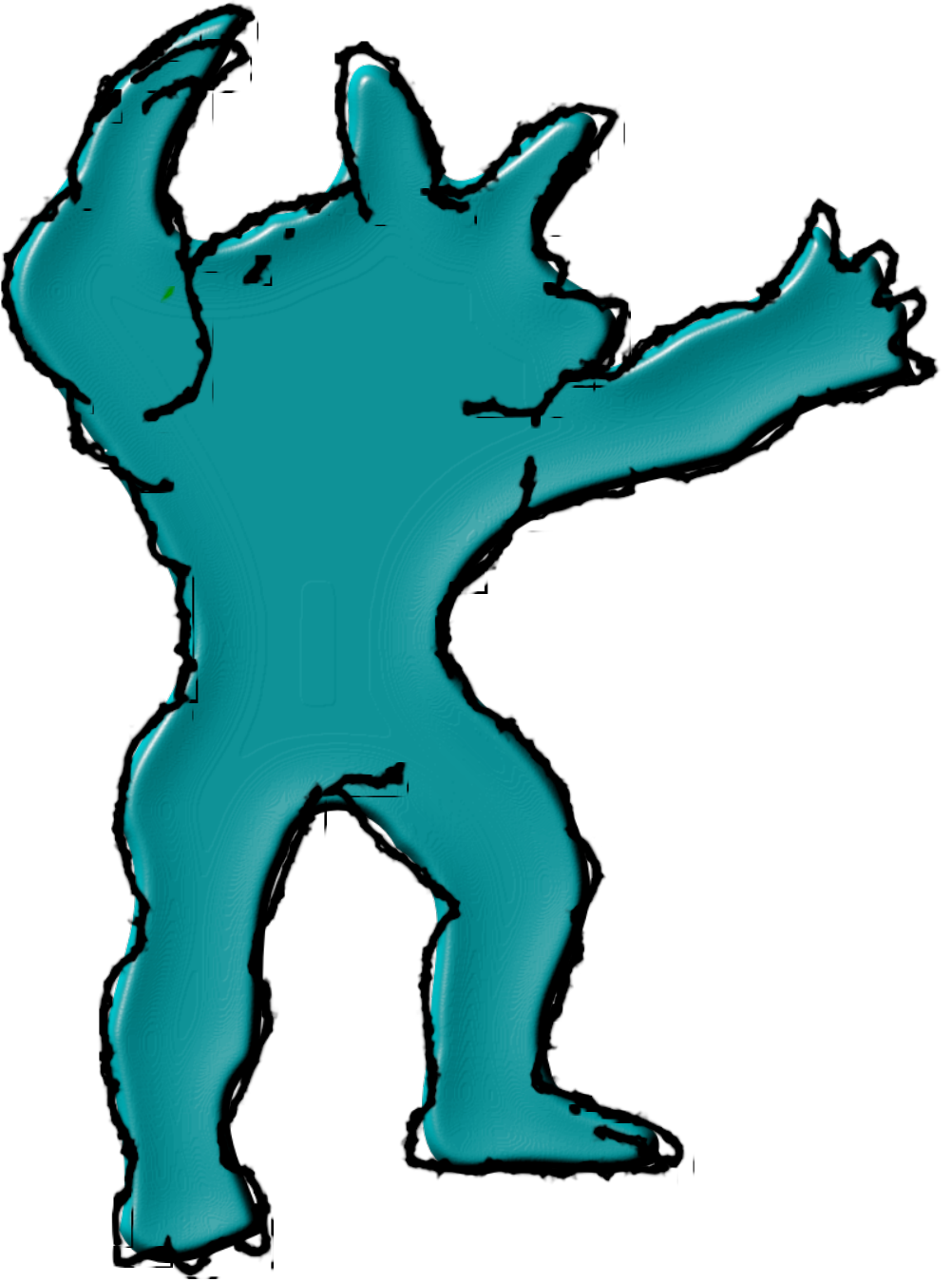}\hspace{2mm}
\includegraphics[height=33mm]{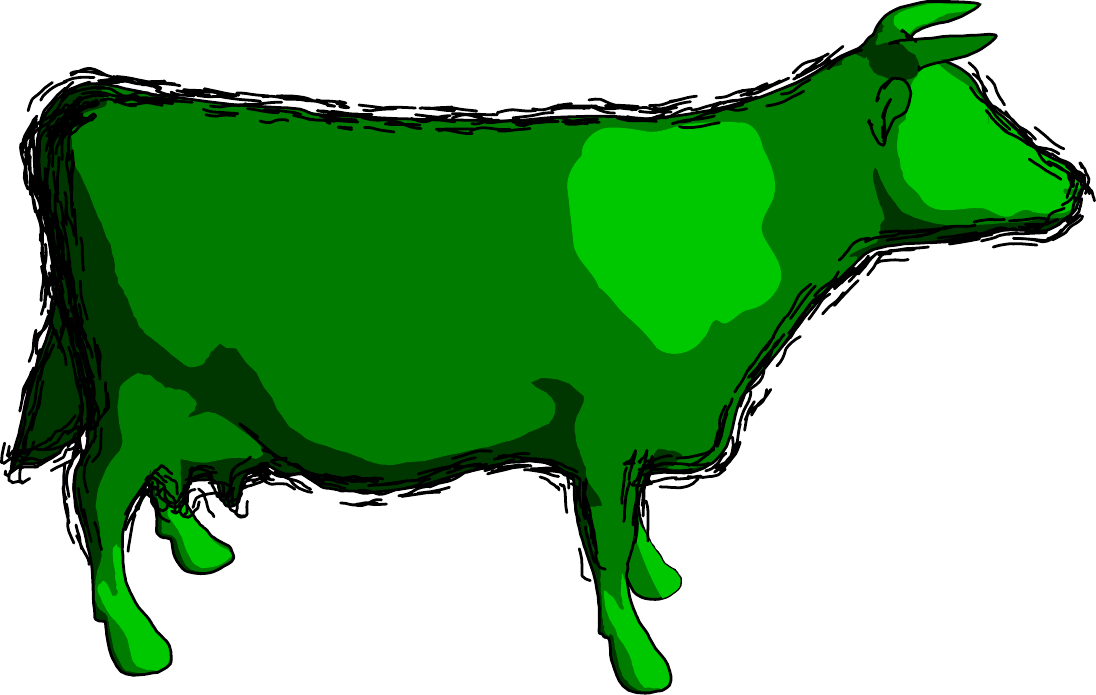}
}
\centerline{
\includegraphics[height=30mm]{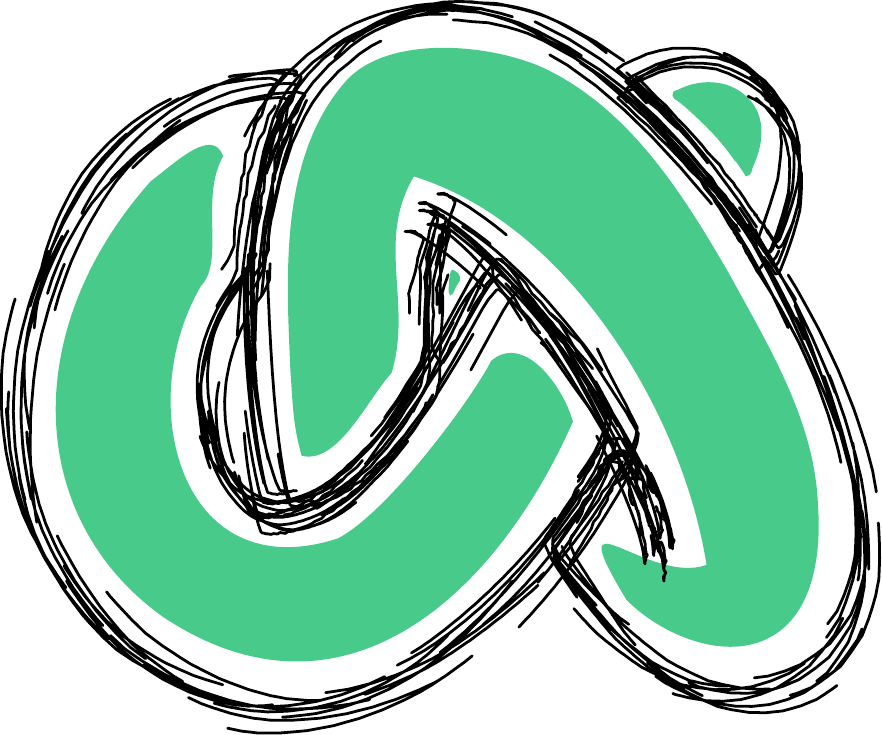}  \hspace{2mm}
\includegraphics[height=30mm]{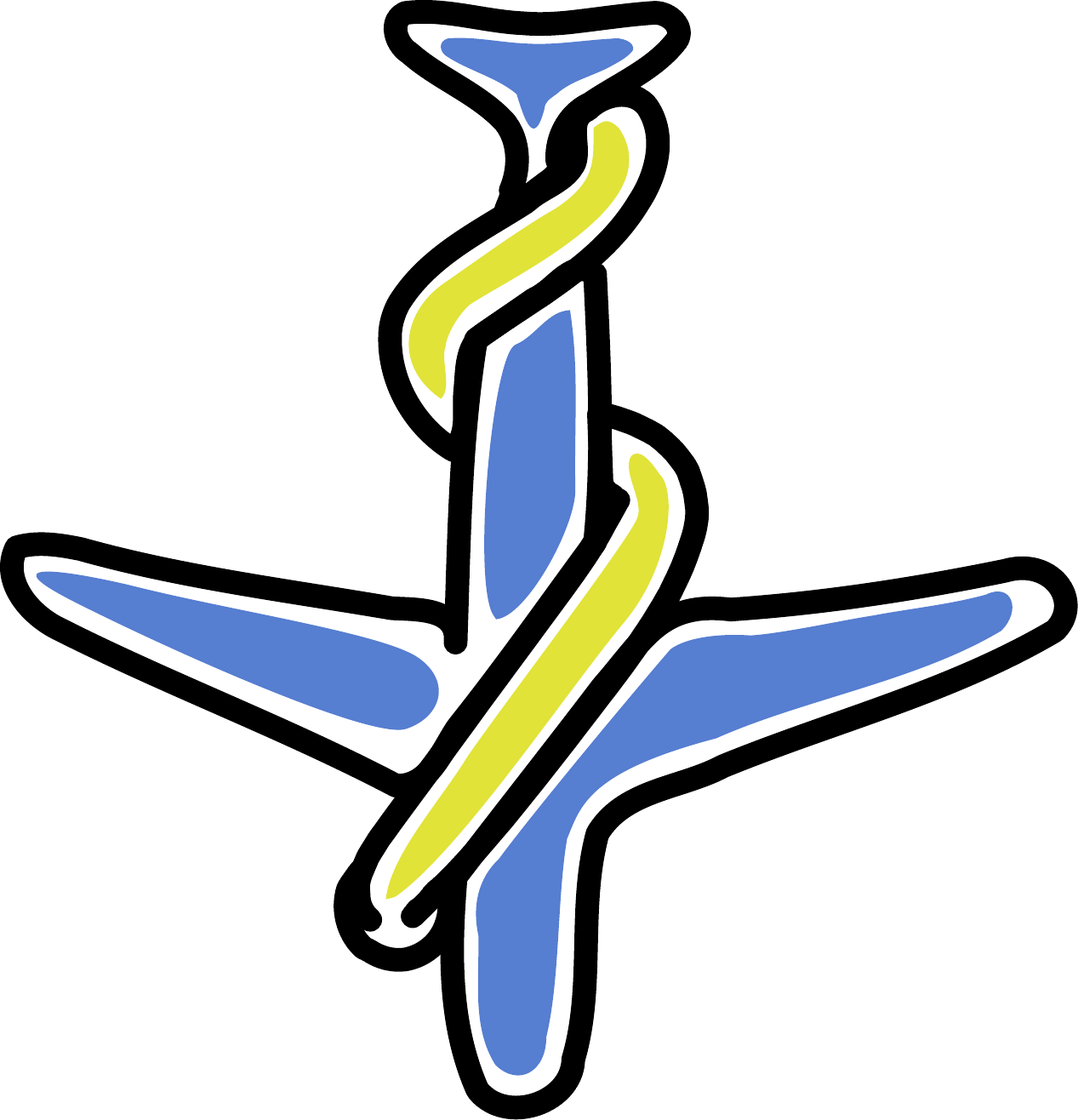}  \hspace{2mm}
\includegraphics[height=30mm]{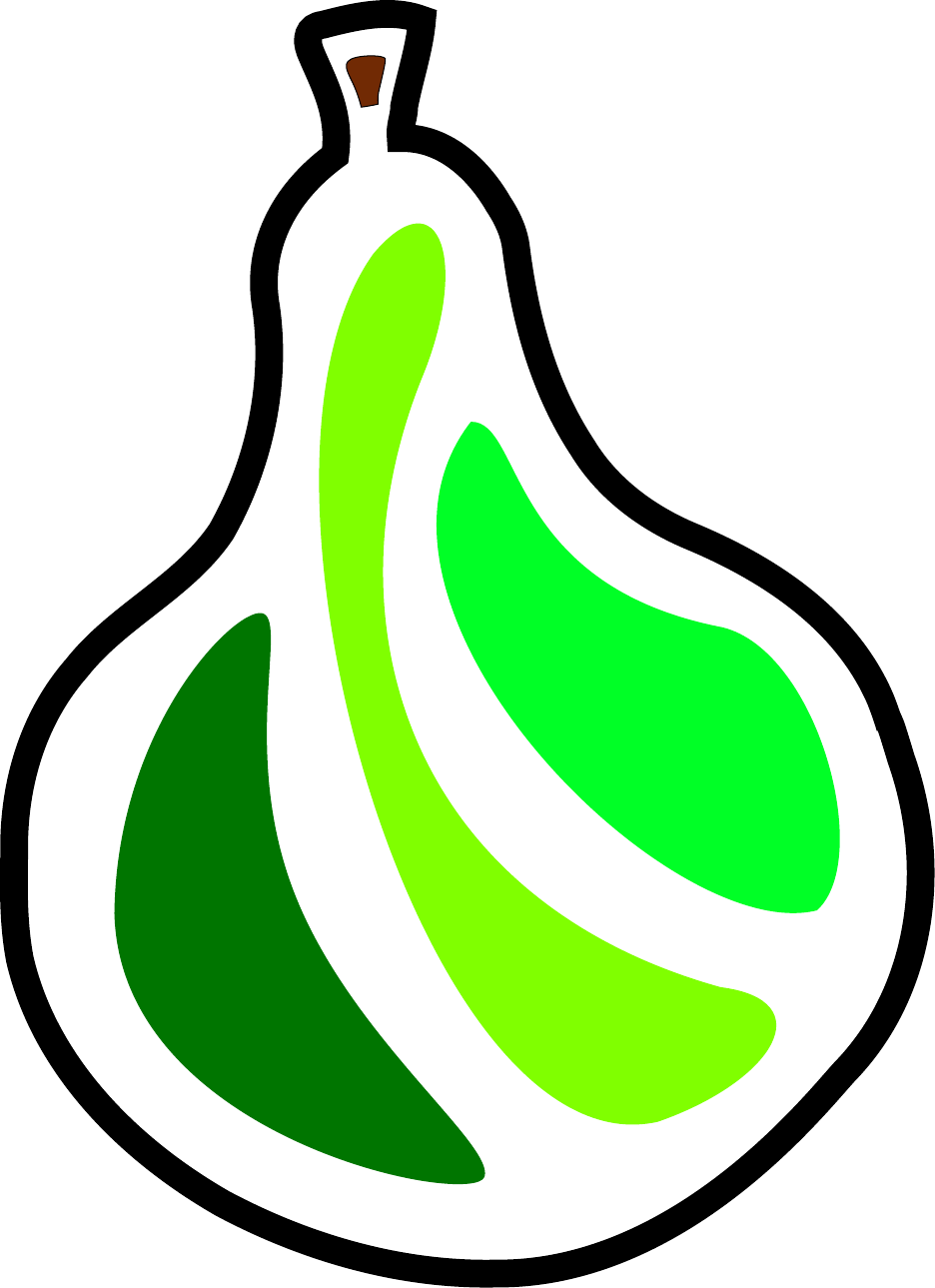}\hspace{2mm}
\includegraphics[height=30mm]{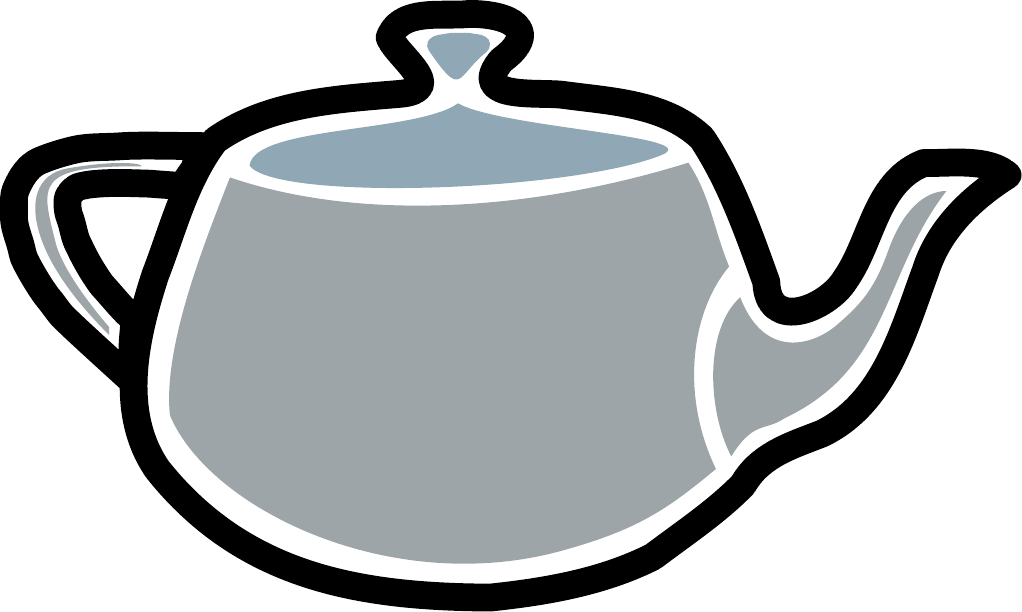}  \hspace{2mm}
\includegraphics[height=30mm]{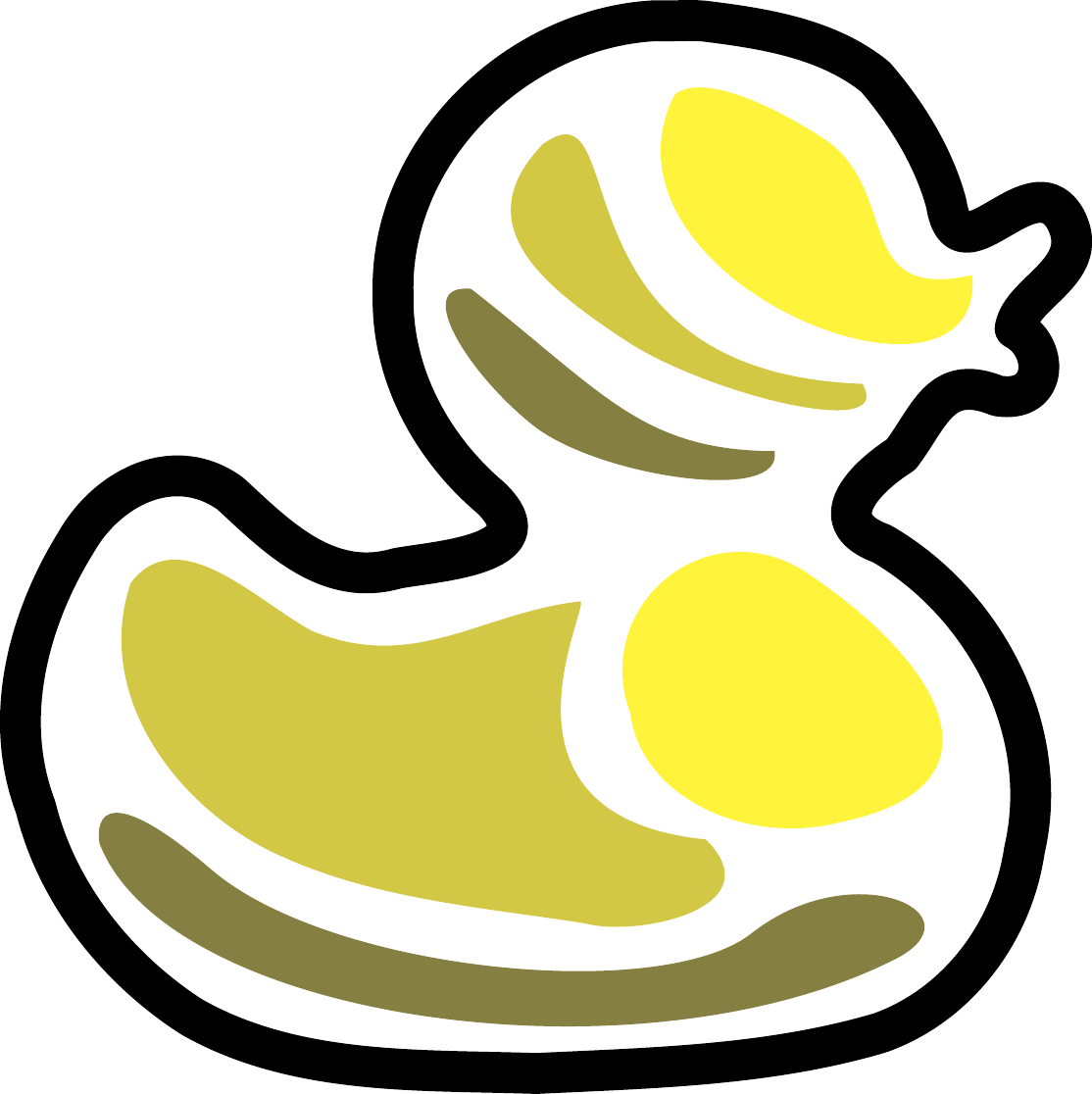} 
}
\caption{Examples of stylized output generated by our system. Snaxels provide a simple mechanism for creating stylized clip-art, such as in Eisemann et al.~\shortcite{Eisemann_cgf08} \emph{(top row)}. The bottom row shows results created with our planar map generation scheme. These results were obtained by morphologically eroding the planar map contours (using the method described in Section~\ref{sec:planar}). \vspace{-1mm}}
\label{fig:stylized}
\end{figure*}

Snaxels also provide a simple and quick mechanism for achieving other 
state-of-the-art results, such as coherent stylized silhouettes 
\cite{kalnins03} (Fig.~\ref{fig:comparison}) and the 
mesh-based stylized vector art of Eisemann et al.~\shortcite{Eisemann_cgf08}
(Fig.~\ref{fig:stylized}).

Perhaps the most compelling aspect of the snaxel formulation is that it can be
expressed with a few simple rules, and is easy to implement on an edge-based
mesh representation. Our CPU implementation achieves interactive speeds
for meshes with up to 100K faces, which sufficed for all results in this paper. 
Larger meshes could be supported by a parallel snaxel
implementation, e.g. on the GPU. The snaxel update process can be directly
parallelized, though fan-out, deletion and topology changes pose challenges
for efficient streaming parallelization due to their irregular control flow.

\begin{figure}
\centerline{ \includegraphics[width=.75\columnwidth]{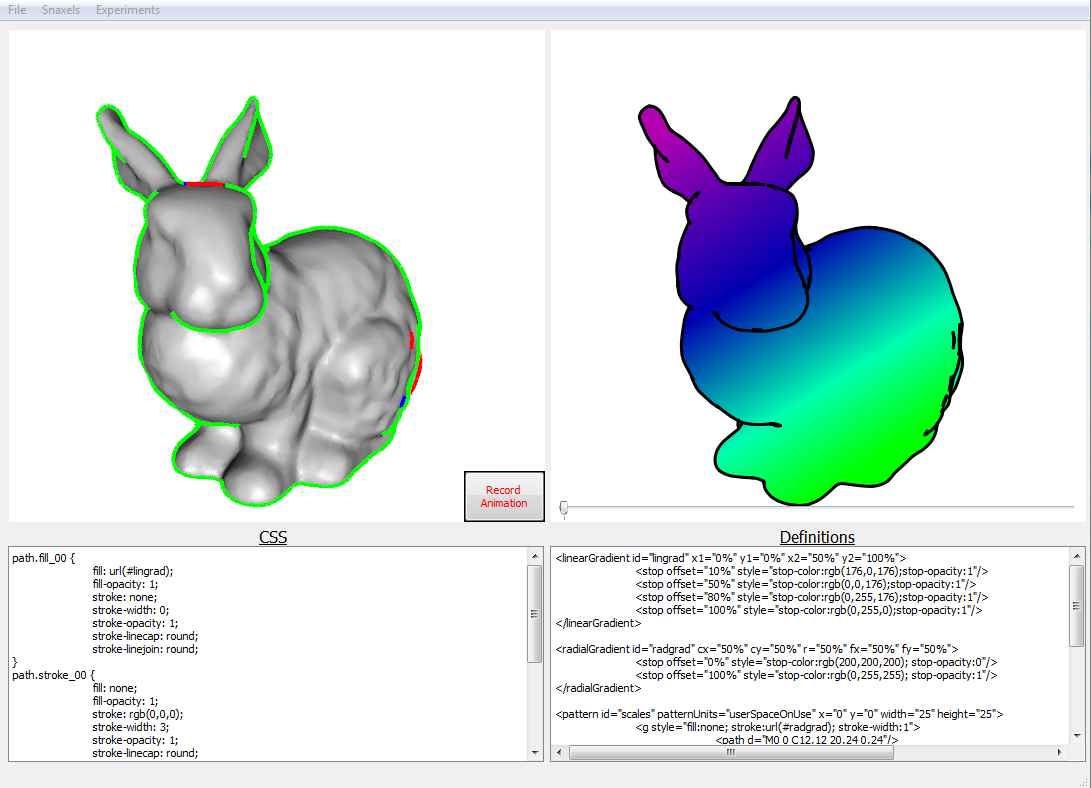} } \vspace{-2mm}
\caption{The user interface for our snaxel-based contour extraction system,
showing the 3-D mesh contours \emph{(upper left)}, their image-plane projection 
and stylization \emph{(upper right)} and editable XML code for dynamically 
specifying stylizations \emph{(bottom)}.\vspace{-3mm}}
\label{fig:interface}
\end{figure}

We developed an intuitive interface to capture mesh-based animations as vector
art, demonstrated in Figure~\ref{fig:interface}. The interface displays both
the 3-D snaxel contours on the surface and their projections (and regions) on
the view plane. The image rendering is SVG based, implemented with the Qt
framework, and includes text windows for editing
and applying various stylizations. Some examples of
our resulting stylizations are demonstrated in Figure~\ref{fig:stylized}.

This investigation of snaxels for NPR contouring has resulted in numerous
ideas for future research. The snaxel rules themselves could be stylized, to
enforce e.g. smoothness or shape constraints. A maximum area constraint 
on snaxel regions might yield a
cellular/scaled/bubble region stylization.

The conversion of 3-D surface contour motion into 2-D curve motion should be
explored further. Projected contours can be simplified from detailed polylines
to simpler, smoother piecewise cubic curves, and the motions of the polyline
vertices can be compiled, reduced and simplified into general control point
motions.

%\todo{Limitations?}

%We did not implement the time coherent stylization of Kalnins et al.
%\shortcite{kalnins03}. Our snaxel reproduction and colocation strategy
%used to create polyline vertex motion is similar to the strategies used for
%coherent stylization, and it would be interesting to re-implement time
%coherent contour stylization using this new snaxel framework.

\section*{Acknowledgements} \vspace{-2mm}
We thank Mahsa Kamali and Victor Lu for their advice and discussions, 
as well as the reviewers for their helpful comments.  This research was 
supported by the NDSEG Fellowship and the NSF Graduate 
Research Fellowship.

\bibliographystyle{acmsiggraph} 
\bibliography{npar11_bib,clipart}

\begin{thebibliography}{\protect\citename{Markosian et~al\mbox{.} }1997}

\bibitem[\protect\citename{Asente et~al\mbox{.} }2007]{Asente:2007}
{\sc Asente, P., Schuster, M., and Pettit, T.}
\newblock 2007.
\newblock Dynamic planar map illustration.
\newblock {\em (Proc. SIGGRAPH) ACM Trans. Graph.\/}.

\bibitem[\protect\citename{Biermann et~al\mbox{.} }2001]{biermann01}
{\sc Biermann, H., Kristjansson, D., and Zorin, D.}
\newblock 2001.
\newblock Approximate boolean operations on free-form solids.
\newblock In {\em Proc. SIGGRAPH}, 185--194.

\bibitem[\protect\citename{Bischoff et~al\mbox{.} }2005]{Bischoff:2005va}
{\sc Bischoff, S., Weyand, T., and Kobbelt, L.}
\newblock 2005.
\newblock {Snakes on triangle meshes}.
\newblock {\em Bildverarbeitung f{\"u}r die Medizin\/}, 208--212.

\bibitem[\protect\citename{Bourdev }1998]{Bourdev:1998wj}
{\sc Bourdev, L.}
\newblock 1998.
\newblock {Rendering nonphotorealistic strokes with temporal and arc-length
  coherence}.
\newblock {\em Master's thesis, Brown University\/}.

\bibitem[\protect\citename{Burns et~al\mbox{.} }2005]{burns05}
{\sc Burns, M., Klawe, J., Rusinkiewicz, S., Finkelstein, A., and DeCarlo, D.}
\newblock 2005.
\newblock Line drawings from volume data.
\newblock In {\em Proc. SIGGRAPH, ACM TOG}, 512--518.

\bibitem[\protect\citename{DeCarlo et~al\mbox{.} }2003]{decarlo2003}
{\sc DeCarlo, D., Finkelstein, A., Rusinkiewicz, S., and Santella, A.}
\newblock 2003.
\newblock Suggestive contours for conveying shape.
\newblock {\em Proc. SIGGRAPH, ACM TOG 22}, 3, 848--855.

\bibitem[\protect\citename{Eisemann et~al\mbox{.} }2008]{Eisemann_cgf08}
{\sc Eisemann, E., Winnem{\"o}ller, H., Hart, J.~C., and Salesin, D.}
\newblock 2008.
\newblock {Stylized Vector Art from 3D Models with Region Support}.
\newblock {\em Computer Graphics Forum 27}, 4 (jun).

\bibitem[\protect\citename{Eisemann et~al\mbox{.} }2009]{Eisemann:2009}
{\sc Eisemann, E., Paris, S., and Durand, F.}
\newblock 2009.
\newblock A visibility algorithm for converting 3d meshes into editable 2d
  vector graphics.
\newblock {\em (Proc. SIGGRAPH) ACM Trans. Graph. 28\/}, 83:1--83:8.

\bibitem[\protect\citename{Elber }1998]{elber98}
{\sc Elber, G.}
\newblock 1998.
\newblock Line art illustrations of parametric and implicit forms.
\newblock {\em IEEE TVCG 4}, 1, 71--81.

\bibitem[\protect\citename{Flato et~al\mbox{.} }2000]{fhhn-dipmc-00}
{\sc Flato, E., Halperin, D., Hanniel, I., Nechushtan, O., and Ezra, E.}
\newblock 2000.
\newblock The design and implementation of planar maps in {CGAL}.
\newblock {\em ACM J. Experimental Algorithmics 5\/}.

\bibitem[\protect\citename{Gangnet et~al\mbox{.} }1989]{Gangnet:1989}
{\sc Gangnet, M., Herv\'{e}, J.-C., Pudet, T., and van Thong, J.-M.}
\newblock 1989.
\newblock Incremental computation of planar maps.
\newblock In {\em Proc. SIGGRAPH}, 345--354.

\bibitem[\protect\citename{Grabli et~al\mbox{.} }2004]{grabli04}
{\sc Grabli, S., Turquin, E., Durand, F., and Sillion, F.}
\newblock 2004.
\newblock Programmable style for {NPR} line drawing.
\newblock In {\em Proc. EGSR}.

\bibitem[\protect\citename{Hertzmann and Zorin }2000]{hertzmann2000}
{\sc Hertzmann, A., and Zorin, D.}
\newblock 2000.
\newblock Illustrating smooth surfaces.
\newblock In {\em Proc. SIGGRAPH}, 517--526.

\bibitem[\protect\citename{Isenberg et~al\mbox{.} }2003]{isenberg03}
{\sc Isenberg, T., Freudenberg, B., Halper, N., Schlechtweg, S., and
  Strothotte, T.}
\newblock 2003.
\newblock {A} {D}eveloper's {G}uide to {S}ilhouette {A}lgorithms for
  {P}olygonal {M}odels.
\newblock {\em IEEE CG\&A 23}, 4, 28--37.

\bibitem[\protect\citename{Judd and Durand }2007]{judd2007}
{\sc Judd, T., and Durand, F.}
\newblock 2007.
\newblock Apparent ridges for line drawings.
\newblock {\em Proc. SIGGRAPH, ACM TOG 26}, 3, Article 19.

\bibitem[\protect\citename{Jung and Kim }2004]{Jung:2004uf}
{\sc Jung, M., and Kim, H.}
\newblock 2004.
\newblock {Snaking across 3d meshes}.
\newblock {\em Computer Graphics and Applications\/}, 87--93.

\bibitem[\protect\citename{Kalnins et~al\mbox{.} }2003]{kalnins03}
{\sc Kalnins, R.~D., Davidson, P.~L., Markosian, L., and Finkelstein, A.}
\newblock 2003.
\newblock Coherent stylized silhouettes.
\newblock {\em Proc. SIGGRAPH, ACM TOG 22}, 3, 856--861.

\bibitem[\protect\citename{Kass et~al\mbox{.} }1988]{Kass:1988tk}
{\sc Kass, M., Witkin, A., and Terzopoulos, D.}
\newblock 1988.
\newblock {Snakes: Active contour models}.
\newblock {\em IJCV 1}, 4, 321--331.

\bibitem[\protect\citename{Liu et~al\mbox{.} }2006]{Liu:2006uu}
{\sc Liu, S., Martin, R., Langbein, F., and Rosin, P.}
\newblock 2006.
\newblock {Segmenting reliefs on triangle meshes}.
\newblock {\em Proceedings of the 2006 ACM symposium on Solid and physical
  modeling\/}, 16.

\bibitem[\protect\citename{Markosian et~al\mbox{.} }1997]{markosian97}
{\sc Markosian, L., Kowalski, M.~A., Goldstein, D., Trychin, S.~J., Hughes,
  J.~F., and Bourdev, L.~D.}
\newblock 1997.
\newblock Real-time nonphotorealistic rendering.
\newblock In {\em Proc. SIGGRAPH, ACM TOG}, 415--420.

\bibitem[\protect\citename{Murta }1999]{Murta}
{\sc Murta, A.}, 1999.
\newblock {GPC:} general polygon clipper.
\newblock Software library.

\bibitem[\protect\citename{Ohtake et~al\mbox{.} }2004]{ohtake04}
{\sc Ohtake, Y., Belyaev, A., and Seidel, H.-P.}
\newblock 2004.
\newblock Ridge-valley lines on meshes via implicit surface fitting.
\newblock {\em (Proc. SIGGRAPH) ACM Trans. Graph. 23\/}, 609--612.

\bibitem[\protect\citename{Olson and Zhang }2006]{olson2006}
{\sc Olson, M., and Zhang, H.}
\newblock 2006.
\newblock {S}ilhouette {E}xtraction in {H}ough {S}pace.
\newblock {\em Proc. Eurographics, CGF 25}, 3 (Sept.), 273--282.

\bibitem[\protect\citename{Orzan et~al\mbox{.} }2008]{Orzan:2008}
{\sc Orzan, A., Bousseau, A., Winnem\"{o}ller, H., Barla, P., Thollot, J., and
  Salesin, D.}
\newblock 2008.
\newblock Diffusion curves: a vector representation for smooth-shaded images.
\newblock {\em (Proc. SIGGRAPH) ACM Trans. Graph. 27\/}, 92:1--92:8.

\bibitem[\protect\citename{Rusinkiewicz }2008]{rusink08}
{\sc Rusinkiewicz, S.}
\newblock 2008.
\newblock Algorithms for extracting lines.
\newblock In {\em Line Drawings from 3D Models}. SIGGRAPH Course Notes.

\bibitem[\protect\citename{Stroila et~al\mbox{.} }2008]{stroila2008}
{\sc Stroila, M., Eisemann, E., and Hart, J.~C.}
\newblock 2008.
\newblock Clip art rendering of smooth isosurfaces.
\newblock {\em IEEE TVCG 14}, 1, 71--81.

\bibitem[\protect\citename{Su and Hart }2005]{su05}
{\sc Su, W.~Y., and Hart, J.~C.}
\newblock 2005.
\newblock A programmable particle system framework for shape modeling.
\newblock In {\em Proc. SMI}, 114--123.

\bibitem[\protect\citename{Vatti }1992]{Vatti92}
{\sc Vatti, B.}
\newblock 1992.
\newblock A generic solution to polygon clipping.
\newblock {\em CACM 35}, 7, 56--63.

\bibitem[\protect\citename{Wein et~al\mbox{.} }2005]{wfzh-aptac-05}
{\sc Wein, R., Fogel, E., Zukerman, B., and Halperin, D.}
\newblock 2005.
\newblock Advanced programming techniques applied to {CGAL}'s arrangement
  package.
\newblock In {\em Proc. Library-Centric Software Design Workshop (LCSD'05)}.

\bibitem[\protect\citename{Winkenbach and Salesin }1994]{winkenbach94}
{\sc Winkenbach, G., and Salesin, D.~H.}
\newblock 1994.
\newblock Computer-generated pen-and-ink illustration.
\newblock {\em Proc. SIGGRAPH, Computer Graphics 28\/}, 91--100.

\bibitem[\protect\citename{Winkenbach and Salesin }1996]{winkenbach96}
{\sc Winkenbach, G., and Salesin, D.~H.}
\newblock 1996.
\newblock Rendering parametric surfaces in pen and ink.
\newblock {\em Proc. SIGGRAPH, Computer Graphics 30\/}, 469--476.

\end{thebibliography}
\end{document}